\documentclass[aps, prl, twocolumn, amsmath, amssymb, nofootinbib]{revtex4-2}
\usepackage{bm}
\usepackage{amsmath}
\usepackage{amsfonts}
\usepackage{dsfont}
\usepackage{graphics}
\usepackage{braket}
\usepackage{float}
\usepackage[colorlinks, allcolors=blue]{hyperref}
\usepackage[caption=false]{subfig}             
\usepackage{overpic}
\usepackage[capitalise]{cleveref}
\usepackage{docmute}

\newcommand{\diff}{\mathrm{d}}
\newcommand{\Diff}{\mathrm{D}}
\newcommand{\mean}[1]{\left\langle#1\right\rangle}

\newcommand{\T}{\mathrm{T}}
\newcommand{\Tr}{\operatorname{Tr}}

\newcommand{\Nabla}{\bm{\nabla}}
\newcommand{\re}{\operatorname{Re}}
\newcommand{\im}{\operatorname{Im}}

\begin{document}
\title{Topological effect on the Anderson transition in chiral symmetry classes}
\author{Pengwei Zhao}
\affiliation{International Center for Quantum Materials, Peking University, Beijing 100871, China}
\author{Zhenyu Xiao}
\affiliation{International Center for Quantum Materials, Peking University, Beijing 100871, China}
\author{Yeyang Zhang}
\affiliation{International Center for Quantum Materials, Peking University, Beijing 100871, China}
\author{Ryuichi Shindou}
\email{rshindou@pku.edu.cn}
\affiliation{International Center for Quantum Materials, Peking University, Beijing 100871, China}
\date{\today}

\begin{abstract}
    In this Letter, we propose a mechanism of an emergent quasi-localized phase in chiral symmetry classes, where wave function along a spatial direction with weak topology is delocalized but exponentially localized along the other directions. The Anderson transition in 2D chiral symmetry classes is induced by the proliferation of vortex-antivortex pairs of a U(1) phase degree of freedom, while the weak topology endows the pair with the Berry phase. We argue that the Berry phase induces spatial polarizations of the pairs along the topological direction through the quantum interference effect, and the proliferation of the polarized vortex pairs results in the quasi-localized phase.
\end{abstract}

\maketitle
\textit{Introduction.}---Topology has played a pivotal role in the development of physics~\cite{wenQuantumFieldTheory2004,nakaharaGeometryTopologyPhysics2018,qiTopologicalInsulatorsSuperconductors2011,chiuClassificationTopologicalQuantum2016,hasanColloquiumTopologicalInsulators2010a}, particularly in understanding topological phases. In contrast to symmetry-breaking phases in Landau’s theory of phase transitions, topological phases are comprehensively elucidated by topological field theory that incorporates crucial topological terms.
Pioneering instances are integer and fractional quantum Hall 
states whose quantized Hall responses are explained 
through $\theta$-terms and Chern-Simons terms~\cite{pruiskenQuasiParticlesTheory1987,zhangEffectiveFieldTheoryModelFractional1989,niemiAxialAnomalyInducedFermionFractionization1983,wenClassificationAbelianQuantum1992,lopezFractionalQuantumHall1991}. Nature of symmetry-protected topological phases in quantum Heisenberg spin chains are portrayed by a Lagrangian with Wess-Zumino terms~\cite{haldaneNonlinearFieldTheory1983,affleck1987}. 
A $\theta$-term in one-dimensional (1D) chiral symmetry class leads to topological phase transitions between distinct topological Anderson insulators~\cite{altlandQuantumCriticalityQuasiOneDimensional2014,altlandTopologyAndersonLocalization2015,altlandSpectralTransportProperties2001}. 

A weak topology is associated with a homotopy group of a field variable of a Lagrangian in a lower dimensional subspace of the spacetime. Valence bond solids 
and deconfined quantum criticality in two-dimensional (2D) quantum magnets 
have been explored through a Lagrangian with a weak topological  term~\cite{haldane1988,read1989,read1990,senthil2004}.  
Recent 
studies investigates the Anderson transition in chiral symmetry classes 
with and without weak topology
~\cite{konigMetalinsulatorTransitionTwodimensional2012,luoCriticalBehaviorAnderson2020,wang2021,karcher2023a,karcher2023b,xiaoAnisotropicTopologicalAnderson2023}, and  
revealed that the weak topology in three-dimensional (3D) chiral symmetry classes universally induces an emergent ``quasi-localized" phase that shows an extreme spatial anisotropy in its transport property~\cite{xiaoAnisotropicTopologicalAnderson2023}. In the quasi-localized phase, an exponential localization length is divergent in a spatial direction with the weak topology (``topological direction''), while it is finite along the other spatial directions. By the Hermitization~\cite{feinberg97,luoUnifyingAndersonTransitions2022}, such quasi-localized phases can also be realized in eigenstates of non-Hermitian operators in the fundamental symmetry classes. Despite its physical relevance, an underlying physical mechanism of the quasi-localization has been veiled in mystery. 

In this Letter, we shed light on a connection between the quasi-localization and the effect of a weak topological term in a field theory. Inspired by a theory 
showing that the Anderson transition in 2D chiral symmetry classes \cite{altlandNonstandardSymmetryClasses1997,chiuClassificationTopologicalQuantum2016,eversAndersonTransitions2008} is instigated by spatial proliferation of vortex excitations associated with the $\mathrm{U}(1)$ symmetry in the non-linear sigma model~\cite{konigMetalinsulatorTransitionTwodimensional2012}, we point out that a weak topology term confers a quantal (``Berry'') phase upon a vortex-antivortex pair, dependent on its spatial polarization relative to the topological direction~\cite{altlandQuantumCriticalityQuasiOneDimensional2014,altlandTopologyAndersonLocalization2015,tanakaShortGuideTopological2015}. The quantal phase results in destructive interference among configurations of those pairs that are not parallel to the topological direction, effectively polarizing the pairs along the topological direction. The proliferation of the polarized vortex-antivortex pairs renders correlations of a  field variable to be weakly and strongly disordered along the topological direction and the other direction(s), respectively. Consequently, the quasi-localized phase with the extreme spatial anisotropy always emerges next to the diffusive metal phase (critical metal phase in 2D case) in a 
global phase diagram for chiral symmetric models with weak topology. 

\begin{figure}[b]
    \centering
    \includegraphics[width=0.5\textwidth]
    {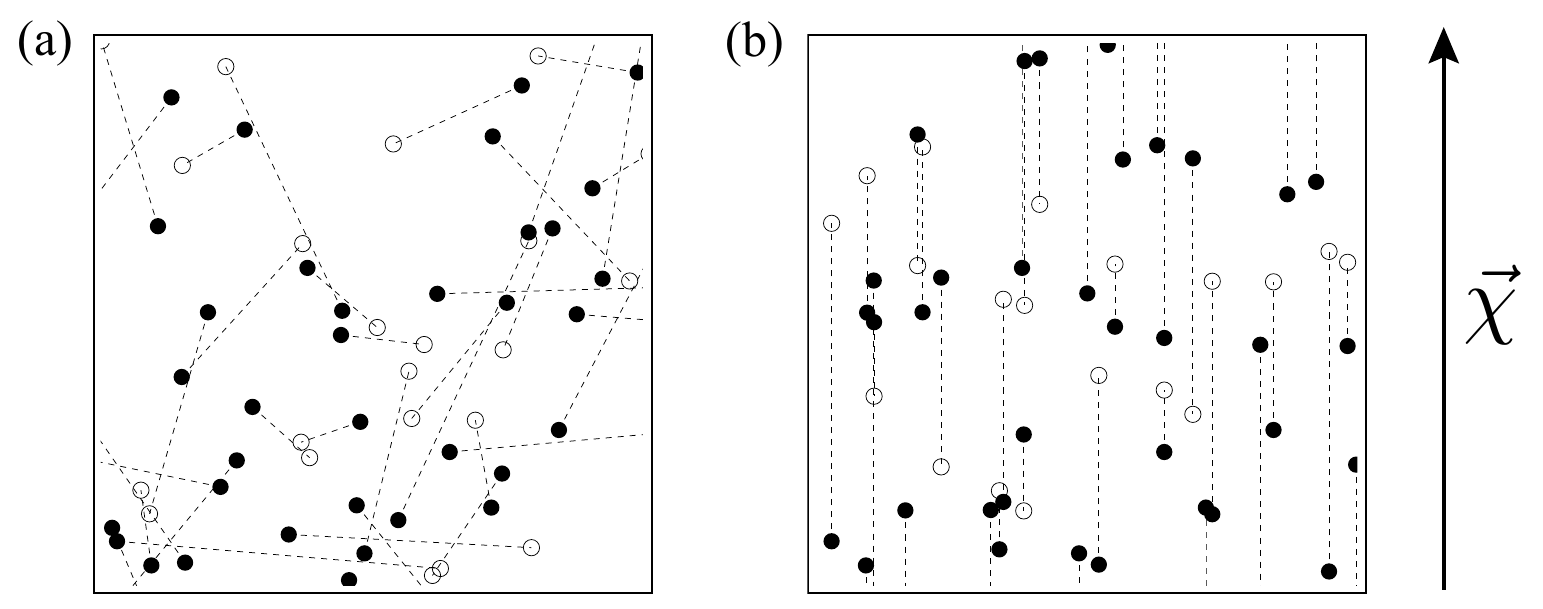}
    \caption{Schematic pictures of a spatial proliferation of vortex-antivortex pairs in 2D space, (a) without and (b) with weak 1D topology. Solid and hollow circles represent vortex and antivortex, respectively, and dashed lines are branchcuts for vortex-antivortex pairs.}
    \label{fig:polarization}
\end{figure}

To study the effect of the weak topological term theoretically, we focus on two-dimensional (2D) systems and carry out a renormalization group (RG) analysis \cite{cardyScalingRenormalizationStatistical1996,altlandCondensedMatterField2010} of a nonlinear sigma model (NLSM) in chiral unitary classes with weak topology. We show that an RG phase diagram is controlled by a stable fixed point for the quasi-localized phase, a stable fixed region for the (critical) metal phase, and a saddle-point fixed point for the quantum phase transition between the two phases. The emergence of the quasi-localized phase in 2D systems is also justified by a numerical simulation on a 2D lattice model in the chiral unitary class.

\emph{Field theory.}---Long-wavelength diffusion properties of disorder systems in chiral symmetry classes are described by the following NLSM~\cite{GADE1991213,GADE1993499,altlandQuantumCriticalityQuasiOneDimensional2014,altlandTopologyAndersonLocalization2015,konigMetalinsulatorTransitionTwodimensional2012} in a replica formalism for the partition function $Z \equiv \int {\rm D}[Q] e^{-S}$,  
\begin{equation}\label{eq:NLSM}
    \begin{aligned}
        S=-\int\frac{\diff^dr}{8\pi}\sum_{\mu=1}^{d}&\left[\sigma_\mu\Tr(Q^{-1}\partial_\mu Q)^2+c_\mu\Tr^2(Q^{-1}\partial_\mu Q)\right.\\
        &\left.-\chi_\mu\Tr(Q^{-1}\partial_\mu Q)\right].
    \end{aligned}
\end{equation}
A field variable $Q$ is an $N\times N$ matrix belonging to the Lie group $\mathrm{U}(N)$, $\mathrm{U}(N)/\mathrm{O}(N)$ and $\mathrm{U}(N)/\mathrm{Sp}(N)$ for chiral unitary, symplectic, and orthogonal classes, respectively. $N$ is a number of replicated fields~\cite{altlandCondensedMatterField2010}. $\sigma_\mu$, $c_\mu$ and $\chi_{\mu}$ are conductivity, Gade constant, and a weak topology index along $r_\mu$-direction, respectively. The Gade term generally appears in the NLSM in chiral symmetry classes~\cite{GADE1991213,GADE1993499}, while $\chi_{\mu}$ is finite only in chiral symmetric models with finite one-dimensional weak topological index $\nu_{\mu}$~\cite{altlandQuantumCriticalityQuasiOneDimensional2014,altlandTopologyAndersonLocalization2015}, and $\chi_{\mu}/8\pi$ in metallic phase takes non-integer values~\cite{SM}.
A perturbative $\beta$ function for the conductivity vanishes completely in chiral symmetry classes~\cite{GADE1991213,GADE1993499}, while it was shown that the 2D metal-insulator transition in chiral symmetry classes is driven by  proliferation of vortex excitations associated with the U(1) subgroup symmetry of the $Q$ field~\cite{konigMetalinsulatorTransitionTwodimensional2012}.

For $N=1$, the $Q$ field reduces to a unit complex number $e^{i\phi}$, where a vortex stands for a phase winding of $\phi$ around some spatial point (vortex core). For $N\ne 1$, eigenvalues of the $Q$ field are unit complex numbers, and a vortex excitation is nothing but the phase winding of one of the eigenvalues. When the others are close to 1, the $Q$ field configuration with a vortex excitation is given by $Q=\mathds{1}+\ket{p}(e^{i\phi}-1)\bra{p}$. An eigenvector $\ket{p}$ generally depends on the spatial coordinates. A $Q$ field configuration with multiple vortices is described by the same ansatz where $\phi$ has phase windings at different points. Due to the periodic boundary condition, vortex and antivortex appear in a pair for 2D ($d=2$), while in 3D, the vortex core forms a closed loop. 
The spatial proliferation of the vortex excitations destroys 
orderings of field variables as in the Berezinskii-Kosterlitz-Thouless  
transition~\cite{kt,Berezinskii71,Berezinskii72}, and the 
disordering of the $Q$ field results in the localization~\cite{altlandCondensedMatterField2010}.

\emph{Quantal phase interference.}---The weak topology term endows the vortex excitation with an imaginary number in the action $S$, yielding a unique interference effect in $Z$, 
\begin{equation}
    S=S_0+\frac{i}{4} \eta \chi_{\mu} \left\{\begin{array}{cc}
    \epsilon_{\mu\nu} {\bm m}_\nu  & (d=2), \\
    \epsilon_{\mu\nu\lambda} {\bm S}_{\nu\lambda} & (d=3). \\
    \end{array}\right.
 \end{equation}
Here $S_0$ comes from the first two terms in the NLSM and it is real-valued. A dipole vector $\bm{m}$ is associated with a vortex-antivortex pair with a vorticity $\eta$ in the 2D case, connecting vortex and antivortex in the 2D space. In the 3D case, a vortex loop with vorticity $\eta$ is characterized by an antisymmetric tensor ${\bm S}$ whose $(\mu,\nu)$-element is a projected area of the loop on the $\mu$-$\nu$ plane. Due to the imaginary number, vortex-antivortex dipoles with different angles and lengths bring about destructive interference mutually in the partition function in 2D, while dipoles parallel to $\vec{\chi}$ are free from the interference. In the partition function for 3D, vortex loops having finite projections in the plane normal to $\vec{\chi}$ induce the interference, and those confined in planes parallel to $\vec{\chi}$ do not. Due to the destructive interference, the vortex excitations polarized along $\vec{\chi}$ dominate the partition function near the Anderson transition in chiral symmetry classes (see Fig.~\ref{fig:polarization}). The spatial proliferation of the polarized vortex excitation does not destroy long-range spatial coherence of the U(1) phase along the topological direction, while it does destroy the phase coherence along the other direction(s)~\cite{SM}. Such spatially-anisotropic phase coherence leads to the strong spatial anisotropy of the diffusion; the transport along the topological direction remains diffusive, while the transport along the other direction(s) becomes insulating.

The consequence of the quantum interference effect is studied  
analytically studied by a renormalization group (RG) analysis. We take the 2D chiral unitary class (AIII) as an example and show that the NLSM with $\vec{\chi}\ne 0$ universally undergoes a phase transition from  (critical) metal phase to quasi-localized phase, the latter of which is characterized by a fixed point with finite conductivity along $\vec{\chi}$ and zero conductivity perpendicular to $\vec{\chi}$. To this end, we map the U($N$) NLSM into its dual theory~\cite{konigMetalinsulatorTransitionTwodimensional2012,SM}. The dual 
theory resembles the 2D sine-Gordon model which facilitates the RG analysis~\cite{konigMetalinsulatorTransitionTwodimensional2012,kogutIntroductionLatticeGauge1979,SM}. 

\emph{Dual theory.}---
To obtain the dual theory, we write the NLSM in terms of $h_{\mu} = -i Q^{-1}\partial_{\mu}Q$ with a local constraint that complies with the vortex excitations. The constraint can be treated by a Lagrange multiplier $\Theta$, that is an $N\times N$ Hermitian matrix field. After integrating over the $h_{\mu}$ field, we obtain the dual theory for the $\Theta$ field \cite{SM}, 
\begin{equation}\label{eq:dual_theory}
    S=S_{\mathrm{u}(1)}+S_{\mathrm{su}(N)}+S_y,
\end{equation}
\begin{equation}\label{eq:S0}
    S_{\mathrm{u}(1)}=2\pi\int\diff^2r\left[\frac{(\partial_x\theta^0)^2}{\sigma_y+Nc_y}+\frac{(\partial_y\theta^0)^2}{\sigma_x+Nc_x}\right],
\end{equation}
\begin{equation}\label{eq:S1}
    S_{\mathrm{su}(N)}=2\pi\int\diff^2r\sum_{a=1}^{N^2-1}\left[\frac{(\partial_x\theta^a)^2}{\sigma_y}+\frac{(\partial_y\theta^a)^2}{\sigma_x}\right]+\mathrm{O}\left(\frac{1}{\sigma_x\sigma_y}\right),
\end{equation}
\begin{equation}\label{eq:Sy}    S_y=-2y\Lambda^2\int\diff^2r\int\diff\ket{p}\cos\left(2\pi\braket{p|\Theta|p}+\epsilon_{\mu\nu}r_\mu\frac{\chi_\nu}{4}\right).
\end{equation}
Here the field variable is expanded by the $\mathrm{u}(N)$ Lie algebra, $\Theta=\sum_{a=0}^{N^2-1}\theta^a T^a$, with  $T^0=\mathds{1}/\sqrt{N}$ and traceless $T^a$ ($a=1,\cdots,N^2-1$). They are orthonormal, $\Tr(T^aT^b)=\delta_{ab}$, and complete, $\sum_{a=1}^{N^2-1}T^a_{ij}T^a_{kl}=\delta_{il}\delta_{jk}-\frac{1}{N}\delta_{ij}\delta_{kl}$. We call $\theta^0$ as a $\mathrm{u}(1)$ part, and $\theta^a$ with $a=1,\dots,N^2-1$ as an $\mathrm{su}(N)$ part; $\mathrm{u}(N)=\mathrm{u}(1)\oplus\mathrm{su}(N)$. The action consists of Gaussian fluctuation terms for the $\mathrm{u}(1)$ part, Eq.~(\ref{eq:S0}) and the $\mathrm{su}(N)$ part, Eq.~(\ref{eq:S1}), and a topological excitation term, Eq.~(\ref{eq:Sy}). The Gaussian term for the $\mathrm{su}(N)$ part is obtained by a leading-order expansion in powers of small $\sigma_x^{-1}$ and $\sigma_y^{-1}$. The expansion is justified for systems in and proximate to a (critical) metal phase. A non-negative real number $y$ in $S_y$ is a fugacity of the vortex excitation with the smallest vorticity. $y$ controls an effect of topological excitations. The ordering/disordering of the $Q$ field in the NLSM is described by the disordering/ordering of the $\Theta$ field in the dual theory, respectively. 
$S_y$ in the dual theory prefers the ordering (``locking'') of the 
$\Theta$ field. When ordering, the $\Theta$ field is uniform along $\vec{\chi}$ direction, while it has a linear slope along a direction perpendicular to $\vec{\chi}$. In the following, we present an RG analysis of the dual theory, \cref{eq:dual_theory}. Without loss of generality, we can choose $\vec{\chi}=(0,\chi_y)$. 

\emph{RG analyses.}---In the RG analysis, the field variable $\Theta$ is decomposed 
into slow and fast modes. Integrating the fast mode and treating $y$ perturbatively, we obtain an effective action for the slow mode~\cite{SM}. From the effective action, 
we obtain RG equations for the coupling constants in the theory~\cite{SM}. In the zero-replica limit ($N\rightarrow 0$)~\cite{altlandCondensedMatterField2010}, the RG equations for $y$, $\chi_y$, $\sigma_{x}$, $\sigma_y$, $c_{x}$ and $c_y$ reduce to differential equations among four coupling constants~\cite{SM}; normalized fugacity $\tilde{y} \equiv y\sqrt{8\pi\sigma_y(\sigma_x+c_x)}$, a stiffness 
parameter $K \equiv (\sigma_xc_y+\sigma_yc_x+2\sigma_x\sigma_y)/(8\sqrt{\sigma_x\sigma_y})$, an anisotropy parameter $\zeta \equiv \sigma_x/\sigma_y$, and normalized weak topology parameter $\tilde{\chi} \equiv \chi_y/(4\sqrt{1+\sigma_y/\sigma_x})$. The equations are given by 
\begin{equation}\label{eq:y_tilde}
    \frac{\diff\tilde{y}}{\diff l}=\left[2-K-\frac{1}{2}\tilde{y}^2(2\tilde{B}_x+\tilde{B}_y)\right]\tilde{y},
\end{equation}
\begin{equation}\label{eq:K}
    \frac{\diff K}{\diff l}=-\tilde{y}^2\frac{\tilde{B}_x(3\zeta+\lambda)+\tilde{B}_y(\zeta+3\lambda)}{2(\zeta+\lambda)}K,
\end{equation}
\begin{equation}\label{eq:zeta}
    \frac{\diff\zeta}{\diff l}=-\tilde{y}^2(\tilde{B}_x-\tilde{B}_y)\zeta,
\end{equation}
\begin{equation}\label{eq:chi_tilde}
    \frac{\diff\tilde{\chi}}{\diff l}=\left[1-\frac{1}{2}\tilde{y}^2\frac{\tilde{B}_x-\tilde{B}_y}{1+\zeta}\right]\tilde{\chi}.
\end{equation}
Here $\lambda \equiv (\sigma_x^2/\sigma^2_y)((\sigma_y+c_y)/(\sigma_x+c_x))$ is one of the anisotropy parameters. $\lambda$ is invariant under the RG equations. $\tilde{B}_x$ and $\tilde{B}_y$ are given by $\zeta$, $\lambda$ and $\tilde{\chi}$, 
{\small\begin{align}
    &\tilde{B}_x=\frac{2\zeta+(3+2\zeta)\tilde{\chi}^2+\frac{\zeta}{\lambda}(1-\zeta-(2+\zeta)\tilde{\chi}^2)}{(1+\tilde{\chi}^2)^4(1+\zeta)^3}, \nonumber \\
    \tilde{B}_y=&\frac{\zeta-1+(13-4\zeta)\tilde{\chi}^2-5(2+\zeta)\tilde{\chi}^4+\frac{\zeta}{\lambda}(2-17\tilde{\chi}^2+5\tilde{\chi}^4)}{(1+\tilde{\chi}^2)^5(1+\zeta)^3}. \nonumber 
\end{align}}
For given $\lambda$, the RG equations establish a global phase diagram in a 4D parameter space subtended by $\tilde{y}$, $K$, $\zeta$, and $\tilde{\chi}$. Transport properties of phases in the phase diagram are determined by RG equations for the conductivity $\sigma_{\mu}$ ($\mu=x,y$), 
\begin{equation}\label{eq:sigmax}
    \frac{\diff\sigma_{\mu}}{\diff l}=-\tilde{y}^2\tilde{B}_{\mu}\sigma_{\mu}. 
\end{equation}

The non-topological and spatially isotropic case corresponds to $\tilde{\chi}=0$ and $\zeta=\lambda=1$, around which the RG equations can be linearized; $\diff \tilde{y}/\diff l = (2-K-3\tilde{y}^2/8) \tilde{y}$, $\diff K/\diff l = - \tilde{y}^2 K/2$, $\diff \zeta/\diff l = (\zeta-1)\tilde{y}^2/4$, and  $\diff \tilde{\chi} /\diff l = \tilde{\chi}$. The linearized equations have a fixed point at ($K,\tilde{y})=(0,4/\sqrt{3})$ and fixed line at $K<2$ and $\tilde{y}=0$ [see Fig.~\ref{fig:phase_diagram}(b)]. The Anderson transition in the chiral symmetry class without the weak topological term has been previously studied in the isotropic case~\cite{konigMetalinsulatorTransitionTwodimensional2012}, where the Anderson localized and (critical) metal phases are described by the fixed point and the fixed line, respectively. The linearized equations show that the weak topology parameter $\tilde{\chi}$ is always a relevant scaling variable around these fixed regions in the 4D parameter space so that the two phases at $\tilde{\chi}=0$ are unstable toward new fixed points at finite $\tilde{\chi}$.

\begin{figure}[t]
    \centering
    \includegraphics[width=0.49\textwidth]
    {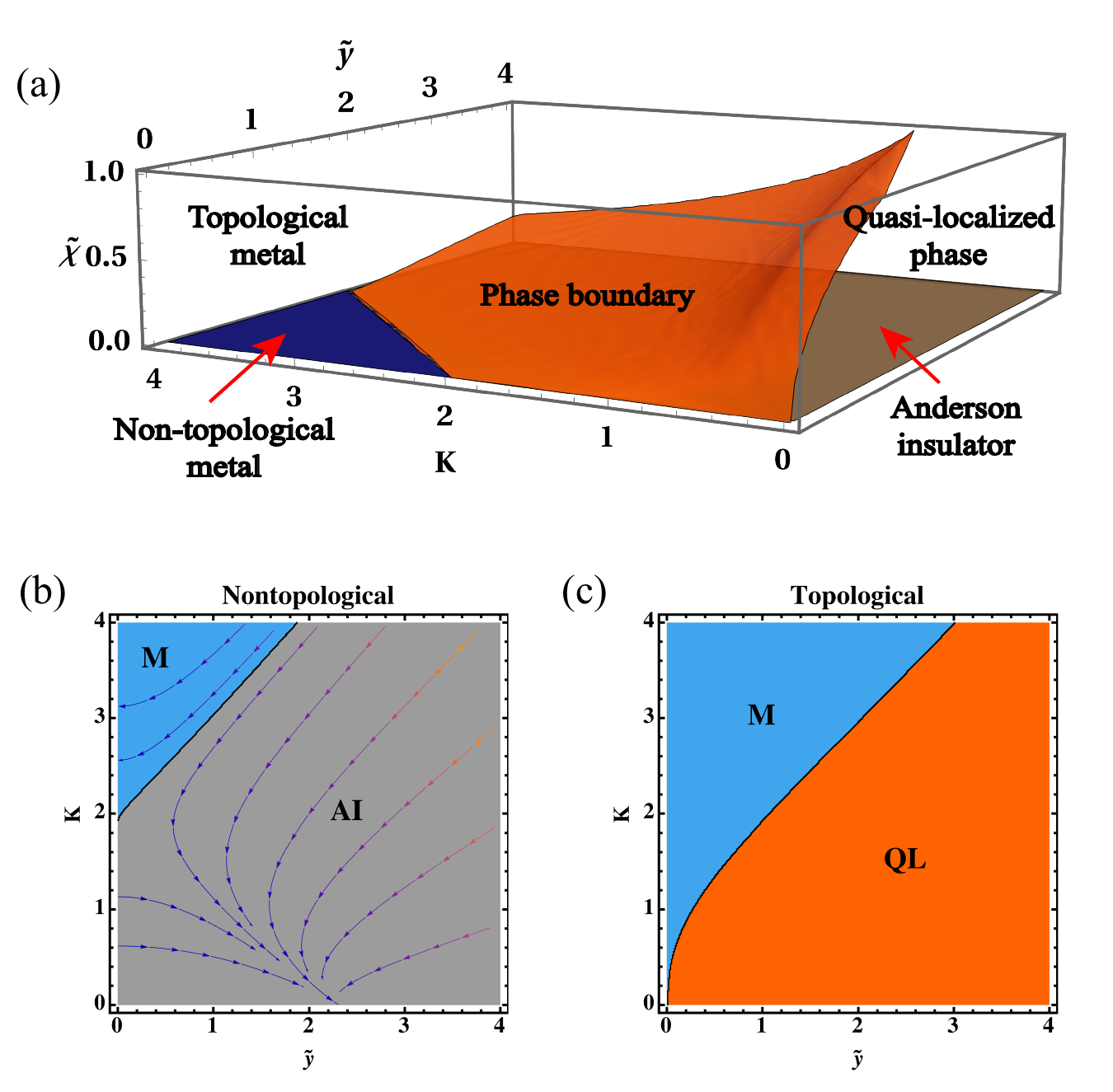}
    \caption{(a) RG phase diagram in the $\tilde{y}$-$K$-$\tilde{\chi}$ space with initial values of $\lambda=\zeta=1$. (b) RG phase diagram in the $\tilde{y}$-$K$ plane at $\tilde{\chi}=0$. (c) RG phase diagram in the $\tilde{y}$-$K$ plane at $\tilde{\chi}=0.1$. The blue regions denote the (critical) metal phase in (b,c), the grey region in (b) is the Anderson localized (AI) phase with $\tilde{\chi}=0$, and the red region in (c) is the quasi-localized (QL) phase with $\tilde{\chi} \ne 0$. In the phase diagram, the metal and quasi-localized phases are separated by a phase boundary (red surface in (a)). The phase diagram with other initial values of $\lambda$ and $\zeta$ takes the same structure as (a) except for the emergence of an unphysical stong-coupling phase in larger $\tilde{y}$ regions~\cite{SM}.}
    \label{fig:phase_diagram}
\end{figure}

\emph{RG phase diagram.}---In the 4D parameter space with $\tilde{y}<\infty$, the RG equations have a stable fixed point and a stable fixed region in an infrared limit ($l \rightarrow \infty$), which characterize quasi-localized and metal phases at $\tilde{\chi}\ne 0$, respectively. At the stable fixed point, $K=\zeta=0$, $\tilde{y}^2 \tilde{B}_x=2$ and $\tilde{y}^2 \tilde{B}_y=0$, around which the conductivity equations reduce to $\diff \ln \sigma_x /\diff l=-2$ and $\diff \ln \sigma_y /\diff l=0$. Thus, all parameter points that contract into the fixed point are in the same thermodynamic phase with vanishing $\sigma_x$ and finite $\sigma_y$; the fixed point characterizes the quasi-localized phase with $\sigma_x=0$ and $\sigma_y \ne 0$. The values of $\tilde{\chi}$ and $\tilde{y}$ at the fixed point are determined by $-1+13\tilde{\chi}^2-10\tilde{\chi}^4=0$ and $\tilde{y} = \sqrt{2/3} (1+\tilde{\chi}^2)^2/|\tilde{\chi}|$. 

The stable fixed region is given by a condition of $\tilde{y}=0$, $2<K$, and $\tilde{\chi}=\infty$, around which the RG equations can be linearized; $\diff \tilde{y}/\diff l=(2-K) \tilde{y}$, $\diff \tilde{\chi}/\diff l = \tilde{\chi}$,  $\diff K/\diff l=\diff \zeta/\diff l =0$. 
All parameter points that contract into the fixed region are in the (critical) metal phase with weak topology. In fact, the fugacity $\tilde{y}$ vanishes, and both $\tilde{B}_x$ and $\tilde{B}_y$ vanish as $1/\tilde{\chi}^6$ toward the fixed region, yielding $\diff \ln \sigma_{\mu} /\diff l=0$ and $\sigma_{\mu} \ne 0$ for $\mu=x,y$ in the infrared limit. 

The RG equations also have a saddle fixed point which characterizes the criticality of a quantum phase transition between the quasi-localized and metal phases. At the fixed point, $\zeta=0$, $K \equiv K_c = 3/4$, $\tilde{y}^2 \tilde{B}_x=3/2$ and $\tilde{y}^2 \tilde{B}_y=-1/2$. 
A scaling dimension of a relevant scaling variable is evaluated as $0.819$. Fig.~\ref{fig:phase_diagram} shows a RG phase diagram subtended by $K$, $\tilde{\chi}$ and $\tilde{y}$ with intial values of isotropic conductivity and Gade constant ($\zeta=\lambda=1$). When $\tilde{\chi} \ne 0$, the RG phase diagram consists of metal phase with $\sigma_\mu \ne 0$ ($\mu=x,y$) and quasi-localized phase with $\sigma_x=0$ and $\sigma_y \ne 0$ [Fig.~\ref{fig:phase_diagram}(a,c)]. From the above scaling analysis, a universal critical exponent of the quantum phase transition between the 
two phases is evaluated as $1/0.819\approx 1.22$.

\emph{Numerical study.}---The RG study predicts that the quasi-localized phase emerges next to the (critical) metal phase in a phase diagram for chiral symmetry class with weak topology. A previous numerical study  showed this in 3D models with weak topology~\cite{xiaoAnisotropicTopologicalAnderson2023}, while in 2D chiral 
symmetry classes, the  Anderson transition has been studied only in models 
without  weak topology. Here we study a 2D model in chiral unitary class with finite weak topological index, and demonstrate the emergence of the quasi-localized phase next to the critical metal phase. We study a $2L_x \times 2L_y$ square-lattice generalization of the Su-Schrieffer-Heeger model~\cite{liTopologicalStatesTwoDimensional2022};   
\begin{equation}
    H=\sum_{x,y}\left(t_{x,y}^{(x)}c_{x+1,y}^\dagger c_{x,y}+t_{x,y}^{(y)}c_{x,y+1}^\dagger c_{x,y}\right)+\text{h.c.},  \label{lattice}
\end{equation}
with integers $x=1,2,\cdots,2L_x$ and $y=1,2,\cdots,2L_y$, and staggered hopping terms $t_{x,y}^{(x)}=(t_x+t)/2+(-1)^{x+y}(t_x-t)/2+\epsilon_{x,y}(1-(-1)^x)/2$ and $t_{x,y}^{(y)}=(t_y+t)/2+(-1)^{x+y}(t_y-t)/2$. $\epsilon_{x,y}$ is a complex number, whose real and imaginary parts are distributed uniformly in a range of $[-W/2,W/2]$. $W$ stands for the disorder strength of the model. The unit cell of the model has four inequivalent sublattices; 1 (even $x$, even $y$), 2 (odd $x$ and odd $y$), 3 (odd $x$, even $y$), and 4 (even $x$ and odd $y$). They are divided into two sublattice groups; (1,2) and (3,4). The model comprises hoppings between the two groups, having the chiral symmetry, $\sigma_3 H \sigma_3 = -H$, where $\sigma_3$ takes $+1$ for one of the two sublattice groups and $-1$ for the other. Finite $\epsilon_{x,y}$ breaks the time-reversal symmetry, and the model belongs to the chiral unitary class. Due to the chiral symmetry, $H$ can be in a block-off-diagonal form, $H = \sigma_{+} \otimes h + \sigma_{-} \otimes h^{\dagger}$, where a non-Hermitian matrix $h$ defines the one-dimensional weak topological indices $\nu_{\mu}$ $(\mu=x,y)$~\cite{SM,mondragon-shemTopologicalCriticalityChiralSymmetric2014,altlandQuantumCriticalityQuasiOneDimensional2014,claes2020}. $\nu_{\mu}$ takes integer in gapped phases, and non-integer values in metallic phases~\cite{SM}. Statistical symmetries of an ensemble of disordered $h$ require the weak topological index along  $\mu$ to be zero for $t_\mu=t$ ($\mu=x,y$)~\cite{SM,xiaoAnisotropicTopologicalAnderson2023}.  
To study a phase diagram of disordered $H$ with $\nu_y\ne 0$ and $\nu_x = 0$, we set the hopping parameters as $(t_x,t_y)=t(1,1.5)$. This corresponds to the NLSM with  $\vec{\chi}=(0,\chi_y)$ and $\chi_y = 8\pi \nu_y$~\cite{SM}. 

\begin{figure}[t]
    \centering
    \includegraphics[width=0.48\textwidth]{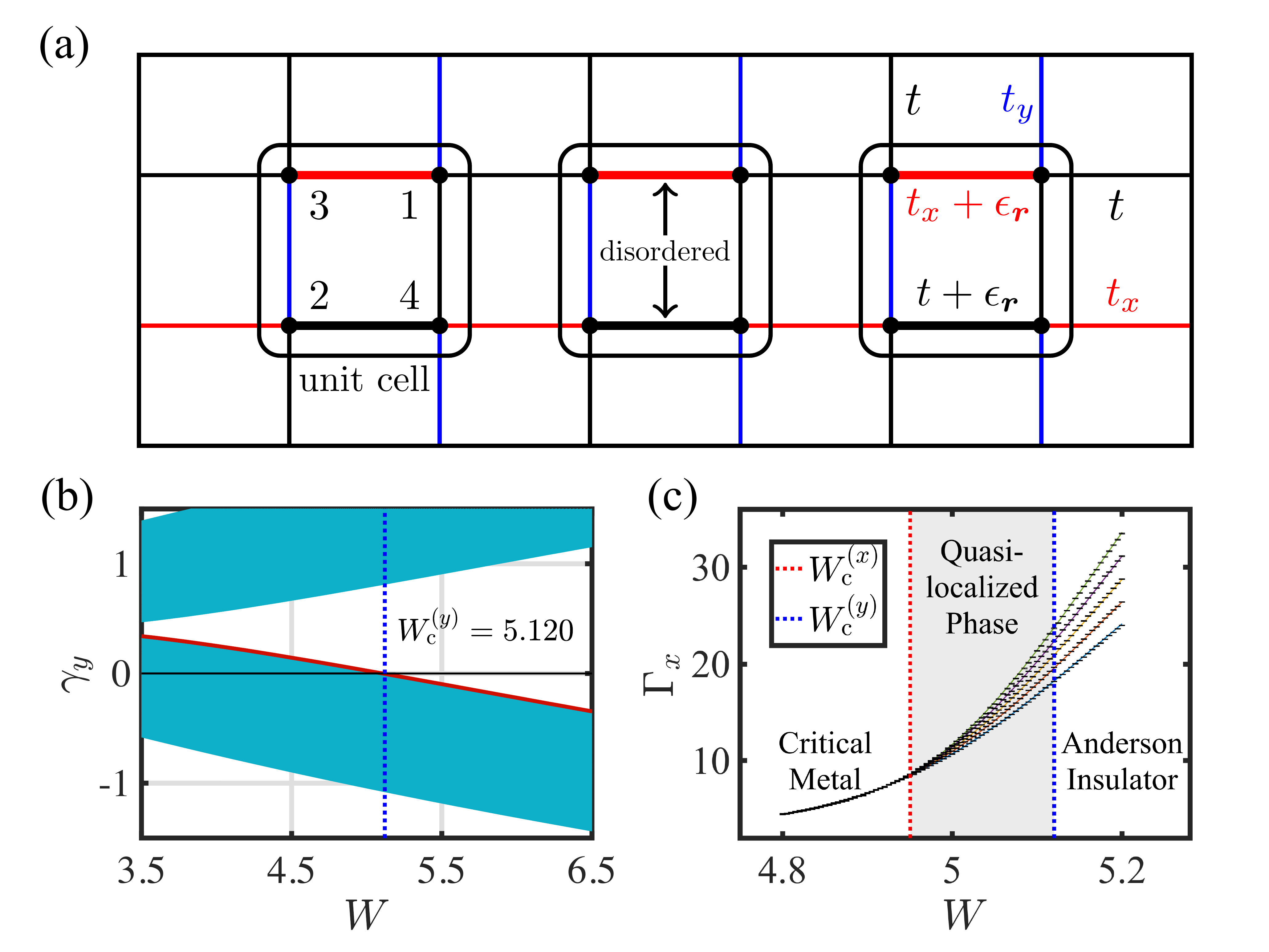}
    \caption{
    (a) Square-lattice model for Eq.~(\ref{lattice}). (b) Numerical distribution of the LEs of $h$ along $y$ as a function of $W$ ($L_x = 240$). The LEs of $h$ form two continuous bands (blue colored bands) for $W \gtrsim 3.5$, and the lower continuous band covers zero in the metal phase ($W < W^{(y)}_{\text{c}}=5.120$). (c) Numerical data with system size $L_y=160\sim 240$ for normalized localization length $L_y/\xi_x$ along $x$, as a function of disorder strength $W$. $W^{(x)}_{\text{c}}$ and $W^{(y)}_{\text{c}}$ determined by finite size scaling analyses are also shown for eye guide.} 
    \label{fig:results}
\end{figure}

The emergence of the quasi-localized phase  
is confirmed by numerical transfer matrix calculations of localization lengths 
$\xi_{\mu}$ of zero-energy wavefunctions of $H$ along $\mu$ ($\mu=x,y$)~\cite{slevinCriticalExponentAnderson2014,markosPhenomenologicalTheoryMetalinsulator1995,luoTransferMatrixStudy2021}. 
Given $H$ or $h$ with $L_x\gg L_y$, for example, the wavefunctions are partitioned into $L_{x}$ slices, where the $x$-th slice and $(x+1)$-th slice are related by a transfer matrix.
The Lyapunov exponents (LEs) along $x$ are eigenvalues of 
$1/(2L_x) \ln M^{\dagger}_x M_x$, where 
$M_x$ is a product of the transfer matrices from $x=1$ to $x=L_x$. The positive (negative) LEs are inverse of decay lengths of the wavefunctions along $+x$ ($-x$) direction, and the LEs of $H$ are a union of the LEs of $h$ and their opposite numbers~\cite{xiaoAnisotropicTopologicalAnderson2023,SM}.
Thus, a localization length of $H$ is an inverse of the smallest absolute value of the 
LEs of $h$.
Fig.~\ref{fig:results}(b,c) show that the model undergoes transitions from metal to 
localized phases in a region of $4<W<6$. In the metal phase, 
the localization length $\xi_y$ along $y$ is divergent; 
the lower half of the LEs of $h$ along $y$ forms a continuous band 
~\cite{markosPhenomenologicalTheoryMetalinsulator1995}, and the band covers zero in the metal phase. The localization length $\xi_x$ along $x$ normalized by a finite system size $L_y$ stays invariant for increasing $L_y$ in the metal phase. In the localized phase, $\xi_y$ is finite in the thermodynamic limit, while $\xi_x/L_y$ decreases for increasing $L_y$, indicating that $\xi_x$ is also finite in the thermodynamic limit $L_y \rightarrow \infty$. We determine a critical disorder strength $W^{(x)}_{\text{c}}$ for the change of $\xi_x/L_y$, and a critical disorder 
strength $W^{(y)}_{\text{c}}$ for the change of $\xi_y$. Note also that 
$\nu_{y}$ is given by a difference between a number $N_{y,+}$ of positive LEs of $h$ along $y$ and a number $N_{y,-}$ of negative LEs; $\nu_{y}=(N_{y,+}-N_{y,-})/(2L_x)$~\cite{xiaoAnisotropicTopologicalAnderson2023,molinari2003,SM}. 
Thus, $\nu_y>0$ for $W<W^{(y)}_{\text{c}}$ and $\nu_y=0$ for $W>W^{(y)}_{\text{c}}$. From finite-size scaling  analyses~\cite{SM,xiaoAnisotropicTopologicalAnderson2023,slevinCriticalExponentAnderson2014,karcher2023a}, the two 
critical disorder strengths are evaluated as $W^{(x)}_{\text{c}}=4.950\pm 0.001$ and $W^{(y)}_{\text{c}}\approx 5.120\pm 0.002$. 
A discrepancy between the two is about $3\%$ of the critical values, and it is much greater than the error bar of the fittings. This suggests a phase diagram with (i) critical metal phase with $\nu_y>0$ ($W<W^{(x)}_{\text{c}}$), (ii) quasi-localized phase with $\nu_y>0$ ($W^{(x)}_{\text{c}}<W<W^{(y)}_{\text{c}}$), and (iii) Anderson localized phase with $\nu_y=0$ ($W^{(y)}_{\text{c}}<W$). 
In the quasi-localized phase, $\nu_y>0$, $\xi_y$ is divergent and $\xi_x$ is finite in the thermodynamic limit. The finite-size scaling analyses also give an critical exponent at $W^{(x)}_{\text{c}}$ with $95\%$ 
confidence interval as $\nu=1.36\pm 0.01$.~\cite{SM}

\emph{Summary.}---
We study a field theory with a weak topological term and clarify that the interplay between weak topology and disorder induces the proliferation of spatially-polarized vortex-antivortex pairs near the metal-insulator transition in chiral symmetry classes. This results in the universal emergence of the quasi-localized phase next to the metal phase, that shows the extreme spatial anisotropy in its transport property. We argue the mechanism by RG analyses of the NLSM for the 2D chiral unitary class with weak topology,  
and test the theory by the simulation of a lattice model. Both analyses confirm the emergence of the quasilocalization next to the metal phase. Meanwhile, the RG analyses predict the critical exponent $\nu=1.22$ for 
the phase transition between metal and quasi-localized phases, while the numerical simulation suggests $\nu = 1.36 \pm0.01$ for the same transition. The discrepancy may come from the nonlinear terms in 
$S_{{\rm su}(N)}$ that are neglected in the RG analyses, while the resolution also needs further development of finite-size scaling theories for 2D phase transitions between critical metal and insulator phases.~\cite{karcher2023a,wangtong2021b}. We expect the analysis in the paper can be also generalized to other chiral symmetry classes (BDI, CII) and higher dimensions.

\emph{Acknowledgement.}---We thank Siyu Pan, Lingxian Kong, Tong Wang, Kohei Kawabata, Xunlong Luo, and Tomi Ohtsuki for discussions and comments. The work was supported by the National Basic Research Programs of China (No. 2019YFA0308401) and the National Natural Science Foundation of China (No. 11674011 and No. 12074008).

\bibliography{ref.bib}

\clearpage
\begin{widetext}
\section*{Supplement for ``Topological effect on the Anderson 
chiral symmetry classes''}

    \setcounter{equation}{0}
    \setcounter{section}{0}
    \setcounter{figure}{0}
    \setcounter{table}{0}
    \setcounter{page}{1}
    \renewcommand{\theequation}{S\arabic{equation}}
    \renewcommand{\thesection}{ \Roman{section}}

    \renewcommand{\thefigure}{S\arabic{figure}}
    \renewcommand{\thetable}{\arabic{table}}
    \renewcommand{\tablename}{Supplementary Table}
    
    \renewcommand{\bibnumfmt}[1]{[S#1]}
    \renewcommand{\citenumfont}[1]{#1}

	\title{Supplemental Material of ``Topological Effect on the Anderson transition in chiral symmetry classes"}
	\author{Pengwei Zhao}
	\affiliation{International Center for Quantum Materials, Peking University, Beijing 100871, China}
	\author{Zhenyu Xiao}
	\affiliation{International Center for Quantum Materials, Peking University, Beijing 100871, China}
	\author{Yeyang Zhang}
	\affiliation{International Center for Quantum Materials, Peking University, Beijing 100871, China}
	\author{Ryuichi Shindou}
        \email{rshindou@pku.edu.cn}
	\affiliation{International Center for Quantum Materials, Peking University, Beijing 100871, China}
	\date{\today}
	
	\maketitle
	\tableofcontents
	
	\section{The effect of weak topological term}
	\subsection{Nonlinear sigma model}

The interplay between disorder and topology gives rise to a novel quantum interference effect in physical systems. A recent numerical work~\cite{xiaoAnisotropicTopologicalAnderson2023} studied the Anderson transition in chiral symmetry classes with a weak topology and revealed that a one-dimensional weak topological index (``winding number") induces an emergent quasi-localized phase between diffusive metal and Anderson insulator phases. The intermediate phase has a divergent localization length in a spatial direction with the non-zero winding number, and finite localization lengths along the other spatial directions. In the quasi-localized phase, the conductance along the divergent length remains finite with a non-Ohmic scaling in the thermodynamic limit, while the conductance along the other directions vanishes exponentially. The physical mechanism of the emergent quasi-localized phase is unknown at this moment. In the present paper, we carry out a field-theoretical study of Anderson localization in chiral classes with weak topological term. We clarify that the interference effect due to the one-dimensional weak topological index results in the emergence of the quasi-localized phase.

Low-energy and long-wavelength behaviors of diffusion in disordered systems are described by a nonlinear sigma model (NLSM). The effective theory is formulated in terms of either a supersymmetry, Keldysh equation approach or a replica method. In this paper, we use the replica formulation, where a partition function $Z$ for the $d$-dimensional chiral symmetric systems is described by the following NLSM, 
\begin{align}
    Z &= \int \Diff [Q] e^{-S}, \nonumber \\
    S &=-\int\frac{\diff^dr}{8\pi}\sum_{\mu=1}^{d}\left[\sigma_\mu\Tr(Q^{-1}\partial_\mu Q)^2+c_\mu\Tr^2(Q^{-1}\partial_\mu Q)-\chi_\mu\Tr(Q^{-1}\partial_\mu Q)\right], \label{eq:SM_NLSM}
\end{align}
with $\mu=1,2,\cdots,d$. The field variable of the NLSM is a $N\times N$ matrix field $Q$, where $N$ is a number of replicated fermion fields in the replica method. $\sigma_\mu$ stands for a physical conductivity along $\mu$-direction. $c_\mu$ is a Gade constant that also has a dependence on spatial direction in general. A vector $(\chi_1,\chi_2,\dots,\chi_d)$ quantifies the one-dimensional weak topology. The direction of the vector specifies the one-dimensional topological direction. The norm of the vector quantifies the winding number along the topological direction averaged over the other directions. The weak topological term can be derived from a lattice model in the chiral class with weak band topology~\cite{altlandQuantumCriticalityQuasiOneDimensional2014}. In the following, a summation over repeated index $\mu$ is always assumed.  

Table~\ref{tab:variable_Q} enumerates symmetries of the field variable $Q$ and its low-energy manifolds (``Goldstone'' manifolds) in the three chiral symmetry classes. As a common feature in the chiral classes, the low-energy manifolds have a $\mathrm{U}(1)$ subgroup symmetry. Due to this $\mathrm{U}(1)$ subgroup symmetry, perturbative $\beta$-functions of the NLSM are zero for the chiral symmetry classes~\cite{GADE1993499,GADE1991213}, while the $\mathrm{U}(1)$ subgroup part of the $Q$ field accommodates vortex excitations. Accordingly, it has been argued that the Anderson transition in the chiral classes is driven by spatial proliferation of the vortex excitations of the $Q$-field~\cite{konigMetalinsulatorTransitionTwodimensional2012,karcherMetalinsulatorTransitionTwodimensional2023}. It has been also argued that strong topological terms with quantized topological number hinder the vortex proliferation through destructive quantum interference effect~\cite{konigMetalinsulatorTransitionTwodimensional2012}. Nonetheless, the quantum interference effect due to the one-dimensional weak topological index has not been discussed previously, and associated consequences on the localized phase proximate to the metal phase remain unclear. We will discuss the quantum interference effect due to the weak topological term, and clarify that the interference alters the localized phase into the quasi-localized phase.

\begin{table}[h!tb]
    \centering
    \begin{tabular}{ccc}
        Symmetry class & Goldstone manifold & Restriction \\
        \hline
        AIII & $\mathrm{U}(N)$ & No restriction \\
            BDI & $\mathrm{U}(N)/\mathrm{Sp}(N)$ & $Q=CQ^\T C,C^2=1,C^\T=-C$ \\ 
        CII & $\mathrm{U}(N)/\mathrm{O}(N)$ & $Q=Q^\T$ 			
    \end{tabular}
    \caption{Goldstone manifolds of the field variable $Q$ and symmetries of the $Q$ field. The matrix $C$ is an arbitrary antisymmetric matrix with condition $C^2=1$.\label{tab:variable_Q}}
\end{table}
	
\subsection{Topological defects}
For further discussion, an exact definition of topological defects is necessary. The vortex excitations can be often introduced as saddle-point configurations of the NLSM, that have a discontinuity around a lower dimensional area. Taking a functional derivative of Eq.(\ref{eq:SM_NLSM}), $S=S_{\text{cond}}+S_{\text{Gade}}+S_{\text{top}}$, one obtain the saddle-point equation. Define a small variation of $Q$ as $Q\rightarrow Q+\delta Q$ and its ivertese $Q^{-1}\rightarrow Q^{-1}-Q^{-1}\delta QQ^{-1}$. One can check the relation holds
\begin{equation}
    \left(Q^{-1}-Q^{-1}\delta QQ^{-1}\right)\left(Q+\delta Q\right)=1+Q^{-1}\delta Q-Q^{-1}\delta Q + {\cal O}(||\delta Q||^2) =1.
\end{equation}
A small matrix $\delta Q$ means $||\delta Q||^2\equiv\Tr\left(\delta Q^\dagger \delta Q\right)\ll 1$. 
Now, we put this replacement into \cref{eq:SM_NLSM} and omit the 2nd order terms of $\delta Q$. For example, 
\begin{equation}
	\delta\Tr\left(Q^{-1}\partial_\mu Q\right)=\Tr\left(-Q^{-1}\delta QQ^{-1}\partial_\mu Q+Q^{-1}\partial_\mu \delta Q\right).
\end{equation}
The weak topological term $S_{\text{top}}$ gives
\begin{equation}
    \delta S_{\text{top}}=-\int\frac{\diff^dr}{8\pi}\,\chi_\mu \Tr\left[\left(Q^{-1}\partial_\mu QQ^{-1}+\partial_\mu Q^{-1}\right)\delta Q\right].
\end{equation}
The Gade term $S_{\text{Gade}}$ yields
\begin{equation}
    \begin{gathered}
        \delta S_{\text{Gade}}=\int\frac{\diff^dr}{8\pi}\,2c_\mu\left(\Tr\left(Q^{-1}\partial_\mu Q\right)\Tr\left(Q^{-1}\partial_\mu QQ^{-1}\delta Q\right)+\Tr\left(\partial_\mu Q^{-1}\partial_\mu Q\right)\Tr\left(Q^{-1}\delta Q\right)\right.\\
        \left.+\Tr\left(Q^{-1}\partial_\mu^2Q\right)\Tr\left(Q^{-1}\delta Q\right)+\Tr\left(Q^{-1}\partial_\mu Q\right)\Tr\left(\partial_\mu Q^{-1}\delta Q\right)\right).
    \end{gathered}
\end{equation}
The conductivity term $S_{\text{cond}}$ can be written in the following form
\begin{equation}
    S_{\text{cond}}=-\int\frac{\diff^d r}{8\pi}\sigma_\mu\Tr\left(Q^{-1}\partial_\mu Q\right)^2=\int\diff^3r\,\sigma_\mu\Tr\left(\partial_\mu Q^{-1}\partial_\mu Q\right), 
\end{equation}
whose variation is given by
\begin{equation}
    \delta S_{\text{cond}}=\int\diff^3r\,\sigma_\mu \Tr\left[\left(Q^{-1}\partial_\mu ^2QQ^{-1}-\partial_\mu ^2Q^{-1}\right)\delta Q\right].
\end{equation}
The stationary condition $\delta S=0$ gives a saddle-point equation
\begin{equation}
    \begin{gathered}
        \sigma_\mu \left(Q^{-1}\partial_\mu ^2QQ^{-1}-\partial_\mu ^2Q^{-1}\right)+2c_\mu\Tr\left(Q^{-1}\partial_\mu Q\right)\left(Q^{-1}\partial_\mu QQ^{-1}+\partial_\mu Q^{-1}\right)\\
        +2c_\mu \left[\Tr\left(\partial_\mu Q^{-1}\partial_\mu Q\right)+\Tr\left(Q^{-1}\partial_\mu ^2Q\right)\right]Q^{-1} 
        + \chi_{\mu} \left(Q^{-1}\partial_\mu QQ^{-1}+\partial_\mu Q^{-1}\right)=0.
    \end{gathered}
\end{equation}
As $\partial QQ^{-1}+Q\partial Q^{-1}=\partial(QQ^{-1})=0$, the stationary condition 
is equivalent to
\begin{equation}\label{eq:SM_saddle_point_equation}
    2\sigma_\mu \partial_{\mu}\left(Q \partial_{\mu}Q^{-1}\right)+2c_\mu \partial_{\mu}\left(\Tr \left(Q \partial_{\mu}Q^{-1}\right) \right)=0
\end{equation}

For $N=1$, $\mathrm{U}(1)$ field is given by $Q=e^{i\phi}$, and the saddle-point equation reduces 
to the Poisson equation 
\begin{equation}
    (\sigma_\mu +c_\mu )\partial_\mu ^2\phi=0.
\end{equation}
With reduced coordinates $\bar{r}_{\mu} \equiv 
r_{\mu}/\sqrt{\sigma_\mu +c_\mu }$,  
the saddle-point equation is  
\begin{equation}\label{eq:SM_saddle-point_equation_phi}
    \bar{\nabla}^2\phi=0.
\end{equation}
Solutions of the saddle-point equation can generally have a phase winding around a $(d-2)$-dimensional vortex core 
\begin{equation}\label{eq:SM_phase-winding}
    \oint_C\diff\bm{\bar{r}}\cdot\bm{\bar{\Nabla}}\phi=2\pi\eta.
\end{equation}
When a one-dimensional loop $C$ encircles the core, $\eta$ on the right-hand side takes an integer value (vortex charge).

For $N \ne 1$, $Q$ has unit complex numbers $e^{i\phi_a}$ as its eigenvalues, $Q = \sum^N_{a=1}\ket{p_a} e^{i\phi_a} \bra{p_a}$. Here $\ket{p_a}$ is a normalized column vector, $\braket{p_a|p_a}=1$, belonging to the complex projective space $\mathbb{CP}^{N-1}$, real projective space $\mathbb{RP}^{N-1}$, and 
quaternion projective space $\mathbb{HP}^{N/2-1}$ for chiral unitary, chiral symplectic, and 
chiral orthogonal classes respectively. Let us consider a uniform 
$Q$-field configuration, $Q({\bm r})= \mathds{1}$, add a vortex exictation. Around the vortex core, one of the eigenvalues has the phase winding, while all the others are close to unit,  
\begin{equation}\label{eq:SM_U(N)_saddle_point}
		Q({\bm r})=\mathds{1}+\ket{p({\bm r})}(e^{i\phi({\bm r})}-1)\bra{p({\bm r})}. 
\end{equation}
Namely, $\phi$ is a solution of $\bar{\nabla}^2\phi=0$ with the phase winding around the core; Eq~(\ref{eq:SM_phase-winding}), and $\ket{p({\bm r})}$ is an eigenvector for the eigenvalue with the phase winding.  $\ket{p({\bm r})}$ generally depends on the spatial coordinate ${\bm r}$. A $Q$-field configuration with multiple vortices at ${\bm r}_1,\cdots,{\bm r}_n$
can be also given by the same ansatz, Eq.~(\ref{eq:SM_U(N)_saddle_point}), where $\phi$ has phase windings around 
these cores, and $\ket{p({\bm r}_i)}$ and $\ket{p({\bm r}_j)}$ are generally different from 
one another because of the ${\bm r}$-dependence of $\ket{p({\bm r})}$. We assume that an eigenvector 
$\ket{p({\bm r})}$ is slowly varying in the space coordinate compared to a typical size of vortex cores. In fact, the ansatz Eq.~(\ref{eq:SM_U(N)_saddle_point}) with 
${\bm r}$-independent $\ket{p}$ can also be introduced as a solution of the saddle-point equation 
for the isotropic case ($\sigma_1=\sigma_2=\cdots=\sigma_d$, $c_1=c_2=\cdots=c_d$) as well as 
an anisotropic case with a fixed ratio between conductivity and Gade constant ($\sigma_1/c_1=\sigma_2/c_2=\cdots=\sigma_d/c_d$). 

The configuration with a vortex is discontinuous at a vortex core with a non-zero commutation of the second-order derivative of the phase $\phi$,  $[\bar{\partial}_{\mu},\bar{\partial}_{\nu}]\phi \ne 0$ for some $\mu \ne \nu$. A line integral along a closed loop $C$ can be written as a surface integral over a region $\Gamma$
\begin{equation}
    \oint_C\diff\bm{\bar{r}}\cdot\bm{\bar{\Nabla}}\phi=\iint_\Gamma\diff\bm{\bar{\Gamma}}\cdot\bm{\bar{\nabla}}\times\bm{\bar{\nabla}}\phi=2\pi\eta,
\end{equation}
with $C=\partial\Gamma$. The surface integral yields a nonzero value if the vortex core penetrates through the surface. The property associates $\bm{\bar{\nabla}}\times\bm{\bar{\nabla}}\phi$ with a Dirac delta function in a way that depends on the spatial dimension. 

\textbf{2D cases}. For $d=2$, the vortex core is a 0-dimensional point. The commutator is a two-dimensional delta function at the core, 
\begin{equation}
    [\bar{\partial}_x,\bar{\partial}_y]\phi=2\pi\eta\delta^2(\bm{\bar{r}}-\bm{\bar{r}}_0).
\end{equation}
Equivalently, 
\begin{equation}\label{eq:SM_vortex}
    [\partial_x,\partial_y]\phi=2\pi\eta\delta^2(\bm{r}-\bm{r}_0), 
\end{equation}
with $\bar{r}_{\mu}= r_{\mu}/\sqrt{\sigma_{\mu}+c_{\mu}}$. 
$\phi$ that satisfies \cref{eq:SM_saddle-point_equation_phi} and \cref{eq:SM_vortex} is given by 
\begin{equation}\label{eq:SM_explicit_vortex}
    \phi=\eta\arctan\left(\frac{\overline{y}-\overline{y}_0}{\overline{x}-\overline{x}_0}\right).
\end{equation}
	
\textbf{3D cases}. For $d=3$, the vortex core takes the form of a one-dimensional line. 
The line can form a closed loop. 
The vortex loop can be parameterized as $\bm{l}(s)$ with a parameter 
$s\in[-1,1]$. For example, a circle in the $xy$-plane is
\begin{equation}
l_x=R\cos(\pi s),\quad l_y=R\sin(\pi s),\quad l_z=0.
\end{equation}
A straight line along $(u_x,u_y,u_z)$ from $(x_0,y_0,z_0)$ is
\begin{equation}
l_x=x_0+u_x {\rm arctanh}(s),\quad l_y=y_0+u_y{\rm arctanh}(s),\quad l_z=z_0+u_z{\rm arctanh}(s).
\end{equation}
Given such parameterizations of a  vortex line $\bm{\bar{l}}(s)$, the non-zero $\bar{\bm \nabla} \times \bar{\bm \nabla}$ is defined by  
\begin{equation}
\epsilon_{ijk}\bar{\partial}_j\bar{\partial}_k\phi=2\pi\eta\int\diff s\,\frac{\diff \bar{l}_i}{\diff s}\delta^3(\bm{\bar{r}}-\bm{\bar{l}}).
\end{equation}
or 
\begin{equation}\label{eq:SM_vortex_loop}
\epsilon_{ijk}\partial_j\partial_k\phi=2\pi\eta\int\diff s\,\frac{\diff l_i}{\diff s}\delta^3(\bm{r}-\bm{l}).
\end{equation}
Importantly, the form of the non-zero commutation of the second-order spatial derivative is not affected 
by the anisotropies of conductivity and Gade constant.
	
\subsection{Weak topological term}
The weak topological term in Eq.~(\ref{eq:SM_NLSM}) is associated with weak topology in the topological band theory, and it is 
quantified by a vector $\vec{\chi}\equiv\bm{\chi}=(\chi_{1},\chi_2,\cdots,\chi_d)$ in the NLSM. In this section, we briefly review the weak topology. Topology in the field theory is defined as the homotopy group of the field variable $Q$ on the $d$-dimensional space, where $Q$ is regarded as a map from the coordinate space $\mathbb{R}^d$ to the low-energy (``Goldstone'') manifold $G$. With some boundary conditions, the coordinate space $\mathbb{R}^d$ becomes a compact space. For a fixed boundary condition $Q(|{\bm r}|=\infty) \equiv 1$, $\mathrm{R}^d$ is compactified into $\mathrm{SO}(d)$, $\mathrm{R}^d\simeq \mathrm{SO}(d)$. For a periodic boundary condition, $\mathrm{R}^d \simeq \mathrm{U}(1)^d$. If two maps, $Q_1$ and $Q_2$, are continuously transformed into each other, they belong to an equivalent class. Inequivalent classes form homotopy group $\pi(\mathrm{R}^d,G)$, and an algebra of the group defines the strong topology of the field variable $Q$ in $\mathrm{R}^d$. A weak topology is associated with a homotopy group of the field variable in a lower dimensional subspace of $\mathrm{R}^d$. For example, $Q$ as a function of one of the coordinates, say $r_1$ with ${\bm r} \equiv (r_1,r_2,\dots,r_{d-1},r_d)$, can be regarded as a map from $\mathbb{R}$ to $G$. For both fixed and periodic boundary conditions, $\mathbb{R}\simeq U(1)$, where the map defines another homotopy group, $\pi(\mathrm{U}(1),G)$. For class AIII, $G=\mathrm{U}(N)$ where the homotopy group in the one-dimensional subspace is classified by an integer;  
\begin{equation}
    \pi(\mathrm{U}(1),\mathrm{U}(N))=\pi(\mathrm{U}(1),\mathrm{U}(1)\times\mathrm{SU}(N))=\pi(\mathrm{U}(1),\mathrm{U}(1))\times \pi(\mathrm{U}(1),\mathrm{SU}(N))=\mathbb{Z}\times 1=\mathbb{Z}.
\end{equation}
The integer manifests itself in the one-dimensional topological term in the NLSM, 
\begin{align}
    \int \frac{dr_1}{2\pi i} \Tr(Q^{-1}\partial_{r_1} Q) = \mathbb{Z}.  
\end{align}
Note that the NLSM with the one-dimensional  
the weak topological term can be derived from a lattice model in the chiral class with a weak topological index of the topological band theory. 
	
\subsection{Effect of the weak topological term on Anderson transition}
The Anderson transition in chiral symmetry classes without the  weak topological term has been studied previously, and the study shows that it is driven 
by the spatial proliferation of 
the vortices of the $Q$-field 
configuration~\cite{konigMetalinsulatorTransitionTwodimensional2012,karcherMetalinsulatorTransitionTwodimensional2023}. 
The weak topological term $S_{\text{top}}$ gives a complex phase factor to the $Q$-field configuration with the vortices. The phase factor induces a quantum interference effect in the configuration space of the $Q$-field, rendering the Anderson transition in chiral symmetry classes into a unique quantum phase transition in a way that depends on the spatial dimension. 

\begin{figure}[t]
    \centering
   \includegraphics[width=0.8\textwidth]{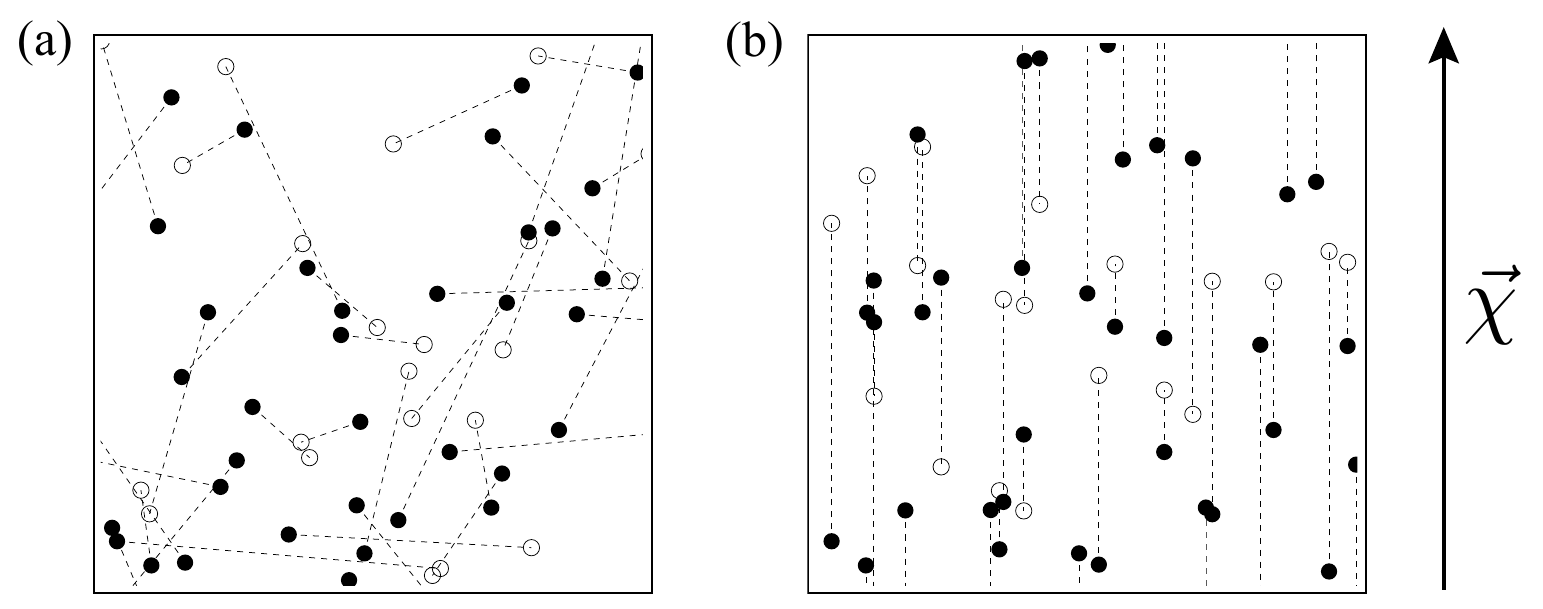}
    \caption{Schematic pictures of a spatial proliferation of vortex-antivortex pairs in 2D space, (a) in the absence of weak topology $\vec{\chi}=0$, (b) in the presence of weak topology $\vec{\chi}\ne 0$. Solid and hollow circles represent vortex and antivortex, respectively, and dashed lines are branchcuts for vortex-antivortex pairs. For given locations of vortices and antivortices in the 2D space, the partition function $Z$ depends neither on the choice of the branch cuts nor on specific pairings of vortices and antivortices.}
    \label{fig:polarization}
\end{figure}

\textbf{2D cases}. To see how the quantum interference effect appears in $d=2$, let us evaluate the weak topological term in the presence of a pair of vortex and antivortex, 
\begin{align}
    Q_{\bm{m}}({\bm r})=\mathds{1}+\ket{p({
    \bm r})}\left(e^{i\phi_{\bm{m}}({\bm r})}-1\right)\bra{p({\bm r})},\quad\phi_{\bm{m}}=\arctan\left(\frac{y-y_1}{x-x_1}\right)-\arctan\left(\frac{y-y_2}{x-x_2}\right),\quad \bm{m}=r_2-r_1.
\end{align}
A relative coordinate between vortex at ${\bm r}_1$ and antivortex at ${\bm r}_2$ can be regarded as dipole vector $\bm{m}$ of the pair. Substituting $Q_{\bm{m}}$ into the action $S_{\text{top}}$, one can see that the weak topological term gives a pure imaginary number that depends on the direction of the dipole vector,  
\begin{equation}
    S_{\text{top}}[Q_{\bm{m}}]=i \eta \chi_{\mu}\int\frac{\diff^2r}{8\pi} \partial_\mu\phi_{\bm{m}} = \frac{i}{4} \eta \big(  \chi_{x} m_y-\chi_{y} m_x\big). 
\end{equation}
Note that the other parts of the action are real-valued. Thus, when the dipole has a finite angle against $(\chi_x, \chi_y)$ in the 2D plane, such dipole configurations with different angles show destructive interference in the partition function. In a simpler case where a ratio between conductivity and Gade constant $\sigma_{\mu}/c_{\mu}$ is independent of the spatial direction $\mu$, $S_{\rm cond}+ G_{\rm Gade}$ with the dipole depends only on $|{\bm m}|$ in the reduced coordination $\overline{r}_{\mu} \equiv r_{\mu}/{\sqrt{\sigma_{\mu}+c_{\mu}}}$. Thereby, the dipole configurations with same $|{\bm m}|$ and different angles against $(\chi_x,\chi_y)$ show the destructive interference. The destructive interference becomes prominent for larger $|{\bm m}|$ and $|\chi_{\mu}|$. On the other hand, dipoles that are parallel to $(\chi_x, \chi_y)$ are free from the interference effect, and naturally dominate the partition function around the Anderson transition. Accordingly, in the presence of the weak topological term, disordered metals in the chiral classes undergo a spatially anisotropic proliferation of vortices where pairs of vortex and antivortex tend to be polarized along $(\chi_x,\chi_y)$ (Fig.~\ref{fig:polarization}). 

\begin{figure}[t]
        \centering
        \includegraphics[width=0.95\textwidth]{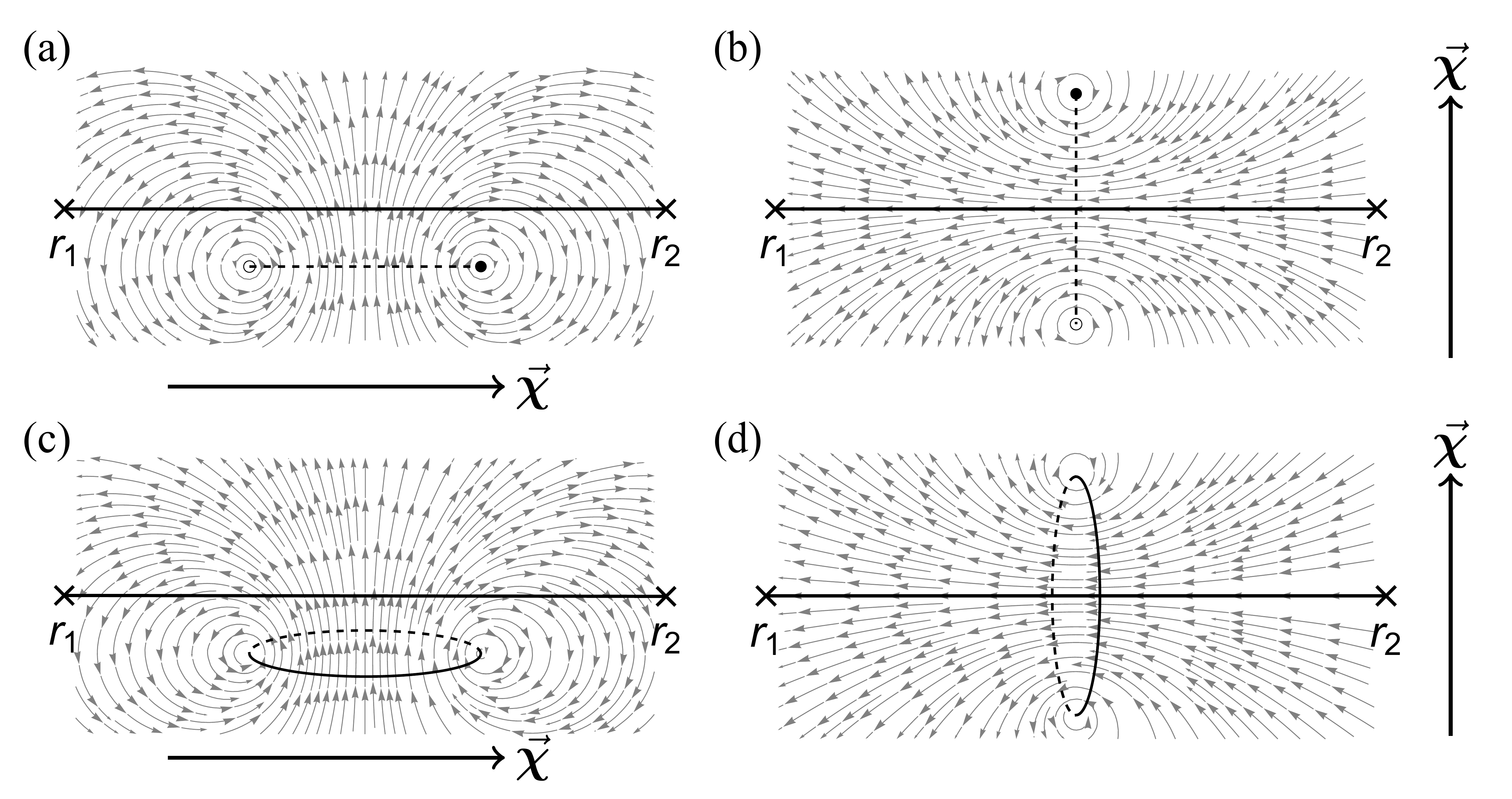}
        \caption{Schematic figures explain how a relative U$(1)$ phase $\phi({\bm r}_1,{\bm r}_2)$ between ${\bm r}_1$ and ${\bm r}_2$ is modified by an introduction of polarized vortex excitation in the two different geometries, (a,c) ${\bm r}_1-{\bm r}_2 \parallel \vec{\chi}$, (b,d) ${\bm r}_1-{\bm r}_2 \perp \vec{\chi}$. In 2D (a,b), the vortex excitation is a pair of vortex (solid circle) and antivortex (hollow circle), and the pair is polarized along $\vec{\chi}$. Grey lines with arrow show magnetic field ${\bm B}({\bm r})$ created by the pair of vortex and 
        antivortex. In 3D (c,d), the vortex excitation is a polarized vortex loop confined in a plane parallel to $\vec{\chi}$. The vortex loop and associated magnetic field are depicted by a black circle and grey lines with arrows, respectively. The relative U(1) phase $\phi({\bm r}_1,{\bm r}_2)$ changes by the introduction of the vortex excitations, and the change 
        $\delta \phi({\bm r}_1,{\bm r}_2)$ is given by the line integral of the magnetic field along a line connecting ${\bm r}_1$ and ${\bm r}_2$ [Eq.~(\ref{phi-B})]. A choice of the line gives an ambiguity of $2\pi$ times integer for $\delta \phi({\bm r}_1,{\bm r}_2)$, while it does not for $\exp[i\delta \phi({\bm r}_1,{\bm r}_2)]$. As in Fig.~(a,c), for ${\bm r}_1-{\bm r}_2 \parallel \vec{\chi}$, the magnetic field from the vortex and that from the antivortex tend to cancel each other in the line integral from ${\bm r}_1$ to ${\bm r}_2$, giving smaller $\delta \phi({\bm r}_1,{\bm r}_2)$. As in Fig.~(b,d), for ${\bm r}_1-{\bm r}_2 \perp \vec{\chi}$, the magnetic field from the vortex and that from the antivortex tends to enhance each other in the line integral, giving larger $\delta \phi({\bm r}_1,{\bm r}_2)$. More importantly, when the polarized vortex excitation changes its size continuously from infinitesimally small size to an infinitely large size, $\exp[i\delta \phi({\bm r}_1,{\bm r}_2)]$ in the perpendicular geometry (Fig.~(b,d)) always rotates around the origin in the complex plane once, while $\exp[i\delta \phi({\bm r}_1,{\bm r}_2)]$ in the parallel geometry (Fig.~(a,c)) usually stays around some point on the unit circle.}
        \label{fig:correlation_function}
\end{figure}

The proliferation of the polarized pairs makes spatial correlations of the $Q$-field to be strongly disordered along $(-\chi_y,\chi_x)$ and weakly disordered along $(\chi_x, \chi_y)$ due to a spatially-anisotropic phase coherence of the U(1) phase  
$\phi({\bm r})$ in Eq.~(\ref{eq:SM_U(N)_saddle_point}). To see this anisotropy, let us consider a relative U(1) phase between the two spatial points ${\bm r}_1$ and ${\bm r}_2$; $\phi({\bm r}_1,{\bm r}_2) \equiv \phi({\bm r}_1)-\phi({\bm r}_2)$ and 
see how the relative phase changes by an introduction of 
a polarized pair of vortex and antivortex between ${\bm r}_1$ and 
${\bm r}_2$ [See Fig.~\ref{fig:correlation_function}]. In 2D, vortex at ${\bm r}={\bm r}_{\rm v}$ and antivortex at 
${\bm r}={\bm r}_{\rm av}$ induce 
magnetic scalar potential $\phi_m({\bm r})$; ${\bm \nabla}\times {\bm \nabla} \phi_{m}({\bm r})
= 2\pi [\delta^2({\bm r}-{\bm r}_{\rm v})-\delta^2({\bm r}-{\bm r}_{\rm av})]$, and the magnetic scalar 
potential changes the relative U(1) phase, 
$\phi ({\bm r}_1,{\bm r}_2) 
\rightarrow \phi({\bm r}_1,{\bm r}_2) 
+ \phi_m({\bm r}_1) - \phi_{m}({\bm r}_2)$. 
In terms of magnetic field ${\bm B}({\bm r}) \equiv \textcolor{red}{-} {\bm \nabla} \phi_m({\bm r})$ associated with the scalar potential, 
the relative phase change 
$\delta \phi({\bm r}_1,{\bm r}_2) \equiv 
\phi_m({\bm r}_1) - \phi_{m}({\bm r}_2)$ is given by a line integral of the field 
along a line connecting ${\bm r}_1$ and ${\bm r}_2$;
\begin{align}
\delta \phi({\bm r}_1,{\bm r}_2) = \int^{{\bm r}_2}_{{\bm r}_1} d{\bm r} \cdot {\bm B}({\bm r}) 
+ 2\pi \mathds{Z}.
\end{align}
A choice of the line gives an ambiguity of $2\pi$ times integer ($\mathds{Z}$) for 
$\delta \phi({\bm r}_1,{\bm r}_2)$, 
while it does not for $\exp [i\delta \phi({\bm r}_1,{\bm r}_2)]$, 
\begin{eqnarray}
e^{i\delta \phi({\bm r}_1,{\bm r}_2)}   
= \exp\bigg[i \int^{{\bm r}_2}_{{\bm r}_1} d{\bm r} \cdot 
{\bm B}({\bm r}) \bigg]. \label{phi-B} 
\end{eqnarray}
According to the Biot-Savart law, 
${\bm B} \equiv (B_x,B_y)$ in 2D  
can be calculated as the magnetic field induced by two antiparallel 
electric current lines along the third dimension ($z$),    
\begin{eqnarray}
\left(\begin{array}{c}
B_x({\bm r}) \\ 
B_y({\bm r}) \\ 
\end{array}\right) = - 
\frac{\hat{z} \times ({\bm r}-{\bm r}_{\rm v})}{|{\bm r}-{\bm r}_{\rm v}|^2} 
+ \frac{\hat{z} \times ({\bm r}-{\bm r}_{\rm av})}{|{\bm r}-{\bm r}_{\rm av}|^2}. 
\end{eqnarray}
with a unit vector $\hat{z}\equiv (0,0,1)$ along $z$. 
Thus, when (i) ${\bm r}_1-{\bm r}_2$ is parallel to the 
dipole vector ${\bm m} \equiv {\bm r}_{\rm v}-{\bm r}_{\rm av}$, the magnetic field 
from the vortex and that from the antivortex tend to cancel each other 
in the line integeral, giving smaller phase change 
[see, for example, the line integral along a straight line connecting ${\bm r}_1$ and ${\bm r}_2$ in  Fig.~\ref{fig:correlation_function}{a}]. 
On the other hand, when (ii) ${\bm r}_1-{\bm r}_2$ is perpendicular to the 
dipole vector, the magnetic 
field from the vortex and that from the antivortex tend to enhance each other in the line integral, 
giving larger phase change [see, for example, the line integral along a straight line connecting ${\bm r}_1$ and ${\bm r}_2$ in  
Fig.~\ref{fig:correlation_function}{b}]. More importantly, when 
the dipole length $|{\bm m}|$ changes from zero to infinite continuously, the unit 
complex number $\exp[i\delta \phi({\bm r}_1,{\bm r}_2)]$ in the perpendicular  
geometry (${\bm r}_1-{\bm r}_2 \perp {\bm m}$) always winds up once around the origin 
in the complex plane. Meanwhile, $
\exp[i\delta \phi({\bm r}_1,{\bm r}_2)]$ in the 
parallel geometry (${\bm r}_1-{\bm r}_2 \parallel {\bm m}$) does not.  This suggests that when 
$\exp[i(\phi({\bm r}_1)-\phi({\bm r}_2))]$ is averaged over 
configuration without vortices and configurations with the 
polarized vortex excitations with different dipole length, the averaged quantity must be 
significantly smaller in the perpendicular geometry than in the parallel geometry, 
due to the cancellation by the phase changes. 
Consequently, when polarized pairs proliferate in space, the U(1) phase correlation along $(\chi_x,\chi_y)$ is less disordered, while 
the long-range phase coherence along $(-\chi_y,\chi_x)$ shall be 
strongly disordered. The strong spatial anisotropy of 
the $Q$-field correlation is consistent with the emergent 
quasi-localized phase with divergent localization length along 
the topological direction, $(\chi_x,\chi_y)$, and finite localization 
length along the other direction, $(-\chi_y,\chi_x)$.

\textbf{3D cases}. For $d=3$, let us evaluate the weak  topological term in the presence of a closed loop $\Gamma$ of the vortex line,
\begin{align}
    Q_{\Gamma}({\bm r})=\mathds{1}+\ket{p({\bm r})}\left(e^{i\phi_{\Gamma}({\bm r})}-1\right)\bra{p({\bm r})}, \quad \epsilon_{ijk} \partial_{j} \partial_k \phi_{\Gamma}({\bm r}) = 2\pi \eta \int^{1}_{-1} ds \!\ \frac{dl_i}{ds} \!\ \delta^3\left({\bm r}-{\bm l}(s)\right).  
\end{align}
Here the loop $\Gamma$ is 
parameterized by $s \in [-1,1]$ as ${\bm l}(s) \equiv \left(l_x(s), l_y(s), l_z(s)\right)$ 
${\bm l}(-1)={\bm l}(1)$. A substitution 
into $S_{\rm top}$ shows that the weak  
topological term is proportional to an area of the closed loop projected onto a plane perpendicular to $(\chi_x,\chi_y,\chi_z)$;
\begin{align}
S_{\rm top}[Q_{\Gamma}]=i \eta \chi_{\mu} 
  \int\frac{\diff^3r}{8\pi} \partial_\mu\phi_{\Gamma} = -\frac{i}{4} \eta \!\  
  \chi_{\mu} \epsilon_{\mu\nu\lambda} \int^1_{-1} 
  l_{\nu} \frac{dl_{\lambda}}{ds} \!\ ds. 
\end{align}
The other parts of the action, $S_{\rm cond}+S_{\rm Gade}$, are real, and depend only on the shape of the vortex loop in the simpler case. Thereby, when vortex loops have finite projections onto the plane, the vortex loops with the same shape and different spatial orientations show destructive interference in the partition function. Consequently, the partition function near the Anderson transition in 3D chiral symmetry classes is dominated by those vortex loops that are confined in planes parallel to the topological direction, $\vec{\chi}=(\chi_x,\chi_y,\chi_z)$. Due to this spatial anisotropic proliferation of vortex loops, a localized phase proximate to a metal phase in the chiral classes with finite $|\vec{\chi}|$ is characterized by the peculiar spatial correlation of the $Q$ field; the correlation is strongly disordered within the plane perpendicular to $\vec{\chi}$, and weakly disordered along $\vec{\chi}$. Like in the 2D case, this can be also understood from a consideration of the relative U(1) phase  
$\phi({\bm r}_1,{\bm r}_2) \equiv \phi({\bm r}_1)-\phi({\bm r}_2)$  
between ${\bm r}_1$ and ${\bm r}_2$ for the two cases; 
(i) ${\bm r}_1-{\bm r}_2 \parallel \vec{\chi}$ and (ii) ${\bm r}_1-{\bm r}_2 \perp \vec{\chi}$. 
Namely, $\phi({\bm r}_1,{\bm r}_2)$ changes upon an introduction of a 
polarized vortex loop between the two points [Fig.~\ref{fig:correlation_function}(c,d)]. Eq.~(\ref{phi-B}) 
relates the change of the relative phase  $\delta \phi({\bm r}_1,{\bm r}_2)$ with 
the magnetic field ${\bm B}({\bm r})$ created by the vortex loop. 
As ${\bm B}({\bm r})$ and the vortex loop $C$ is related by the Ampere's equation, 
\begin{eqnarray}
{\bm \nabla} \times {\bm B}({\bm r}) = \textcolor{red}{-}2\pi \oint_{C} d{\bm l} \!\ \delta^3({\bm r}-{\bm l}), 
\end{eqnarray} 
the Biot-Savart law gives ${\bm B}=(B_x,B_y,B_z)$ as follows, 
\begin{eqnarray}
\left(\begin{array}{c}
B_x({\bm r}) \\ 
B_y({\bm r}) \\ 
B_z({\bm r}) \\ 
\end{array}\right) = - \frac{1}{2}\oint_C d{\bm l} \times  
\frac{({\bm r}-{\bm l})}{|{\bm r}-{\bm l}|^3}. 
\end{eqnarray}
Thereby, when (i) ${\bm r}_1-{\bm r}_2$ is parallel to a flat plane with the polarized 
vortex loop, the magnetic fields at different points 
along the line for the integral tend to cancel each other in the line integral, resulting in smaller 
$\delta \phi({\bm r}_1,{\bm r}_2)$ [see, for example, the line integral along a straight line connecting ${\bm r}_1$ and ${\bm r}_2$ shown in  Fig.~\ref{fig:correlation_function}(c)]. 
When (ii) ${\bm r}_1-{\bm r}_2$ is perpendicular to the plane with the polarized 
loop, the magnetic fields at different points  
along the line tend to enhance each other in the integral, resulting in larger 
$\delta \phi({\bm r}_1,{\bm r}_2)$ [see, for example, the line integral along a straight line connecting ${\bm r}_1$ and ${\bm r}_2$ shown in  Fig.~\ref{fig:correlation_function}(d)]. Moreover, when the size of the polarized vortex loop changes 
continuously from a tiny size to infinitely large size, 
$\exp[i\delta \phi({\bm r}_1,{\bm r}_2)]$ in the perpendicular geometry winds up around 
the origin in the complex plane, while  $\exp[i\delta \phi({\bm r}_1,{\bm r}_2)]$ in the parallel 
geometry does not. Thus, when 
$\exp[i(\phi({\bm r}_1)-\phi({\bm r}_2))]$ is averaged over 
configuration without vortex loops and configurations with the 
polarized vortex loop with different sizes, the averaged quantity must be 
significantly smaller in the perpendicular geometry than in the parallel geometry. 
This suggests that the proliferation of the polarized vortex loops makes the U(1) phase correlation perpendicular to $\vec{\chi}$ to be strongly disordered, while leaving intact the long-range phase coherence along $\vec{\chi}$. The spatially anisotropic $Q$-field correlations are expected to 
result in a spatially anisotropic transport property as in the quasi-localized phase.

Motivated by these intuitive arguments, we carry out a field-theoretical analysis of Eq.~(\ref{eq:SM_NLSM}), and show that the 
weak topological term indeed changes 
the nature of the localized phase proximate to the metal phase 
in chiral symmetry class with finite $\vec{\chi}$. Specifically, our renormalization group analysis demonstrates that Eq.~(\ref{eq:SM_NLSM}) with finite $\vec{\chi}$ has only two stable fixed points, where one of them has finite conductivity only along the topological direction $\vec{\chi}$ and vanishing conductivity in the other direction, giving a characterization of the quasi-localized phase. The other stable fixed point has finite conductivities along all directions, characterizing a metal phase. We also find a saddle-point fixed point that controls the criticality of a phase transition between these two phases. For simplicity, in this paper, we only focus on the two-dimensional chiral unitary class, while leaving other dimensions and chiral symmetry classes for future study.

\section{Dual theory for 2D AIII nonlinear sigma model} 
The field variable $Q$ in the chiral unitary (AIII) class belongs to $\mathrm{U}(N)$, satisfying  $Q^\dagger Q=\mathds{1}$. The dual transformation can be generalized 
from the $\mathrm{U}(1)$ case to the $\mathrm{U}(N)$ case. Define a Hermitian matrix 
$h_\mu \equiv -iQ^{-1}\partial_\mu Q$ and rewrite $S$ of \cref{eq:SM_NLSM} in terms of $h_\mu$ 
	\begin{equation}
		S=\int\frac{\diff^2r}{8\pi}\sum_{\mu=x,y}\left(\sigma_\mu\Tr h_\mu^2+c_\mu\Tr^2 h_\mu+i\chi_\mu \Tr h_\mu\right).
	\end{equation}
When rewriting the integral measure $\Diff[Q]$ in terms of $\Diff[h_\mu]$, 
we include $Q$-field configurations with vortex excitations, using 
a field strength associated with the Hermitian matrix. Specifically,  
the field strength tensor is given by 
	\begin{equation}
		F_{\mu\nu} \equiv \partial_\mu h_\nu-\partial_\nu h_\mu+i[h_\mu,h_\nu].
	\end{equation}
When $Q$-field is a smooth function over 
the spatial coordinate ${\bm r}$, the 2nd-order spatial derivatives 
of $Q$ is commutative, $[\partial_\mu,\partial_\nu]Q=0$, leading to 
$F_{\mu\nu}=0$.

When $Q$-field configuration includes the vortex excitation, 
$Q$ around the vortex core takes the form of 
\begin{align}
Q({\bm r})=\mathds{1}+\ket{p({\bm r})}\left(e^{i\phi({\bm r})}-1\right)\bra{p({\bm r})}, \label{eq:SM_ansatz-vortex}
\end{align}
where the U(1) phase $\phi({\bm r})$ has the quantized phase winding around the core. 
The loop integral of the spatial gradient of $\phi$ around the core gives the   
winding number, resulting in a non-zero commutator of the 2nd-order spatial 
derivatives of the phase, 
	\begin{equation}
		[\partial_x,\partial_y]\phi=2\pi\eta\delta^2(\bm{r}-\bm{r}_0).
	\end{equation}
The integer $\eta$ stands for vorticity and $\bm{r}_0$ is the location of the core. 
This yields a condition for $F_{xy}$
	\begin{equation}
		F_{xy}=2\pi\eta\delta^2(\bm{r}-\bm{r}_0)\ket{p({\bm r})}\bra{p({\bm r})}. 
	\end{equation}
When $Q$-field configurations include $n$ vortices, $\phi$ in Eq.~(\ref{eq:SM_ansatz-vortex}) has vortices at $n$
different spatial points, ${\bm r}_1, \cdots ,{\bm r}_n$. Then,  
the condition for $F_{xy}$ can be generalized into the 
$n$-vortices case, 
	\begin{equation}
		F_{xy}=J^{(n)},\quad J^{(n)}=\sum_{i=1}^n2\pi\eta_{i}\delta\left(\bm{r}-\bm{r}_i\right)\ket{p_i}\bra{p_i}.
	\end{equation}
Namely, $\ket{p_i}$ is the eigenvector $\ket{p({\bm r})}$ 
at ${\bm r}={\bm r}_i$, $\ket{p_i} \equiv \ket{p({\bm r}_i)}$, and 
positive (negative) integer $\eta_i$ is the vorticity at the $i$-th 
vortex (antivortex). 

The path integral for the paritition 
function $Z$ in Eq.~(\ref{eq:SM_NLSM}) can include all the possible 
multiple-vortices configurations through its integral measure;  
	\begin{equation}
		\int\Diff[Q]=\sum_{n=0}^{\infty}\frac{\Lambda^{2n}}{n!}\prod_{i=1}^n\left(\int \Diff[p_i] \int \diff^2r_i \sum_{\eta_{i}=-\infty}^{+\infty}y_{\eta_i}\right) 
  \int\Diff[h]\,\delta[F_{xy}-J^{(n)}].
	\end{equation}
 $y_{\eta}$ is fugacity of single vortex with the vorticity $\eta$. $\delta(F_{xy}-J^{(n)})$ imposes the delta function condition for every matrix element of the $N$ by $N$ Hermitian matrix $F_{xy}$. $\Lambda^{-1}$ is a linear dimension of a vortex core size, which plays the role of an infrared cutoff of the theory. A factor of $``n!"$ in the right-hand side is a symmetry factor that takes into account $n!$
 equivalent configurations with $n$-vortices. Here we consider that $n$-vortices are spatially separated from each other, and treat $\ket{p_i}=
 \ket{p({\bm r}_i)}$ for different $i$ as independent integral variables; 
 $\prod^n_{i=1} \int \Diff[p_i]$.  When ${\bm r}_i$ becomes equal to ${\bm r}_j$ in the integrals over ${\bm r}_i$ and ${\bm r}_j$, $\ket{p_i}$ must be identical to $\ket{p_j}$ because they are from the same $\ket{p({\bm r})}$. Thereby, $\int \Diff[p_i] \int \Diff [p_j]$ reduces to $\int \Diff[p_i]$ for the ${\bm r}_i={\bm r}_j$ case. Likewise, a double sum $\sum_{\eta_i}\sum_{\eta_j}$ reduces to a single sum over $\eta_m \equiv \eta_i+\eta_j$ for the ${\bm r}_i={\bm r}_j$ case.  
Note that the $n=0$ term corresponds to a continuous configuration of the $Q$ field without vortices. 
For simplicity, let's define the following notation
\begin{equation}\label{eq:SM_integral-measure}
\int\Diff[J^{(n)}] \equiv \prod_{i=1}^n\left(\int \Diff[p_i] \int \diff^2r_i\sum_{\eta_i=-\infty}^{+\infty}y_{\eta_i}\right), 
	\end{equation}
and 
\begin{equation}
\int\Diff[Q]=\int\Diff[h]\sum_{n=0}^\infty 
\frac{\Lambda^{2n}}{n!}\int\Diff[J^{(n)}]\delta[F_{xy}-J^{(n)}]. 
\end{equation}
In terms of the integral measure, the partition function is given by 
	\begin{equation}
		Z= \int \Diff [Q] e^{-S} = \int\Diff[h]\sum_{n=0}^{\infty}\frac{\Lambda^{2n}}{n!}\int\Diff[J^{(n)}]\delta[F_{xy}-J^{(n)}]\exp\left(-S\right).
	\end{equation}
The $\delta$ function can be exponentiated  in terms of an auxiliary field 
$\Theta({\bm r})$ ($N$ times $N$ Hermitian matrix)
	\begin{equation}
		Z=\int\Diff[h]\int\Diff[\Theta]\sum_{n=0}^{\infty}\frac{\Lambda^{2n}}{n!}\int\Diff[J^{(n)}]\exp\left[-S-i\int\diff^2r\,\Tr\left[\Theta({\bm r}) 
  \left(F_{xy}({\bm r})-J^{(n)}({\bm r})\right)\right]\right].
	\end{equation}

The partition function can be separated into two parts; one is given by an integral over $h$ field, and the other is given by 
an integral over vortices degrees of freedom. 
The two integrals lead to an effective action for 
the auxiliary field $\Theta$,
\begin{equation}
Z=\int\Diff[\Theta]\,\exp\left(-S_{\text{smooth}}-S_{\text{vortex}}\right),
\end{equation}
\begin{equation}
S_{\text{smooth}} \equiv -\ln\int\Diff[h]\,\exp\left[-S-i\int\diff^2r\,\Tr\left(\Theta F_{xy}\right)\right],
\end{equation}
\begin{align}
e^{-S_{\text{vortex}}}
&= \sum_{n=0}^{\infty} \frac{\Lambda^{2n}}{n!} 
\prod^n_{i=1} \bigg( \int \Diff [p_i] \int \diff^2 {\bm r}_i 
\sum^{+\infty}_{\eta_i=-\infty} \bigg) 
\exp\left[ i 2\pi \sum^n_{i=1} \eta_i \bra{p_i} \Theta({\bm r}_i) \ket{p_i} \right], \label{exp-Sv} \\
& ``="  \sum_{n=0}^{\infty}\frac{\Lambda^{2n}}{n!} 
\bigg(\int \Diff [p] \int \diff^2 {\bm r} 
\sum^{+\infty}_{eta=-\infty} e^{i2\pi \eta \bra{p} \Theta({\bm r}) \ket{p}}\bigg)^n = \exp \bigg[\Lambda^2 \int \Diff[p] \int 
\diff^2 r \!\ \sum^{\infty}_{\eta=-\infty} 
e^{i 2\pi \eta \bra{p} \Theta({\bm r}) \ket{p}} \bigg],  \nonumber 
\end{align}
or equivalently, 
\begin{align}
S_{\text{vortex}}= -\Lambda^2 \int \Diff[p] \int 
\diff^2 r \!\ \sum^{\infty}_{\eta=-\infty} 
e^{i 2\pi \eta \bra{p} \Theta({\bm r}) \ket{p}}. \label{ln-Sv}
\end{align} 
Note that the multiple integrals over 
${\bm r}_i$ and $\ket{p_i}$ for different $i$ in Eq.~(\ref{exp-Sv}) 
are taken in such  a way that any double integral over $\ket{p_i}$ 
and $\ket{p_j}$ reduces to a single integral $\ket{p_i}$ whenever 
${\bm r}_i={\bm r}_j$ in the integrals over ${\bm r}_i$ and ${\bm r}_j$. 
In this sense, an exponential of Eq.~(\ref{ln-Sv}) is not exactly the same as Eq.~(\ref{exp-Sv}). For simplicity of presentation, 
we write $S_{\rm vortex}$ and its associated function $S_y$ (see below) as in Eq.~(\ref{ln-Sv}), while their exponentials are always defined with the same 
definition of the multiple integrals as Eq.~(\ref{exp-Sv}).  
  
\emph{Smooth part}. The integral over $h$ field is a partition function from 
$Q$-field configurations without vortices (`smooth part');
\begin{equation}
\begin{gathered}
S_{\text{smooth}}=-\ln\int\Diff[h]\,\exp\left[-\int\frac{\diff^2r}{8\pi}\sum_{\mu=x,y}\left(\sigma_\mu\Tr h_\mu^2+c_\mu\Tr^2 h_\mu+i\chi_\mu \Tr h_\mu\right)\right.\\
\left.-i\int\diff^2r\,\Tr\left(\Theta \left(\partial_x h_y-\partial_y h_x+i[h_x,h_y]\right)\right)\right].
\end{gathered}
\end{equation}
To carry out the Gaussian integral over $h_{\mu}$, use $\mathrm{u}(N)$ Lie algebra, $T^a$ ($a=0,1,\cdots,D-1  \equiv N^2-1)$, as a basis of the $N$ by $N$ Hermitian matrices $h_{\mu}$ and $\Theta$. The Lie algebra 
$T_a$ satisfies
\begin{equation}
\Tr\left(T^a T^b\right)=\delta_{ab},\quad \sum_{\tilde{a}=1}^{D-1} 
  T_{ij}^{\tilde{a}}T_{kl}^{\tilde{a}}= 
\delta_{il}\delta_{jk}-\frac{1}{N}\delta_{ij}\delta_{kl}
,\quad \Tr\left(T^{\tilde{a}}\right)=0,
\quad T^0=\frac{1}{\sqrt{N}}\mathds{1},  
\label{eq:SM_Ta_identity}
\end{equation}
with a  a structure factor constant of the $\mathrm{u}(N)$ Lie algebra, 
\begin{equation}
f_{abc} \equiv -i\Tr\left([T^a,T^b]T^c\right).
\end{equation} 
By expending $h_\mu$ and $\Theta$ in terms of the basis,
\begin{equation}
h_\mu=\sum^{D-1}_{a=0,1,\cdots}\pi_\mu^aT^a,\quad \Theta=\sum^{D-1}_{a=0}\theta^aT^a, 
\end{equation}
we have 
\begin{equation}
-i\Tr\left[\Theta(\partial_x h_y-\partial_y h_x+i[h_x,h_y])\right]=
-i\left(\theta^a\partial_x \pi_y^a-\theta^a\partial_y \pi_x^a 
-\pi^a_x\pi^b_y\theta^cf_{abc}\right), 
\end{equation}
\begin{equation}
-i\int\diff^2r\left(\theta^a\partial_x \pi_y^a-\theta^a\partial_y \pi_x^a-\pi^a_x\pi^b_y\theta^cf_{abc}\right)=
-i\int\diff^2r\left(\pi_x^a\partial_y\theta^a-\pi_y^a\partial_x\theta^a
-\pi^a_x\pi^b_y\theta^cf_{abc}\right),
\end{equation}
and 
\begin{equation}
\begin{gathered}
S_{\text{smooth}}=-\ln\int\Diff[h]\exp\left[-\int\frac{\diff^2r}{8\pi}\left[(\sigma_\mu+Nc_\mu\delta_{a0})\left(\pi^a_\mu\right)^2+i\sqrt{N}\chi_\mu\pi_\mu^0\right]\right]\\
\times\exp\left[-i\int\diff^2r 
\left(\pi_x^a\partial_y\theta^a-\pi_y^a\partial_x\theta^a-\pi^a_x\pi^b_y\theta^cf_{abc}\right)\right]. 
\end{gathered}
\end{equation}

The Gaussian integral over $h_{\mu}$ can be separated into the integral over the U(1) part $\pi^0$ and the 
integral over the SU($N$) part $\pi^{\tilde{a}}$ ($\tilde{a}=1,2,\cdots, D-1$),
\begin{equation}
	\begin{gathered}
			S_{\text{smooth}}^0=-\ln\int\Diff[\pi^0]\exp\left[-\int\frac{\diff^2r}{8\pi}\begin{pmatrix}
				\pi^0_x & \pi^0_y
			\end{pmatrix}\begin{pmatrix}
				\sigma_x+Nc_x & 0 \\
				0 & \sigma_y+Nc_y
			\end{pmatrix}\begin{pmatrix}
				\pi^0_x \\ \pi^0_y
			\end{pmatrix}\right]\\
			\times\exp\left[\int\frac{\diff^2r}{8\pi}\begin{pmatrix}
				-i\sqrt{N}\chi_x-8\pi i\partial_y\theta^0 & 
    -i\sqrt{N}\chi_y+8\pi i\partial_x\theta^0
			\end{pmatrix}\begin{pmatrix}
				\pi^0_x \\
				\pi^0_y
			\end{pmatrix}\right],
		\end{gathered}
	\end{equation}
	\begin{equation}
		\begin{gathered}
			S_{\text{smooth}}^1=-\ln\int\prod_{\tilde{a}=1}^{D-1}\Diff[\pi^{\tilde{a}}]\exp\left[-\int\frac{\diff^2r}{8\pi}\begin{pmatrix}
				\pi^{\tilde{a}}_x & \pi^{\tilde{a}}_y
			\end{pmatrix}\begin{pmatrix}
				\sigma_x\delta_{\tilde{a}\tilde{b}} & -4\pi i f_{\tilde{a}\tilde{b}\tilde{c}}\theta^{\tilde{c}} \\
				-4\pi i f_{\tilde{b}\tilde{a}\tilde{c}}\theta^{\tilde{c}} & \sigma_y\delta_{\tilde{a}\tilde{b}}
			\end{pmatrix}\begin{pmatrix}
				\pi^{\tilde{b}}_x \\ \pi^{\tilde{b}}_y
			\end{pmatrix}\right]\\
			\times\exp\left[\int\frac{\diff^2r}{8\pi}\begin{pmatrix}
				-8\pi i\partial_y\theta^{\tilde{a}} & 8\pi i\partial_x\theta^{\tilde{a}}
			\end{pmatrix}\begin{pmatrix}
				\pi^{\tilde{a}}_x \\
				\pi^{\tilde{a}}_y
			\end{pmatrix}\right].
		\end{gathered}
	\end{equation}
An integration over $\pi^0$ leads to a free theory of $\theta^0$; 
	\begin{equation}
		\begin{gathered}
S^{0}_{\text{smooth}}=2\pi\int\diff^2r\left[\frac{\left(\partial_x\theta^0 -\sqrt{N}\chi_y/8\pi\right)^2}{\sigma_y+Nc_y}+\frac{\left(\partial_y\theta^0 +\sqrt{N}\chi_x/8\pi\right)^2}{\sigma_x+Nc_x}\right].
		\end{gathered}
	\end{equation}
An integration over $\pi^{\tilde{a}}$ ($\tilde{a}=1,\cdots, D-1$) 
leads to an interacting action for $\theta^{\tilde{a}}$, where the interaction 
appears through a metric tensor of SU$(N)$ matrices. To see this, it is convenient to  
define anti-symmetric Hermitian matrix $P$ and real anti-symmetric matrix $Z$ as follows; 
	\begin{equation}
		P_{\tilde{a}\tilde{b}} \equiv -4\pi if_{\tilde{a}\tilde{b}\tilde{c}}\theta^{\tilde{c}},\quad Z_{\tilde{a}\tilde{b}} \equiv f_{\tilde{a}\tilde{b}\tilde{c}}\theta^{\tilde{c}}.
	\end{equation}
The matrix of quadratic part in $S^1_{\rm smooth}$ and its inverse  
can be put in block forms, 
\begin{equation}
\begin{pmatrix}
			\sigma_x\mathds{1} & P \\
			P^\T & \sigma_y\mathds{1}
		\end{pmatrix}=\begin{pmatrix}
			\sqrt{\sigma_x}\mathds{1} & 0 \\
			0 & \sqrt{\sigma_y}\mathds{1}
		\end{pmatrix}\begin{pmatrix}
			\mathds{1} & \frac{P}{\sqrt{\sigma_x\sigma_y}} \\
			\frac{P^\T}{\sqrt{\sigma_x\sigma_y}} & \mathds{1}
		\end{pmatrix}\begin{pmatrix}
			\sqrt{\sigma_x}\mathds{1} & 0 \\
			0 & \sqrt{\sigma_y}\mathds{1}
\end{pmatrix}, 
\end{equation} 
\begin{align}
		\begin{pmatrix}
				\sigma_x\mathds{1} & \bar{P} \\
				\bar{P}^\T & \sigma_y\mathds{1}
			\end{pmatrix}^{-1}  
   &=\begin{pmatrix}
				\frac{1}{\sqrt{\sigma_x}}\mathds{1} & 0 \\
				0 & \frac{1}{\sqrt{\sigma_y}}\mathds{1}
			\end{pmatrix}
			\begin{pmatrix}
				\mathds{1}+\bar{P}(\mathds{1}-\bar{P}^\T\bar{P})^{-1}\bar{P}^\T & -\bar{P}(\mathds{1}-\bar{P}^\T\bar{P})^{-1} \\
				-(\mathds{1}-\bar{P}^\T\bar{P})^{-1}\bar{P}^\T & (\mathds{1}-\bar{P}^\T\bar{P})^{-1}
			\end{pmatrix}
			\begin{pmatrix}
				\frac{1}{\sqrt{\sigma_x}}\mathds{1} & 0 \\
				0 & \frac{1}{\sqrt{\sigma_y}}\mathds{1}
			\end{pmatrix}\\ 
   &=\begin{pmatrix}
				\frac{1}{\sqrt{\sigma_x}}\mathds{1} & 0 \\
				0 & \frac{1}{\sqrt{\sigma_y}}\mathds{1}
			\end{pmatrix}
			\begin{pmatrix}
				\mathds{1}-\bar{P}(\mathds{1}+\bar{P}^2)^{-1}\bar{P} & -\bar{P}(\mathds{1}+\bar{P}^2)^{-1} \\
				(\mathds{1}+\bar{P}^2)^{-1}\bar{P} & (\mathds{1}+\bar{P}^2)^{-1}
			\end{pmatrix}
			\begin{pmatrix}
				\frac{1}{\sqrt{\sigma_x}}\mathds{1} & 0 \\
				0 & \frac{1}{\sqrt{\sigma_y}}\mathds{1}
      \end{pmatrix}, 
\end{align}
with $\bar{P} \equiv P/\sqrt{\sigma_x\sigma_y}$. Note that following 
identities hold true for an arbitrary matrix $A$, 
\begin{align}
		\frac{(1-A)^{-1} + (1+A)^{-1}}{2} 
  =\frac{(1-A)^{-1}\left[(1+A) + (1-A)\right](1+A)^{-1}}{2}
  = (1-A^2)^{-1},
\end{align}
and 
\begin{align}
		\frac{(1-A)^{-1} - (1+A)^{-1}}{2} 
  =\frac{(1-A)^{-1}\left[(1+A) - (1-A)\right](1+A)^{-1}}{2}
  = A (1-A^2)^{-1}.
\end{align}
By substituting $A=i\bar{P}$ into the first line, we obtain
$\mathds{1}-\bar{P}(\mathds{1}+\bar{P}^2)^{-1}\bar{P}=(\mathds{1}+\bar{P}^2)^{-1}=\frac{(\mathds{1}-i\bar{P})^{-1}+(\mathds{1}+i\bar{P})^{-1}}{2}$. By substituting $A=i\bar{P}$ 
into the second line, we obtain 
$-\bar{P}(\mathds{1}+\bar{P}^2)^{-1}=i\frac{(\mathds{1}-i\bar{P})^{-1}-(\mathds{1}+i\bar{P})^{-1}}{2}$. In terms of $g \equiv (\mathds{1}-i\bar{P})^{-1}$, the inverse is given by  
\begin{align}
	\begin{pmatrix}
			\sigma_x\mathds{1} & \bar{P} \\
			-P & \sigma_y\mathds{1}
		\end{pmatrix}^{-1}=\begin{pmatrix}
			\frac{1}{\sqrt{\sigma_x}}\mathds{1} & 0 \\
			0 & \frac{1}{\sqrt{\sigma_y}}\mathds{1}
		\end{pmatrix}
		\begin{pmatrix}
			\frac{g+g^\T}{2} & i\frac{g-g^\T}{2} \\
			-i\frac{g-g^\T}{2} & \frac{g+g^\T}{2}
		\end{pmatrix}
		\begin{pmatrix}
			\frac{1}{\sqrt{\sigma_x}}\mathds{1} & 0 \\
			0 & \frac{1}{\sqrt{\sigma_y}}\mathds{1}
		\end{pmatrix}.
\end{align}. 
Thus, an effective action for  $\theta^{\tilde{a}}$ 
is given by 
	\begin{equation}
		S_{\text{smooth}}^1=2\pi \int\diff^2r\,g_{\tilde{a}\tilde{b}}\left(\frac{\partial_x\theta^{\tilde{a}}\partial_x\theta^{\tilde{b}}}{\sigma_y}+\frac{\partial_y\theta^{\tilde{a}}\partial_y\theta^{\tilde{b}}}{\sigma_x}+i\frac{\partial_x\theta^{\tilde{a}}\partial_y\theta^{\tilde{b}}}{\sqrt{\sigma_x\sigma_y}}-i\frac{\partial_y\theta^{\tilde{a}}\partial_x\theta^{\tilde{b}}}{\sqrt{\sigma_x\sigma_y}}\right)-\int\diff^2r\,\ln \det \sqrt{\sigma_x \sigma_y} g.
	\end{equation}
Note that $g$ is the metric tensor of the $\mathrm{su}(N)$ part, and 
$\ln \det g$ in the right-hand side can be interpreted  as an integral measure of the su$(N)$ 
part of $\Theta$ (see below). 
As they depend on $\theta^{\tilde{a}}$, $S^{1}_{\rm smooth}$ is generally not a free theory.  
	
	\emph{Vortex part}. In terms of the u($N$) Lie algebra, the vortex part can be given by
	\begin{equation}
		S_{\text{vortex}}=-\Lambda^2 \int\diff^2r \int\Diff\ket{p}\sum_\eta y_{\eta} \exp\left[
   2\pi i \eta \bigg(\theta^0 \frac{1}{\sqrt{N}} + \sum^{D-1}_{\tilde{a}=1} \theta^{\tilde{a}}
  \bra{p}T^{\tilde{a}}\ket{p} \bigg)\right].
	\end{equation}
	
  Now we obtained the dual theory in terms of u(1) part $\theta^0$ and su($N$) part $\theta^a$ ($a=1,\cdots, D-1$), 
\begin{align}
S&=S_0+S_\theta+S_y, \nonumber \\  
S_0&=\frac{2\pi}{\overline{\sigma+Nc}^2}\int\diff^2r\,(\sigma_\mu+Nc_\mu)\left(\partial_\mu\theta^0-\frac{\sqrt{N}}{8\pi}\epsilon_{\mu\nu}\chi_\nu\right)^2, \nonumber \\ 
S_\theta&=\frac{2\pi}{\overline{\sigma}^2}\int\diff^2r\,\sqrt{\sigma_\mu\sigma_\nu}(\delta_{\mu\nu}+i\epsilon_{\mu\nu})g_{ab}\partial_\mu\theta^a\partial_\nu\theta^b, \nonumber \\
S_y&=-\Lambda^2\int\diff^2r\int\diff\ket{p}\sum_{\eta=-\infty}^{+\infty}y_{\eta}\exp\left[2\pi i\eta\left(\theta^0\frac{1}{\sqrt{N}}+\sum_{a=1}^{N^2-1}\braket{p|T^a|p}\theta^a\right)\right]. \label{eq:SM_Sy}
\end{align}
$\overline{\sigma}$ in $S_0$ and $S_\theta$ is the geometric average of $\sigma_{\mu}$ ($\mu=x,y$), $\overline{\sigma}=\sqrt{\sigma_x\sigma_y}$, and $\overline{\sigma + Nc} = \sqrt{(\sigma_x + Nc_x)(\sigma_y + N c_y)}$. 
The summation is assumed for repeated indices $\mu,\nu=x,y$ $a,b=1,\cdots,D-1$. $g_{ab}$ in $S_\theta$ is a nonlinear term of the su($N$) part and it is defined in terms of the structure factor constant $f_{abc}$ of the su($N$) Lie algebra, 
	\begin{equation}
		g=\left(\mathds{1}-\frac{4\pi}{\overline{\sigma}}Z\right)^{-1},\quad Z_{ab}=f_{abc}\theta^c.
	\end{equation}
	$g$ and $Z$ are both $(N^2-1)\times(N^2-1)$ matrices. 
$\ket{p}$ in $S_y$ is a column vector from $\mathbb{CP}^{N-1}$, where $\ket{p}$ and $e^{i\phi}\ket{p}$ are regarded as the same quantity. Fugacity $y_{\eta}$ is 
a non-negative dimensionless quantity. The partition function is given by
$\mathcal{Z}=\int\Diff[\Theta]\,e^{-S_0-S_\theta-S_y}$, with the integral measure, 
\begin{equation}
		\int\Diff[\Theta]=\int_{-\infty}^{+\infty}\prod_{\bm{r}}\diff\theta^0\prod_{a=1}^{N^2-1}\diff\theta^a\det g(\theta(\bm{r})).
	\end{equation}
The measure is flat for $\mathrm{u}(1)$ part and it is curved for 
 $\mathrm{su}(N)$ part.  Note that the mapping between 2D NLSM of class 
 AIII and the dual theory described above are exact, and no 
 approximation has been made so far. 
	
\section{Renormalization of the dual theory}
\subsection{Approximations}
Vortex terms with lower vorticity are more relevant than those with higher vorticity. In fact, it is easier for a metal phase to put up vortices with a lower winding number than those with a higher winding number. Thereby, a transition from metal to insulator phase must be primarily driven by the proliferation of vortices with $\eta=\pm 1$. Since the theory must be invariant under $\eta\to-\eta$, $y_{-1}=y_{+1}=y$ and, 
	\begin{equation}
		S_y=-2\Lambda^2 y \!\ \int\diff^2r\int\diff\ket{p} \cos\left[2\pi \left(\theta^0\frac{1}{\sqrt{N}}+\sum_{a=1}^{N^2-1}\braket{p|T^a|p}\theta^a\right)\right].
	\end{equation}
For $N=1$, this dual theory reduces to the sine-Gordon theory. 

In the metal phase, the $g$-matrix can be expanded in the power of the conductivity,  $g = \mathds{1}+\mathrm{O}(1/\sqrt{\sigma_x\sigma_y})$. When $\overline{\sigma} \equiv \sqrt{\sigma_x\sigma_y} \gg 1$, $g$ matrix can be approximated by the identity matrix,  and $S_\theta$ becomes a free action of $\theta$ with the flat integral measure of the su$(N)$ part,
	\begin{equation}
		S_\theta=2\pi\int\diff^2r\sum_{a=1}^{N^2-1}\left[\frac{(\partial_x\theta^a)^2}{\sigma_y}+\frac{(\partial_y\theta^a)^2}{\sigma_x}\right].
	\end{equation}
When fugacity $y$ is much smaller than the inverse of the conductivity, $y\ll 1/\overline{\sigma}$, we can further treat $y$ as the perturbation, and treat $S_0+S_\theta$ as the free Gaussian theory (perturbative RG analysis). 

The weak 
topological term ($\vec{\chi} \ne 0$) brings about a non-uniform saddle-point state of 
the Gaussian theory of $\theta^0({\bm r})$; $\theta^0_{\text{sd}}({\bm r})=\frac{\sqrt{N}}{8\pi}\epsilon_{\mu\nu}r_\mu\chi_\nu$. 
A fluctuation around the non-uniform state can be included by 
$\theta^0= \theta^0_{\mathrm{sd}}+\tilde{\theta}^0$. Thereby, the fluctuation part $\tilde{\theta}^0$ satisfies the 
Born-von Karman periodic boundary 
condition. Due to the non-uniform background, an argument of 
the vortex term acquires a term that depends linearly on 
the spatial coordinate. For $\vec{\chi}=(0,\chi_y)$, $S_0$ and $S_y$ 
for the fluctuation part are given by
	\begin{equation}
		S_0=2\pi\int\diff^2r\left[\frac{(\partial_x\theta^0)^2}{\sigma_y+Nc_y}+\frac{(\partial_y\theta^0)^2}{\sigma_x+Nc_x}\right],
	\end{equation}
	\begin{equation}
		S_y=-2y\Lambda^2\int\diff^2r\int\diff\ket{p}\cos\left[2\pi \left(\theta^0\frac{1}{\sqrt{N}}+\sum_{a=1}^{N^2-1}\braket{p|T^a|p}\theta^a\right)+\frac{\chi_y}{4}x\right].
	\end{equation} 
Here we call the fluctuation part  $\tilde{\theta}^0$ as $\theta^0$. For convenience, we use $\theta^0$ for the fluctuation part henceforth. 
 
\subsection{Momentum-space decomposition}
To study a phase diagram of the dual theory, we use a momentum-shell renormalization group analysis. To this end, we introduce a UV cutoff of the dual theory in the momentum space and decompose the field variable $\Theta$ into slow modes $\Theta_<$ with smaller momenta and fast modes $\Theta_>$ with larger momenta. By integrating the fast modes in the partition function, we obtain an effective action for the slow modes. Then, we rescale the momenta ${\bm k}$, such that the field variable of the effective action shares the same UV cutoff as the original theory. A comparison between coupling constants in the action before and after integration and rescaling gives a set of RG equations for the coupling constants.
	
When a free theory is isotropic in space, the UV cutoff is introduced in the momentum space. Since the free part of the theory is spatially anisotropic, we introduce the cutoff in energy.  
For the su($N$) part, the energy and its cutoff are given by $\epsilon_{\bm k} \rightarrow \epsilon_{0,{\bm k}} \equiv k^2_x/\sigma_y + k^2_y/\sigma_x$ and 
$\epsilon_{0,\Lambda} \equiv \Lambda^2/\sigma_y + \Lambda^2/\sigma_x$. For the u($1$) part, they are given by $\epsilon_{{\bm k}} \rightarrow \epsilon_{N,\bm k} \equiv k^2_x/(\sigma_y + Nc_y) + k^2_y/(\sigma_x+N c_x)$ and  
$\epsilon_{N,\Lambda} \equiv \Lambda^2/(\sigma_y + Nc_y) + \Lambda^2/(\sigma_x+N c_x)$. With these cutoffs, we 
decompose the fast and slow mode as follows, 
\begin{align}
\theta^a({\bm r}) = \theta^a_{<}({\bm r}) + \theta^a_{>}({\bm r}) 
\equiv \frac{1}{V}\sum_{|\epsilon_{0,\bm k}|<\epsilon_{0,\Lambda} e^{-2dl}} e^{i{\bm k}\cdot {\bm r}} 
\!\ 
\theta^{a}({\bm k}) + \frac{1}{V}\sum_{\epsilon_{0,\Lambda} e^{-2dl}< \epsilon_{0,\bm k}<
\epsilon_{0,\Lambda}} e^{i{\bm k}\cdot{\bm r}} \!\ \theta^{a}({\bm k}),   
\end{align}
for $a=1,\cdots,D-1$. For $\theta^0$, we replace  $\epsilon_{0,{\bm k}}$ and $\epsilon_{0,\Lambda}$ by $\epsilon_{N,{\bm k}}$ and $\epsilon_{N,\Lambda}$, respectively. After the integration of the fast modes, we will change ${\bm k}$ into ${\bm k}^{\prime} = {\bm k} e^{dl}$, putting  
$\epsilon_{N,\Lambda} e^{-2dl}$ back into the original UV energy cutoff (rescaling). The free theory can be decomposed into fast and slow mode parts, 
\begin{align}
        S_{0} + S_{\theta} &= S_{0,<} + S_{\theta,<} + S_{0,>} + S_{\theta,>}, \nonumber \\
		S_{\theta,<} &=\frac{2\pi}{V} \sum^{D-1}_{a=1}
  \sum_{\epsilon_{0,\bm k} <\epsilon_{0,\Lambda} e^{-2dl}} 
  \left(\frac{k_x^2}{\sigma_y}+\frac{k_y^2}{\sigma_x}\right)\tilde{\theta}^a(-\bm{k})\theta^a(\bm{k}), \nonumber \\
  S_{\theta,>}&=\frac{2\pi}{V} \sum^{D-1}_{a=1}\sum_{\epsilon_{0,\Lambda}e^{-2dl}<\epsilon_{0,\bm k}<\epsilon_{0,\Lambda}} 
  \left(\frac{k_x^2}{\sigma_y}+\frac{k_y^2}{\sigma_x}\right)\theta^a(-\bm{k})\theta^a(\bm{k}). 
\end{align}
For $S_{0,<}$ and $S_{0,>}$, one has only to change $\epsilon_{0,\bm k}$ and 
$\epsilon_{0,\Lambda}$ by $\epsilon_{N,{\bm k}}$ and $\epsilon_{N,\Lambda}$ respectively.  

The integration of the fast modes in the partition function gives an effective action 
for the slow modes, 
\begin{align}
Z = \int {\rm D}[\Theta] e^{-S_0 - S_{\theta} - S_y} =  Z_{0,>}
\int {\rm D}[\Theta_{<}] e^{-S_{0,<} - S_{\theta,<} - \langle S_y\rangle 
+ \frac{1}{2}\big(\langle S^2_y \rangle - \langle S_y\rangle^2\big)+ {\cal O}(y^3)},  \label{exp:Z-rg}
\end{align}
where  
\begin{equation}\label{eq:SM_Scouple}
S_y=-2y\Lambda^2\int\diff^2r\int\diff\ket{p}\cos\left[2\pi \braket{p|\left(\Theta_<+\Theta_>\right)|p}+\frac{\chi_y}{4}x\right], 
\end{equation}
with  
\begin{align}
&\mean{\cdots}=\frac{\int\Diff[\Theta_>]\cdots e^{-S_{0,>}-S_{\theta_>}}}{\int\Diff[\Theta_>]e^{-S_{0,>}-S_{\theta_>}}}, \ \ \ Z_{0,>} \equiv \int\Diff[\Theta_>]e^{-S_{0,>}-S_{\theta_>}}, \nonumber \\
&\Theta_{<} \equiv \theta^0_{<} \frac{1}{\sqrt{N}} \mathds{1} 
+ \sum^{D-1}_{a=1} \theta^{a}_{<} T^a, \quad 
\Theta_{>} \equiv \theta^0_{>} \frac{1}{\sqrt{N}} \mathds{1} 
+ \sum^{D-1}_{a=1} \theta^{a}_{>} T^a. 
\end{align}

In the following, we will calculate the effective action for the slow mode; 
\begin{equation}
	S_{\text{eff}}=S_{0,<}+S_{\theta,<}+\mean{S_y}-\frac{1}{2}\left(\mean{S_y^2}-\mean{S_y}^2\right)+\mathrm{O}(y^3).
\label{effective-S}
\end{equation}

Before closing this section, we note some remarks on the multiple integrals over ${\bm r}_i$ and $\ket{p_i}$ that appear in $e^{-\langle S_y\rangle }$, and $\langle S^2_y\rangle-\langle S_y\rangle^2$; 
\begin{align}
e^{-\langle S_y \rangle} \equiv 
\sum^{\infty}_{n=0} \frac{(2y \Lambda^2)^n}{n!} 
\prod^n_{i=1} \bigg(\int \Diff p_i \int \diff^2 {\bm r}_i 
\bigg) \prod^n_{i=1} \bigg\langle 
\cos \left[ 2\pi \bra{p_i} 
\left(\Theta_{<}({\bm r}_i) + \Theta_{>}({\bm r}_i)\right)  
\ket{p_i} + \frac{\chi_y}{4} x_i \right] \bigg\rangle,  \label{esy}
\end{align}
\begin{align}
&\langle S^2_y \rangle 
- \langle S_y \rangle^2 \equiv 
\left(2y \Lambda\right)^2 \int \diff r_1 \int \diff r_2 
\int \Diff p_1 \int \Diff p_2 \nonumber \\
&\hspace{0.8cm} 
\bigg\{ 
\Big\langle \cos\left[2\pi \bra{p_1} 
\left(\Theta_{<}({\bm r}_1) + \Theta_{>}({\bm r}_1)\right)  
\ket{p_1} + \frac{\chi_y}{4} x_1\right] 
\cos\left[2\pi \bra{p_2} 
\left(\Theta_{<}({\bm r}_2) + \Theta_{>}({\bm r}_2)\right)  
\ket{p_2} + \frac{\chi_y}{4} x_2\right] \Big\rangle \nonumber \\
&\hspace{1.2cm}- \Big\langle \cos\left[2\pi \bra{p_1} 
\left(\Theta_{<}({\bm r}_1) + \Theta_{>}({\bm r}_1)\right)  
\ket{p_1} + \frac{\chi_y}{4} x_1\right] \Big\rangle 
\Big\langle \cos\left[2\pi \bra{p_2} 
\left(\Theta_{<}({\bm r}_2) + \Theta_{>}({\bm r}_2)\right)  
\ket{p_2} + \frac{\chi_y}{4} x_2\right] \Big\rangle 
\bigg\}.  \label{s2y-sy2}
\end{align}
The multiple integrals in Eqs.~(\ref{esy},\ref{s2y-sy2}) 
must be taken in the same sense as 
in Eq.~(\ref{exp-Sv}). Namely, any pair of integrals over $\ket{p_i}$ and $\ket{p_j}$ reduce to a single integral over $\ket{p_i}$, whenever ${\bm r}_i$ becomes equal to ${\bm r}_j$ in the integrals 
over ${\bm r}_i$ and ${\bm r}_j$. As shown below, $e^{-\langle S_y \rangle}$ thus defined leads to 
$e^{-S_y}$ for the slow mode with the same definition of the multiple integral as Eq.~(\ref{exp-Sv}). Meanwhile, Eq.~(\ref{s2y-sy2}) reduces to a term that 
depends on neither $\ket{p_1}$ nor $\ket{p_2}$, so we can safely re-exponentiate 
$\langle S^2_y \rangle - \langle S_y \rangle^2$ as in Eq.~(\ref{exp:Z-rg}).

	\subsection{Calculations of renormalizations}
	\subsubsection{Correlation functions}
To this end, it is useful to calculate the following correlation function, 
	\begin{equation}
		G(\bm{r}_1-\bm{r}_2,p_1,p_2)=\mean{\braket{p_1|2\pi\Theta_>(\bm{r}_1)|p_1}\braket{p_2|2\pi\Theta_>(\bm{r}_2)|p_2}}. \label{correlation1}
	\end{equation}
Here $\mean{\cdots}$ is defined by the integration of the fast mode with its 
free theory $S_{0,>}+S_{\theta,>}$. In terms of the basis of the u($N$) Lie algebra, 
$\Theta_>=\sum_{a=0}^{N^2-1}\theta_>^aT^a$, the function is given by 
\begin{equation}
G(\bm{r}_1-\bm{r}_2,p_1,p_2)=4\pi^2\sum_{a=0}^{N^2-1}\mean{\theta_>^a(\bm{r}_1)\theta_>^a(\bm{r}_2)} 
\braket{p_1|T^a|p_1}\braket{p_2|T^a|p_2}. 
\end{equation}
Since $\mean{\theta_>^a(\bm{r}_1)\theta_>^a(\bm{r}_2)}$ is same for $a=1,2,\dots,N^2-1$, 
a usage of \cref{eq:SM_Ta_identity} simplifies the function as follows,   
\begin{equation}
	G(\bm{r}_1-\bm{r}_2,p_1,p_2)=4\pi^2\left[\frac{\mean{\theta_>^0(\bm{r}_1)\theta_>^0(\bm{r}_2)}}{N}+\left(|\braket{p_1|p_2}|^2-\frac{1}{N}\right)\mean{\theta_>^1(\bm{r}_1)\theta_>^1(\bm{r}_2)}\right].
	\end{equation}
Note that $\mean{\theta_>^0(\bm{r}_1)\theta_>^0(\bm{r}_2)}$ can be obtained from $\mean{\theta_>^1(\bm{r}_1)\theta_>^1(\bm{r}_2)}$ 
by a replacement of $\sigma_x$ and $\sigma_y$ by $\sigma_x+Nc_x$ and $\sigma_y+Nc_y$ 
respectively. In terms of   
\begin{equation}
S_{\theta,>}=\frac{4\pi}{V}\sum_{\epsilon_{0,\Lambda} e^{-2dl} 
< \epsilon_{0,\bm k} < \epsilon_{0,\Lambda}}^{k_x>0}  
\left(\frac{ k_x^2}{\sigma_y}+\frac{k_y^2}{\sigma_x}\right)\left[\re^2 \theta_>^1(\bm{k})+\im^2 \theta_>^1(\bm{k})\right],  
\end{equation}
with $\theta^a(\bm{k})=\theta^{a*}(-\bm{k})$, $\mean{\theta_>^1(\bm{r}_1)\theta_>^1(\bm{r}_2)}$ 
is given by an ${\bm k}$-integral within a momentum shell region,  
\begin{align}
\mean{\theta_>^1(\bm{r}_1)\theta_>^1(\bm{r}_2)}
&=\frac{1}{V^2}\sum_{\bm{k}}\mean{\re^2 \theta_>^1(\bm{k})+\im^2 \theta_>^1(\bm{k})}e^{i\bm{k}\cdot(\bm{r}_1-\bm{r}_2)}, \nonumber \\
&=\frac{1}{4\pi } \int \frac{d^2{\bm k}}{(2\pi)^2} 
\frac{\theta\left(\epsilon_{0,\bm k}-\epsilon_{0,\Lambda} e^{-2dl}\right) 
\theta\left(\epsilon_{0,\Lambda}-\epsilon_{0,\bm k}\right)}{k_x^2/\sigma_y+k_y^2/\sigma_x} 
e^{i\bm{k}\cdot(\bm{r}_1-\bm{r}_2)}, 
\end{align}
with the Heviside step function $\theta(x)$.  
Note that due to the ``hard" cutoff nature of the momentum shell region, the integral in the right-hand 
side gives oscillating terms. The unphysical oscillation can be removed when the momentum integral is 
evaluated with a soft cutoff function $h(\epsilon_{0,\bm k},\epsilon_{0,\Lambda})$;
\begin{align} 
 \int\frac{d^2{\bm k}}{(2\pi)^2}  
 \frac{\theta\left(\epsilon_{0,\bm k}-\epsilon_{0,\Lambda} e^{-2dl}\right) 
\theta\left(\epsilon_{0,\Lambda}-\epsilon_{0,\bm k}\right)}{k_x^2/\sigma_y+k_y^2/\sigma_x} 
e^{i\bm{k}\cdot(\bm{r}_1-\bm{r}_2)} 
\simeq \int \frac{d^2{\bm k}}{(2\pi)^2} \frac{h\left(\epsilon_{0,\bm k},\epsilon_{0,\Lambda}\right) - 
h\left(\epsilon_{0,\bm k},\epsilon_{0,\Lambda} e^{-2dl}\right)}{k_x^2/\sigma_y+k_y^2/\sigma_x} 
e^{i\bm{k}\cdot(\bm{r}_1-\bm{r}_2)}.    
\end{align}
Here the soft cutoff function has only to satisfy 
\begin{equation}
h(\epsilon_{0,\bm k},\epsilon_{0,\Lambda})\ge 0,\quad\lim_{\bm{k}\to \bm{0}}h(\epsilon_{0,\bm k},\epsilon_{0,\Lambda})=1,\quad \lim_{\bm{k}\to \bm{\infty}}h(\epsilon_{0,\bm k},\epsilon_{0,\Lambda})=0.
\end{equation}
In the following derivation, we use 
\begin{equation}
h(\epsilon_{0,\bm k},\epsilon_{0,\Lambda}) \equiv 
\frac{\epsilon_{0,\Lambda}}{\epsilon_{0,\bm{k}}+\epsilon_{0,\Lambda}}, 
\end{equation}
and evaluate the correlation function,  
\begin{equation}
\mean{\theta_>^1(\bm{r})\theta_>^1(\bm{0})}=\frac{1}{4\pi}\int\frac{\diff^2k}{(2\pi)^2}
\left(\frac{1}{k_x^2/\sigma_y+k_y^2/\sigma_x + \epsilon_{0,\Lambda} e^{-2dl}}
- \frac{1}{k_x^2/\sigma_y+k_y^2/\sigma_x + \epsilon_{0,\Lambda}}\right)e^{i\bm{k}\cdot\bm{r}}.
\end{equation}
In terms of scaled momentum $\overline{\bm{k}} \equiv (k_x/\sqrt{\sigma_y},k_y/\sqrt{\sigma_x})$ and coordinate $\overline{\bm{r}} \equiv (\sqrt{\sigma_y}x,\sqrt{\sigma_x}y)$, the correlation function is given by the modified Bessel function, 
\begin{align}
\mean{\theta_>^1(\bm{r})\theta_>^1(\bm{0})}
&=\frac{\sqrt{\sigma_x\sigma_y}}{4\pi}\frac{1}{(2\pi)^2}\int\diff^2\overline{k}\,\left[\frac{1}{\overline{k}^2+\epsilon_{0,\Lambda}e^{-2\diff l}}-\frac{1}{\overline{k}^2+\epsilon_{0,\Lambda}}\right]e^{i\overline{\bm{k}}\cdot\overline{\bm{r}}}, \nonumber \\
& =\frac{\sqrt{\sigma_x\sigma_y}}{2}\frac{1}{(2\pi)^2}\left[K_0(\sqrt{\epsilon_{0,\Lambda}}e^{-\diff l}\overline{r})-K_0(\sqrt{\epsilon_{0,\Lambda}}\overline{r})\right].
\end{align}
Since $\diff l$ is infinitesimally small, we can expand the right-hand side in $\diff l$,
\begin{align}
\mean{\theta_>^1(\bm{r})\theta_>^1(\bm{0})}=\frac{\sqrt{\sigma_x\sigma_y}}{2}\frac{1}{(2\pi)^2}\sqrt{\epsilon_{0,\Lambda}}\overline{r}K_1(\sqrt{\epsilon_{0,\Lambda}}\overline{r})\diff l+\mathrm{O}(\diff l^2).
\end{align}
Similarly, the correlation function for the u$(1)$ part is obtained by the replacement, 
\begin{align}
\mean{\theta_>^0(\bm{r})\theta_>^0(\bm{0})}=\frac{\sqrt{(\sigma_x+Nc_x)(\sigma_y+Nc_y)}}{2}\frac{1}{(2\pi)^2}\sqrt{\epsilon_{N,\Lambda}}\overline{r}_N 
K_1(\sqrt{\epsilon_{N,\Lambda}}\overline{r}_N)\diff l+\mathrm{O}(\diff l^2), 
\end{align}
with $\epsilon_{N,\Lambda} \equiv \Lambda^2/(\sigma_y+Nc_y)+\Lambda^2/(\sigma_x+Nc_x)$ and $\overline{\bm{r}}_N \equiv (\sqrt{\sigma_y+Nc_y}x,\sqrt{\sigma_x+Nc_x}y)$. 
Thus, the correlation function for 
general $N$ is given by,
\begin{equation}
G(\bm{r}_1-\bm{r}_2,p_1,p_2)=\frac{1}{2}\left[\frac{g_N(\bm{r}_1-\bm{r}_2)}{N}+\left(|\braket{p_1|p_2}|^2-\frac{1}{N}\right)g_0(\bm{r}_1-\bm{r}_2)\right]\diff l, \label{correlation2}
\end{equation}
with 
\begin{equation}
g_N(\bm{r})= \sqrt{(\sigma_x+N c_x)(\sigma_y+N c_y)}
\sqrt{\epsilon_{N,\Lambda}}\overline{r}_NK_1(\sqrt{\epsilon_{N,\Lambda}}\overline{r}_N). \label{correlation3}
\end{equation}
Note that the zero-replica limit $(N\rightarrow 0)$ of the correlation function is well-defined in terms of a partial derivative of $N$,  
\begin{equation}
\lim_{N\to 0}G(\bm{r}_1-\bm{r}_2,p_1,p_2)=\frac{1}{2}\left(\left.\frac{\partial g_N(\bm{r}_1-\bm{r}_2)}{\partial N}\right|_{N=0}+|\braket{p_1|p_2}|^2g_0(\bm{r}_1-\bm{r}_2)\right) {\rm d}l.
	\end{equation}
	
\subsubsection{Integral over \texorpdfstring{$\mathbb{CP}^{N-1}$}{} elements} 
 In this subsection, we define an integral with respect to the $\mathbb{CP}^{N-1}$ element  $\ket{p}$ and provide a formula for the integral. Each element $\ket{p}$ in $\mathbb{CP}^{N-1}$ is a vector with $N$ complex components $p_i$. These components are subjected to a constraint $\braket{p|p}=\sum_{i=1}^{N}|p_i|^2=1$. The integral over $\ket{p}$ is thus defined as
 \begin{equation}
     \int\diff\ket{p}\cdots=\frac{1}{2\pi}\int_{-\infty}^{+\infty}\prod_{i=1}^N\diff\operatorname{Re} p_i\diff\operatorname{Im} p_i\,2\delta\left(\sum_{i=1}^N|p_i|^2-1\right)\cdots, 
 \end{equation}
 with $2\delta(x^2-1)=\delta(\sqrt{x^2}-1)$. A factor $1/2\pi$ is because $\ket{p}$ and $e^{i\phi}\ket{p}$ are regarded as the same element in $\mathbb{CP}^{N-1}$. The $\delta$ function constrains $({\rm Re} p_1,{\rm Im}p_1, \cdots,{\rm Re}p_{N},{\rm Im}p_N)$ on the $2N$-dimensional sphere. The integral with the delta function gives an area of the sphere measured by the unit of $2\pi$
 \begin{equation}\label{eq:SM_volume}
     \int\diff\ket{p}=\frac{\pi^{N-1}}{\Gamma(N)},
 \end{equation}
where
 \begin{equation}
    \begin{aligned}
        \int\diff\ket{p}&=\frac{1}{\pi}\int_{-\infty}^{+\infty}\prod_{i=1}^N\diff\operatorname{Re} p_i\diff\operatorname{Im}p_i\,\delta\left(\sum_{i=1}^N|p_i|^2-1\right)\\
        &=\frac{1}{\pi}\int_{-\infty}^{+\infty}\prod_{i=1}^N\diff\operatorname{Re} p_i\diff\operatorname{Im}p_i\int_{-\infty}^{+\infty}\frac{\diff\lambda}{2\pi}\,\exp\left[-i\lambda\left(\sum_{i=1}^N|p_i|^2-1\right)\right]\\
        &=\frac{1}{\pi}\int_{-\infty}^{+\infty}\prod_{i=1}^N\diff\operatorname{Re} p_i\diff\operatorname{Im}p_i\int_{-\infty}^{+\infty}\frac{\diff\lambda}{2\pi}\,\lim_{\epsilon\to+0}\exp\left[-(i\lambda+\epsilon)\left(\sum_{i=1}^N|p_i|^2-1\right)\right]\\
        &=\frac{1}{\pi}\lim_{\epsilon\to+0}\int_{-\infty}^{+\infty}\frac{\diff\lambda}{2\pi}\int_{-\infty}^{+\infty}\prod_{i=1}^N\diff\operatorname{Re} p_i\diff\operatorname{Im}p_i\exp\left[-(i\lambda+\epsilon)\left(\sum_{i=1}^N|p_i|^2-1\right)\right]\\
        &=\frac{1}{\pi}\lim_{\epsilon\to+0}\int_{-\infty}^{+\infty}\frac{\diff\lambda}{2\pi}\,e^{i\lambda+\epsilon}\left(\int_{-\infty}^{+\infty}\diff x\,e^{-(i\lambda+\epsilon)x^2}\right)^{2N}\\
        &=\frac{1}{\pi}\lim_{\epsilon\to+0}\int_{-\infty}^{+\infty}\frac{\diff\lambda}{2\pi}\left(\frac{\pi}{i\lambda+\epsilon}\right)^Ne^{i\lambda+\epsilon}.
    \end{aligned}
 \end{equation}
The above derivation indicates that the integral over $\ket{p}$ is a Gaussian integral. By using the Wick's theorem, one can prove the following formula
\begin{equation}
    \int\diff\ket{p}\braket{p|A|p}\braket{p|B|p} 
    =\frac{\pi^{N-1}}{N(N+1)\Gamma(N)}\left(\Tr A\Tr B+\Tr AB\right),
\end{equation}
where $A$ and $B$ are arbitrary $N\times N$ matrices.
 

\subsubsection{Calculation of \texorpdfstring{$\mean{S_y}$}{} and renormalization of fugacity \texorpdfstring{$y$}{}}
The first order term in Eq.~(\ref{esy}) is calculated as follows, 
\begin{align}
\mean{S_y}&=
-2y\Lambda^2\int\diff^2r\int\diff\ket{p} \bigg[\cos\left(2\pi \braket{p|\Theta_<({\bm r})|p}+\frac{\chi_y}{4}x\right)\mean{\cos\left(2\pi \braket{p|\Theta_>{\bm r}|p}\right)} \nonumber \\
 &\hspace{5cm} 
 -\sin\left(2\pi \braket{p|\Theta_<({\bm r})|p}+\frac{\chi_y}{4}x\right)\mean{\sin\left(2\pi \braket{p|\Theta_>({\bm r})|p}\right)} \bigg]
\end{align}
where $\mean{\sin(2\pi\braket{p|\Theta_>({\bm r})|p})}$ vanishes because of symmetry. $\mean{\cos(2\pi\braket{p|\Theta_>({\bm r})|p})}$ can be evaluated in terms of Wick's theorem and the correlation function Eq.~(\ref{correlation1}), 
\begin{equation}
\mean{\cos(2\pi\braket{p|\Theta_>({\bm r})|p})}=\sum_{n=0}^{\infty}\frac{(-1)^n}{(2n)!}\mean{(2\pi\braket{p|\Theta_>({\bm r})|p})^{2n}}=\sum_{n=0}^{\infty}\frac{(-1)^n}{(2n)!}(2n-1)!!G(\bm{0},p,p)^n=e^{-\frac{1}{2}G(\bm{0},p,p)}.
\end{equation}
Thus, we have 
\begin{equation}
\mean{S_y}=-2y\Lambda^2e^{-\frac{1}{2}G(\bm{0})}\int\diff^2r\int\diff\ket{p}\cos\left(2\pi\braket{p|\Theta_<|p}+\frac{\chi_y}{4}x\right).
\end{equation}
Note that from Eq.~(\ref{correlation2}), $G({\bm 0},p,p) \equiv G({\bm 0})$ is independent from $\ket{p}$,   
\begin{equation}
G({\bm 0}) =\frac{1}{2}\left(\frac{\sqrt{(\sigma_x+N c_x)(\sigma_y+N c_y)}}{N}+\left(1-\frac{1}{N}\right)\sqrt{\sigma_x\sigma_y}\right)\diff l.
\end{equation}
The factor of $e^{-\frac{1}{2}G({\bm 0})}$ in $\mean{S_y}$ gives rise to renormalization of the fugacity $y$. Namely, after the rescaling of the coordinate and the field operator, 
${\bm r}^{\prime}={\bm r} e^{-dl}$, 
$\theta({\bm r
}^{\prime}) = \Theta_{<}({\bm r})$, 
we obtain a renormalization group equation of $y$;
\begin{align}
\frac{\diff y}{\diff l} 
= \left[ 2 - \frac{1}{4} 
\bigg(\frac{\sqrt{(\sigma_x+Nc_x)(\sigma_y + Nc_y)}}{N} 
+ \Big(1-\frac{1}{N} 
\Big) \sqrt{\sigma_x \sigma_y} 
\bigg) \right] y + {\cal O}(y^3). \label{eq:SM_rg-y}
\end{align}
In the zero replica limit, 
\begin{equation}
\lim_{N\to 0}G(\bm{0})=\frac{1}{2}\frac{\sigma_xc_y+\sigma_yc_x+2\sigma_x\sigma_y}{2\sqrt{\sigma_x\sigma_y}} {\rm d}l, 
\end{equation}
the equation reduces to,  
	\begin{equation}
		\frac{\diff y}{\diff l}=\left(2-\frac{1}{4}\frac{\sigma_xc_y+\sigma_yc_x+2\sigma_x\sigma_y}{2\sqrt{\sigma_x\sigma_y}}\right)y+\mathrm{O}(y^3). \label{eq:SM_rg-y-zero}
	\end{equation}
 The rescaling also gives the RG equation for $\chi_y$,
 \begin{align}
 \frac{\diff\chi_y}{\diff l} = \chi_y. \label{eq:SM_rg-chi-y}
 \end{align}
For isotropic case $\sigma_x=\sigma_y=\sigma$ and $c_x=c_y=c$, Eq.~(\ref{eq:SM_rg-y}) reduces to
	\begin{equation}
		\frac{\diff y}{\diff l}=\left(2-\frac{\sigma+c}{4}\right)y+\mathrm{O}(y^3),
	\end{equation}
which is consistent with \cite{konigMetalinsulatorTransitionTwodimensional2012}.
	
\subsubsection{Calculation of \texorpdfstring{$\mean{S_y^2}$}{} and renormalization of \texorpdfstring{$\sigma_x,\sigma_y,c_x,c_y$}{}}
$\mean{S_y^2}-\mean{S_y}^2$ in Eq.~(\ref{s2y-sy2}) is calculated as follows, 
\begin{align}
\mean{S_y^2} - \mean{S_y}^2& = 4y^2\Lambda^4e^{-G(\bm{0})}\int\diff^2r_1 \int \diff^2r_2 
\int\diff\ket{p_1}\int \diff \ket{p_2} \nonumber \\
& \hspace{-0.5cm} \times \bigg\{ 
\big(e^{-G(\bm{r}_1-\bm{r}_2,p_1,p_2)}-1\big)  
\cos\left[2\pi\braket{p_1|\Theta_<(\bm{r}_1)|p_1}+2\pi\braket{p_2|\Theta_<(\bm{r}_2)|p_2}+\frac{\chi_y}{4}(x_1+x_2)\right] \nonumber \\
&\hspace{1.5cm} + 
\big(e^{G(\bm{r}_1-\bm{r}_2,p_1,p_2)}-1\big)
\cos\left[2\pi\braket{p_1|\Theta_<(\bm{r}_1)|p_1}-2\pi\braket{p_2|\Theta_<(\bm{r}_2)|p_2}+\frac{\chi_y}{4}(x_1-x_2)\right]\bigg\}, \label{sy2}
\end{align}
where we use 
\begin{align}
\mean{\cos(2\pi\braket{p_1|\Theta_>(\bm{r}_1)|p_1})\cos(2\pi\braket{p_2|\Theta_>(\bm{r}_2)|p_2})}
&=e^{-G(\bm{0})}\cosh G(\bm{r}_1-\bm{r}_2,p_1,p_2), \nonumber \\
\mean{\sin(2\pi\braket{p_1|\Theta_>(\bm{r}_1)|p_1})\sin(2\pi\braket{p_2|\Theta_>(\bm{r}_2)|p_2})}
&=e^{-G(\bm{0})}\sinh G(\bm{r}_1-\bm{r}_2,p_1,p_2), \nonumber \\
\mean{\sin(2\pi\braket{p_1|\Theta_>(\bm{r}_1)|p_1})\cos(2\pi\braket{p_2|\Theta_>(\bm{r}_2)|p_2})} 
&= 0. \nonumber 
\end{align}
Since the integrant in Eq.~(\ref{sy2}) has  
$G({\bm r}_1-{\bm r}_2,p_1,p_2)$ that is short-ranged in ${\bm r}_1-{\bm r}_2$, we introduce the relative coordinate ${\bm R}$ between 
${\bm r}_1$ and ${\bm r}_2$, $\bm{r}_1=\bm{r}+\bm{R}/2$ and $\bm{r}_2=\bm{r}-\bm{R}/2$, 
together with $\ket{p_1}=\ket{p({\bm r}+\bm{R}/2)}$, and $\ket{p_2}=\ket{p({\bm r}-\bm{R}/2)}$, 
and expand in power of small ${\bm R}$. 
Note that in the leading order expansion around ${\bm R}=0$, 
the double integral
over $\ket{p_1}$ and $\ket{p_2}$ reduces to a single integral over 
$\ket{p} \equiv \ket{p({\bm r})}$ (see the comment below Eq.~(\ref{s2y-sy2})). From the expansion,  
the first line of Eq.~(\ref{sy2}) yields terms with $\cos(4\pi \bra{p}\Theta_{<}\ket{p}+\chi_y r_x/2)$ or $\sin(4\pi \bra{p}\Theta_{<}\ket{p}+\chi_y r_x/2)$. These are 
higher-vorticity terms and irrelevant near the metal phase. The second line of Eq.~(\ref{sy2}) 
yields the following terms up to the 2nd order gradient expansion,   
\begin{align}
&\cos\left[2\pi\braket{p_1|\Theta_<(\bm{r}_1)|p_1}-2\pi\braket{p_2|\Theta_<(\bm{r}_2)|p_2}+\frac{\chi_y}{4}(x_1-x_2)\right] \nonumber \\
& \hspace{2cm} \simeq \cos \left[
2\pi R_{\mu} \partial_{\mu}\left(\braket{p|\Theta_<(\bm{r})|p}\right) \right]
\cos\left[\frac{\chi_y}{4}R_x \right] - \sin \left[
2\pi R_{\mu} \partial_{\mu}\left(\braket{p|\Theta_<(\bm{r})|p}\right)\right]
\sin\left[\frac{\chi_y}{4}R_x\right] \nonumber \\
& \hspace{2cm} \simeq  - R_{\mu} \partial_{\mu}\left(\braket{p|\Theta_<(\bm{r})|p}\right) 
\sin \left[\frac{\chi_y}{4}R_x\right] - \frac{1}{2} R_{\mu} R_{\nu}  
\partial_{\mu}\left(\braket{p|\Theta_<(\bm{r})|p}\right)
\partial_{\nu}\left(\braket{p|\Theta_<(\bm{r})|p}\right)
\cos\left[\frac{\chi_y}{4}R_x\right] + \cdots. 
\end{align}
Under the integral over ${\bm r}$ and ${\bm R}$, the first term 
can be integrated over ${\bm r}$, and the second term gives the following 
\begin{equation}
\mean{S_y^2}-\mean{S_y}^2 
  \simeq - 2 \pi^2y^2\Lambda^4e^{-G(\bm{0})} 
\int\diff^2r\int\diff^2R\int\diff\ket{p}\big(e^{G(\bm{R},p,p)}-1\big) \sum_{\mu=x,y}R_\mu^2\cos\left(\frac{\chi_y}{4} R_x \right)  \braket{p|\partial_\mu\Theta_<(\bm{r})|p}^2. 
	\end{equation}
Here we ignore the spatial derivatives of 
$\ket{p}$; $\partial_{\mu}\left(\braket{p|\Theta_<(\bm{r})|p}\right) \simeq 
\braket{p|\partial_{\mu}\Theta_<(\bm{r})|p}$. 
Then, we have, 
\begin{align}
-\frac{1}{2}\left(\mean{S_y^2}-\mean{S_y}^2\right) 
&=   \pi^2y^2\Lambda^4\int\diff^2r\int\diff\ket{p}\int\diff^2R\,G(\bm{R},p,p)\sum_{\mu=x,y}R_\mu^2\cos\left(\frac{\chi_y}{4}X\right)
  \braket{p|\partial_\mu\Theta_<(\bm{r})|p}^2, \nonumber \\
  & = 2\pi \sum_{\mu=x,y} \int d^2 r \int d\ket{p} 
  \braket{p|\partial_\mu\Theta_<(\bm{r})|p}^2 
  y^2 B_{\mu} \diff l.  
\end{align}
On the right-hand side, 
we define the following integrals
\begin{align}
B_{y} \diff l &=\frac{1}{2}\pi\Lambda^4\int\diff^2R\,R^2_x\cos\bigg(\frac{\chi_y}{4} R_x\bigg) G(\bm{R},p,p) 
= \diff l 
\bigg(\frac{I_{y,N}}{N} + \Big(1-\frac{1}{N}\Big) I_{y,0} \bigg), \nonumber \\
B_x \diff l &= \frac{1}{2}\pi\Lambda^4 
\int\diff^2R\,R^2_y \cos \bigg(\frac{\chi_y}{4} R_x\bigg)G(\bm{R},p,p) 
= \diff l \bigg(\frac{I_{x,N}}{N} + \Big(1-\frac{1}{N}\Big) I_{x,0} \bigg), \nonumber 
\end{align}
with 
\begin{equation}\label{eq:SM_IyIx}
I_{y,N} \equiv \frac{\pi\Lambda^4}{4}\int\diff^2R\, R^2_x\cos\left(\frac{\chi_y}{4} R_x\right)g_N(\bm{R}), \quad 
I_{x,N} \equiv \frac{\pi\Lambda^4}{4} \int\diff^2R\,R^2_y\cos\left(\frac{\chi_y}{4}R_x\right)g_N(\bm{R}). 
\end{equation}
As for the integral of $I_{\mu,N}$ for $N=0$, we introduce rescaled coordinates $(\tilde{R}_x,\tilde{R}_y) \equiv (\sqrt{\epsilon_{0,\Lambda}} \overline{R}_x,\sqrt{\epsilon_{0,\Lambda}}\overline{R}_y) 
= (\sqrt{\epsilon_{0,\Lambda}\sigma_y} R_x,  \sqrt{\epsilon_{0,\Lambda} \sigma_x} R_y)$ together with $\tilde{\chi}_0 \equiv \chi_y/(4\sqrt{\epsilon_{0,\Lambda} \sigma_y})$; 
\begin{align}
I_{y,0}&=\frac{\pi\Lambda^4}{4\sigma_y\epsilon_{0,\Lambda}^2}\int\diff^2\tilde{R}\,\tilde{R}^2_x\cos\left(\tilde{\chi}_0\tilde{X}\right)\tilde{R}K_1(\tilde{R})=\frac{\pi\Lambda^4}{4\sigma_y\epsilon_{0,\Lambda}^2}\int_0^\infty\diff\tilde{R}\int_0^{2\pi}\diff\theta\, 
\tilde{R}^4\cos^2\theta 
\cos\left(\tilde{\chi}_0\tilde{R}\cos\theta\right)K_1(\tilde{R}), \nonumber \\
I_{x,0}&=\frac{\pi\Lambda^4}{4\sigma_x\epsilon_{0,\Lambda}^2}\int\diff^2\tilde{R}\,
\tilde{R}^2_y   \cos\left(\tilde{\chi}_0\tilde{X}\right)\tilde{R}K_1(\tilde{R})=\frac{\pi\Lambda^4}{4\sigma_y\epsilon_{0,\Lambda}^2}\int_0^\infty\diff\tilde{R}\int_0^{2\pi}  
\diff\theta\,\tilde{R}^4\sin^2\theta \cos\left(\tilde{\chi}_0\tilde{R}\cos\theta\right)K_1(\tilde{R}).  
\end{align}
In terms of the Bessel functions of the first kind $J_n(a)$ ($n=0,1,\cdots$), 
\begin{align}
\int_0^{2\pi}\diff\theta\cos\left(a\cos\theta\right) 
&=2\pi J_0(a),\quad a\ge 0,  \nonumber \\
\int_0^{2\pi}\diff\theta\cos^2\theta \cos\left(a\cos\theta\right) 
&=-\frac{\diff^2}{\diff a^2}\int_0^{2\pi}\diff\theta\cos\left(a\cos\theta\right)
=2\pi\left[J_0(a)-\frac{J_1(a)}{a}\right], \nonumber \\
\int_0^{2\pi}\diff\theta\sin^2\theta \cos\left(a\cos\theta\right) 
& = 2\pi \frac{J_1(a)}{a}, \nonumber \\
\end{align}
we have, 
\begin{align}
I_{y,0}&=\frac{\pi^2\Lambda^4}{2\sigma_y\epsilon_{0,\Lambda}^2}\int_0^\infty\diff\tilde{R}\,\tilde{R}^4K_1(\tilde{R})\left[J_0(\tilde{\chi}_0\tilde{R})-\frac{J_1(\tilde{\chi}_0\tilde{R})}{\tilde{\chi}_0\tilde{R}}\right]=\frac{4\pi^2\Lambda^4}{\sigma_y\epsilon_{0,\Lambda}^2}\frac{1-5\tilde{\chi}^2_0}{(1+\tilde{\chi}^2_0)^4}  
\nonumber \\
I_{x,0}&=\frac{\pi^2\Lambda^4}{2\sigma_x\epsilon_{0,\Lambda}^2}\int_0^\infty\diff\tilde{R}\,\tilde{R}^3K_1(\tilde{R})\frac{J_1(\tilde{\chi}_0\tilde{R})}{\tilde{\chi}_0}=\frac{4\pi^2\Lambda^4}{\sigma_x\epsilon_{0,\Lambda}^2}\frac{1}{(1+\tilde{\chi}^2_0)^3}. \nonumber 
\end{align}
As for $I_{\mu,N}$ for general $N$, we have only to replace $\tilde{\chi}_0$, $\epsilon_{0,\Lambda}$, and $\sigma_{\mu}$ by 
$\tilde{\chi}_N \equiv \chi_y/(4\sqrt{\epsilon_{N,\Lambda}(\sigma_y+Nc_y)})$,  $\epsilon_{N,\Lambda}$ and $\sigma_{\mu}+N c_{\mu}$ respectively,  
\begin{align}
I_{y,N}=\frac{4\pi^2 \Lambda^4}{(\sigma_y+Nc_y)\epsilon_{N,\Lambda}^2}\frac{1-5\tilde{\chi}^2_N}{(1+\tilde{\chi}^2_N)^4}, \quad  
I_{x,N}=\frac{4\pi^2 \Lambda^4}{(\sigma_x+Nc_x)\epsilon_{N,\Lambda}^2}\frac{1}{(1+\tilde{\chi}^2_N)^3}. \label{eq:SM_I_mu_N}
\end{align} 
Since  
\begin{align}
\int \diff\ket{p} \bra{p} \partial_{\mu} \Theta_{<}({\bm r}) \ket{p}^2 
= \frac{\pi^{N-1}}{N(N+1)\Gamma(N)} 
\Big( \big({\rm Tr} [\partial_{\mu}\Theta_{<}({\bm r})]\big)^2 
+ {\rm Tr}[\partial_{\mu} \Theta_{<}({\bm r}) \partial_{\mu} \Theta_{<}({\bm r})] \Big), 
\end{align}
we obtain 
\begin{align}
-\frac{1}{2}\left(\mean{S_y^2}-\mean{S_y}^2\right)=
2\pi y^2 \diff l \!\ 
\int\diff^2r\sum_{\mu=x,y} 
B_\mu\frac{\pi^{N-1}} {N(N+1)\Gamma(N)}\left[(N+1)(\partial_\mu\theta^0)^2+\sum_{a=1}^{N^2-1}(\partial_\mu\theta^a)^2\right]. 
\end{align}
This leads to a set of RG equations for conductivity and Gade constant,
\begin{align}
\frac{\diff}{\diff l}\frac{1}{\sigma_\mu+Nc_\mu} = \frac{\pi^{N-1}}{N\Gamma(N)}y^2B_\mu, \quad 
\frac{\diff}{\diff l}\frac{1}{\sigma_\mu} = \frac{\pi^{N-1}} {N(N+1)\Gamma(N)}y^2B_\mu.  \label{eq:SM_rg-sigma-c}
\end{align} 
Equivalently, 
\begin{equation}\label{eq:SM_raw_eq_sigma}
\frac{\diff\sigma_\mu}{\diff l}=-\frac{\pi^{N-1}}{N(N+1)\Gamma(N)}y^2\sigma_\mu^2B_\mu,
\end{equation}
\begin{equation}\label{eq:SM_raw_eq_c}
\frac{\diff c_\mu}{\diff l}=-\frac{\pi^{N-1}}{N\Gamma(N)}y^2B_\mu\frac{1}{N}\left[(\sigma_\mu+Nc_\mu)^2-\frac{\sigma_\mu^2}{N+1}\right].
\end{equation}
Eqs.~(\ref{eq:SM_rg-y},\ref{eq:SM_rg-chi-y},\ref{eq:SM_raw_eq_sigma},\ref{eq:SM_raw_eq_c}) 
comprise a set of RG equations for the coupling constants in the dual theory.  
The effect of the weak 
topological term on the 2D Anderson transition in 
the chiral unitary class can be clarified by a study of the RG equation in the zero replica limit ($N\rightarrow 0$). 
Before studying the zero replica limit, we will see first whether 
the RG equations give consistent results with previous works for the case of $N=1$. 
	
\subsection{\texorpdfstring{$N=1$}{} case: commensurate-incommensurate transition}
The dual theory for $N=1$ takes exactly the same form as the sine-Gordon theory in the presence of the   
incommensurability effect. 
When $N=1$, matrix field $\Theta$ 
reduces to $\theta^0$, where  
the second term in Eq.~(\ref{eq:SM_rg-y}) vanishes and the second equation in Eq.~(\ref{eq:SM_rg-sigma-c}) is absent. 
In terms of $\kappa_\mu=\sigma_\mu+c_\mu$, the RG equations for $N=1$ are given by
	\begin{equation}
		\frac{\diff y}{\diff l}=\left(2-\frac{\sqrt{\kappa_x\kappa_y}}{4}\right)y,
	\end{equation}
	\begin{equation}
		\frac{\diff\kappa_x}{\diff l}=-8\pi^2y^2\frac{\kappa_x^3\kappa_y^2}{(\kappa_x+\kappa_y)^2}\frac{1}{(1+\tilde{\chi}_1^2)^3},
	\end{equation}
	\begin{equation}
		\frac{\diff\kappa_y}{\diff l}=-8\pi^2y^2\frac{\kappa_x^2\kappa_y^3}{(\kappa_x+\kappa_y)^2}\frac{1-5\tilde{\chi}_1^2}{(1+\tilde{\chi}_1^2)^4},
	\end{equation}
	\begin{equation}
		\frac{\diff \chi_y}{\diff l}=\chi_y, 
	\end{equation}
with $\tilde{\chi}_1 \equiv \frac{\chi_y}{4\Lambda\sqrt{1+\kappa_y/\kappa_x}}$. 
The RG equation for normalized $\tilde{\chi}_1$ is 
given by, 
	\begin{equation}
		\frac{\diff\tilde{\chi}_1}{\diff l}=\left[1-20\pi^2y^2\frac{\kappa_x^2\kappa_y^3}{(\kappa_x+\kappa_y)^3}\frac{\tilde{\chi}_1^2}{(1+\tilde{\chi}_1^2)^3}\right]\tilde{\chi}_1.
	\end{equation}
In terms of normalized fugacity $\tilde{y} \equiv \sqrt{8\pi^2}y\frac{\kappa_x\kappa_y}{\kappa_x+\kappa_y}$, 
stiffness parameter 
$K \equiv \sqrt{\kappa_x\kappa_y}/4$, and
an anisotropic factor $\eta \equiv 
\kappa_x /\kappa_y$, the coupled RG equations are
simplified, 
	\begin{equation}
		\frac{\diff\tilde{y}}{\diff l}=\left(2-K-\tilde{y}^2\frac{1+\tilde{\chi}_1^2+\eta(1-5\tilde{\chi}_1^2)}{(1+\tilde{\chi}_1^2)^4(\eta+1)}\right)\tilde{y},
	\end{equation}
	\begin{equation}
		\frac{\diff K}{\diff l}=-\tilde{y}^2 K\frac{1-2\tilde{\chi}_1^2}{(1+\tilde{\chi}_1^2)^4},
	\end{equation}
	\begin{equation}
		\frac{\diff \eta}{\diff l}=-6\tilde{y}^2\eta\frac{\tilde{\chi}_1^2}{(1+\tilde{\chi}_1^2)^4},
	\end{equation}
	\begin{equation}
		\frac{\diff\tilde{\chi}_1}{\diff l}=\left[1-3\tilde{y}^2\frac{1}{\eta+1}\frac{\tilde{\chi}_1^2}{(1+\tilde{\chi}_1^2)^4}\right]\tilde{\chi}_1.
	\end{equation}

The coupled RG equations have s stable fixed point, stable fixed region, and saddle-point fixed point in a four-dimensional (4D) parameter space subtended by $\tilde{y}$, $K$, $\eta$, and $\tilde{\chi}_1$. Two stable fixed regions characterize strong and weak coupling phases with non-zero incommensurability, while the saddle point characterizes a universality class of a phase transition between these two phases;
\begin{enumerate}
\item 2D stable fixed region at $\tilde{y}=0,\,\tilde{\chi}_1=\infty$ with arbitrary $K$ and $\eta$. It characterizes the weak coupling phase with 
$\tilde{\chi}_1 \ne 0$, 
\item stable fixed point: $\tilde{y}=\frac{12}{5}\sqrt{\frac{3}{5}},\,K=0,\,\eta=0,\,\tilde{\chi}_1=\pm\frac{1}{\sqrt{5}}$. It characterizes the strong coupling phase with $\tilde{\chi}_1 \ne 0$. 
\item saddle-point fixed point: $\tilde{y}=\frac{3}{2}\sqrt{\frac{3}{2}},\,K=1,\,\eta=0,\,\tilde{\chi}_1=\pm\frac{1}{\sqrt{2}}$. The fixed point has one relevant scaling variable whose scaling dimension is $\lambda=\frac{2}{3}$. Thereby, the critical exponent of the phase transition at 
$\tilde{\chi}_1 \ne 0$ is given by $\nu=\frac{1}{\lambda}=\frac{3}{2}$. 
\end{enumerate}
Part of the RG flow diagram at $\tilde{\chi}_1$ is shown in Fig.~\ref{fig:3D_RGflow_N1}. In addition, the RG equations have two other fixed points (region) on a plane of $\tilde{\chi}_1=0$, both of which are unstable against small $\tilde{\chi}_{1}$. They characterize strong and weak coupling phases at $\tilde{\chi}_1=0$. Specifically, the plane of $\tilde{\chi}_1=0$ has a 2D fixed region at $\tilde{y}=\tilde{\chi}_1=0$. The region comprises a part with $K>2$ and the other with $K<2$. The 2D region with $K>2$ characterizes the weak coupling phase with $\tilde{\chi}_1=0$, where vortex fugacity is renormalized to zero, and the phase is described by the free critical theory of $\theta^0$. The other with $K<2$ corresponds to the strong coupling phase with $\tilde{\chi}_1=0$. The strong coupling phase is renormalized into a 1D fixed region at $\tilde{y}=\frac{1} {2\pi},\,K=0,\,\tilde{\chi}_1=0$.  

\begin{figure}[htb]
    \centering
    \includegraphics[width=0.65\textwidth]{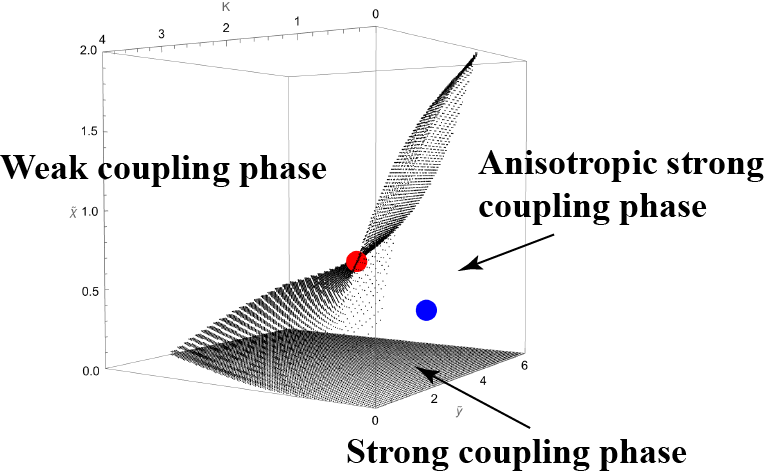}
    \caption{3D RG phase diagram for $N=1$ case at subspace $\eta=0$. The blue point is the stable fixed point of the anisotropic strong coupling phase. The red point is the saddle-point fixed point that characterizes the weak-strong coupling phase transition. There are three phases: the weak coupling phase, the anisotropic strong coupling phase, and the strong coupling phase. The black dots form the boundary of these phases.}
    \label{fig:3D_RGflow_N1}
\end{figure}

\begin{figure}[htb]
	\centering
    \subfloat[subspace $\tilde{\chi}=0$]{\includegraphics[width=0.4\textwidth]{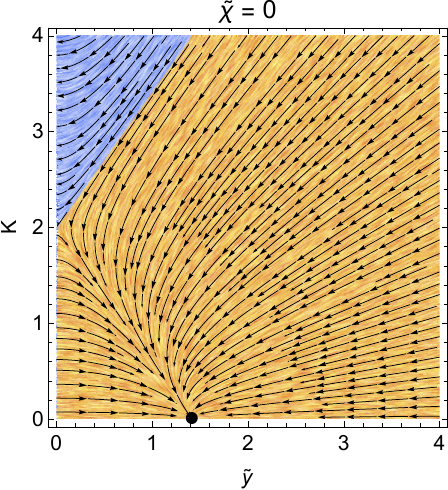}}
    \hspace{10mm}
    \subfloat[subspace $K=\eta=0$]{\includegraphics[width=0.4\textwidth]{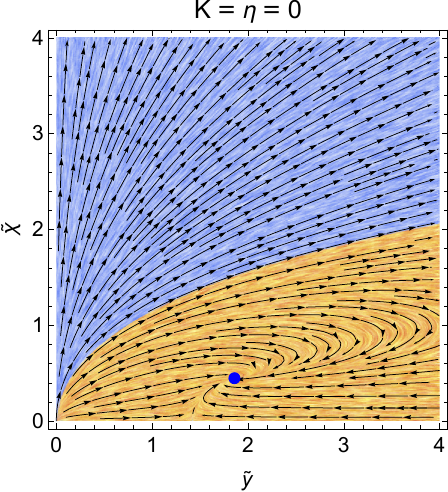}}
    \caption{2D cross-sections with RG flows of the 3D RG phase diagrams in Fig.~\ref{fig:3D_RGflow_N1}. (a) 2D RG phase diagram at $\tilde{\chi}=0$. The black point is a fixed point of a strong coupling phase with $\tilde{\chi}_1=0$. (b) 2D RG phase diagram at $K=\eta=0$. The blue point is a stable fixed point for the strong coupling phase with $\tilde{\chi}_1\ne 0$. Blue and orange regions are weak and strong coupling phases, respectively, and arrows stand for RG flow in the infrared limit ($l\rightarrow \infty$).}
    \label{fig:RGflow_crosssection_N1}
\end{figure}

\subsection{\texorpdfstring{$N\to 0$}{} case: Anderson transition and quasi-localization}
	
The coupled RG equations in the zero replica limit are among six coupling constants, $y$, $\sigma_{x}$, $\sigma_y$, $c_{x}$, $c_y$ and $\chi_y$, 
\begin{equation}
		\frac{\diff y}{\diff l}=\left(2-\frac{1}{4}\frac{\sigma_xc_y+\sigma_yc_x+2\sigma_x\sigma_y}{2\sqrt{\sigma_x\sigma_y}}\right)y, \label{eq:SM_RG0-y}
	\end{equation}
	\begin{equation}
		\frac{\diff\sigma_\mu}{\diff l}=-\pi^{-1}y^2\sigma_\mu^2\lim_{N\to 0}B_\mu, \label{eq:SM_RG0-smu}
	\end{equation}
	\begin{equation}
		\frac{\diff c_\mu}{\diff l}=-\pi^{-1}y^2\sigma_\mu(\sigma_\mu+2c_\mu)\lim_{N\to 0}B_\mu,
  \label{eq:SM_RG0-cmu}
	\end{equation}
	\begin{equation}
		\frac{\diff \chi_y}{\diff l}=\chi_y, 
  \label{eq:SM_RG0-chi}
	\end{equation}
with $\mu=x,y$. $\lim_{N \rightarrow 0} B_{\mu}$ is defined by 
\begin{align}
\lim_{N\rightarrow 0} B_{\mu} \equiv  
\frac{\diff I_{\mu,N}}{\diff N}\Big|_{N=0} + I_{\mu,0}, 
\end{align}
with Eqs.~(\ref{eq:SM_I_mu_N}). Here we use, 
\begin{align}
\lim_{N\to 0}\frac{1}{N\Gamma(N)}=1,\quad \lim_{N\to 0}\frac{1}{N}\left[(\sigma_\mu+Nc_\mu)^2-\frac{\sigma_\mu^2}{N+1}\right]=\sigma_\mu(\sigma_\mu+2c_\mu). 
\end{align}
$\lim_{N\to 0}B_\mu$ can be calculated from Eqs.~(\ref{eq:SM_I_mu_N}) with 
redefinitions of coupling constants, 
\begin{align}
\zeta &\equiv \frac{\sigma_x}{\sigma_y}, 
\quad 
\eta \equiv \frac{\sigma_x+c_x}{\sigma_y+c_y}, 
\quad \lambda \equiv \frac{\zeta^2}{\eta} = \frac{\sigma^2_x}{\sigma^2_y}\frac{\sigma_y+c_y}{\sigma_x+c_x}, 
\quad \tilde{\chi} \equiv 
\frac{\chi_y}{4\Lambda \sqrt{1+\zeta}}.  
\label{eq:SM_zeta-eta-lambda-chi}
\end{align} 
In terms of these new constants, the limits are calculated as follows, 
\begin{align}
\tilde{B}_x\equiv \frac{1}{4\pi^2}\frac{\zeta} {\sigma_x+c_x}\frac{1}{\lambda}\lim_{N\to 0}B_x =\frac{2\zeta+(3+2\zeta)\tilde{\chi}^2+\frac{\zeta}{\lambda}(1-\zeta-(2+\zeta)\tilde{\chi}^2)}{(1+\tilde{\chi}^2)^4(1+\zeta)^3}, \label{eq:SM_tildeBx}
\end{align}
\begin{align}
\tilde{B}_y\equiv \frac{1}{4\pi^2}\frac{1}{\sigma_x+c_x}\frac{1}{\lambda}\lim_{N\to 0}B_y =\frac{\zeta-1+(13-4\zeta)\tilde{\chi}^2-5(2+\zeta)\tilde{\chi}^4+\frac{\zeta}{\lambda}(2-17\tilde{\chi}^2+5\tilde{\chi}^4)}{(1+\tilde{\chi}^2)^5(1+\zeta)^3}. \label{eq:SM_tildeBy}
\end{align}
Note that from Eqs.~(\ref{eq:SM_RG0-smu},\ref{eq:SM_RG0-cmu},\ref{eq:SM_tildeBx},\ref{eq:SM_tildeBy}), 
$\lambda$ is invariant under the renormalization, 
\begin{align}
\frac{\diff \ln \lambda}{\diff l} = 2  \bigg(\frac{\diff \ln \sigma_x}{\diff l} 
- \frac{\diff \ln \sigma_y}{\diff l}\bigg) + 
\bigg(\frac{\diff \ln (\sigma_y+c_y)}{\diff l} 
- \frac{\diff \ln (\sigma_x+c_x)}{\diff l} \bigg) = 0,  
\end{align}
because  
\begin{align}
2 \frac{\diff \ln \sigma_{\mu}}{\diff l} 
= \frac{\diff \ln (\sigma_{\mu}+c_{\mu})}{\diff l} 
= - 8\pi y^2 (\sigma_y + c_y) \frac{\sigma^2_x}{\sigma_y} \tilde{B}_{\mu},  \label{eq:SM_lns-lns+c}  
\end{align}
for $\mu=x,y$. Eq.~(\ref{eq:SM_lns-lns+c}) also shows that 
$\sigma^2_x/(\sigma_x+c_x)$ as well as $\sigma^2_y/(\sigma_y+c_y)$ are invariant unon the renormalization. Thus, the RG equations among the six coupling constants can be reduced into closed-coupled 
differential equations among four coupling constants. The right hand side of Eq.~(\ref{eq:SM_lns-lns+c}) suggests a usage of  
following normalized vortex fugacity as one of these four constants, 
\begin{align}
\tilde{y} \equiv y \sqrt{4\pi\lambda \sigma_y (\sigma_x+c_x)} = y \sqrt{\frac{4\pi (\sigma_y + c_y)\sigma^2_x}{\sigma_y}}. \label{eq:SM_tildey}
\end{align}
In fact, one can derive closed coupled RG equations among  $\tilde{y}$, $\zeta$, $\tilde{\chi}$ and a 
stiffness parameter $K$,  
\begin{align}
K=\frac{1}{4}\frac{\sigma_xc_y+\sigma_yc_x+2\sigma_x\sigma_y}{2\sqrt{\sigma_x\sigma_y}}. \label{eq:SM_K-def}
\end{align}
The equations are given as follows, 
\begin{equation}\label{eq:SM_y_tilde}
\frac{\diff\tilde{y}}{\diff l}=\left[2-K-\frac{1}{2}\tilde{y}^2(2\tilde{B}_x+\tilde{B}_y)\right]\tilde{y},
\end{equation}
\begin{equation}\label{eq:SM_K}
\frac{\diff K}{\diff l}=-\tilde{y}^2\frac{\tilde{B}_x(3\zeta+\lambda)+\tilde{B}_y(\zeta+3\lambda)}{2(\zeta+\lambda)}K,
\end{equation}
\begin{equation}\label{eq:SM_zeta}
\frac{\diff\zeta}{\diff l}=-\tilde{y}^2(\tilde{B}_x-\tilde{B}_y)\zeta,
\end{equation}
\begin{equation}\label{eq:SM_chi_tilde}
\frac{\diff\tilde{\chi}}{\diff l}=\left[1-\frac{1}{2}\tilde{y}^2\frac{\tilde{B}_x-\tilde{B}_y}{1+\zeta}\right]\tilde{\chi}. 
\end{equation}
Together with these closed equations, we have  
secondary equations for $\sigma_x$ and $\sigma_y$, 
\begin{equation}\label{eq:SM_lnsigma}
\frac{\diff \ln \sigma_{\mu}}{\diff l}=-\tilde{y}^2\tilde{B}_{\mu},
\end{equation}
for $\mu=x,y$. In summary, Eqs.~(\ref{eq:SM_y_tilde},\ref{eq:SM_K},\ref{eq:SM_zeta},\ref{eq:SM_chi_tilde}) together with Eqs.~(\ref{eq:SM_tildeBx},\ref{eq:SM_tildeBy}) determine a RG phase diagram in a four-dimensional parameter space subtended by $\tilde{y}$, $K$, $\zeta$ and $\tilde{\chi}$. The RG phase diagram is characterized by stable and saddle-point fixed points, where stable fixed points of the differential equations characterize thermodynamic phases in the phase diagram and saddle-point fixed points characterize phase transitions among phases. In the following, we first enumerate all the stable fixed points in the four-dimensional parameter space. Using Eq.~(\ref{eq:SM_lnsigma}), we further characterize the transport properties of these fixed points, and clarify fixed points of non-topological metal, topological metal, Anderson insulator, and quasi-localized phases, respectively. Note that the RG equations are derived perturbatively in the vortex fugacity $y$, so an application of the RG equations should be limited to weak and intermediate coupling regimes of 
$y$, e.g. $y < 1/\sqrt{\sigma_x\sigma_y}$. For this reason, 
we do not discuss a structure of RG flows and fixed points in large $\tilde{y}$ region; $\tilde{y} \equiv y \sqrt{4\pi \lambda} \sqrt{\sigma_y(\sigma_x+c_x)}$. 
The following argument applies to arbitrary finite 
(initial) $\lambda$. 

\subsection{Fixed points at finite \texorpdfstring{$\tilde{y}$}{}} 
Fixed points at finite $\tilde{y}$ must satisfy 
the following equations simultaneously. 
\begin{equation}
2-K-\frac{1}{2}\tilde{y}^2(2\tilde{B}_x+\tilde{B}_y)=0,\quad \frac{\tilde{B}_x(3\zeta+\lambda)+\tilde{B}_y(\zeta+3\lambda)}{2(\zeta+\lambda)}K=0,\quad (\tilde{B}_x-\tilde{B}_y)\zeta=0,\quad \left[1-\frac{1}{2}\tilde{y}^2\frac{\tilde{B}_x-\tilde{B}_y}{1+\zeta}\right]\tilde{\chi}=0.
\end{equation}
They are satisfied by the following points
\begin{itemize}
\item{\bf stable FP at 
$\tilde{y}=3.64...,\,K=0,\,\zeta=0,\,\tilde{\chi}=\pm1.10...$ (quasi-localized phase).}
\item{\bf saddle-point FP at $\tilde{y}=3.56...,\,K=\frac{3}{4},\,\zeta=0,\,\tilde{\chi}=\pm1.22...$ (quantum phase transition).}
\item{saddle-point FP at $\tilde{y}=3.34...,\,K=0,\,\zeta=0,\,\tilde{\chi}=\pm0.287...$.}
\item{unstable FP at $\tilde{y}=\frac{2(1+\lambda)}{\sqrt{3}},\,K=0,\,\zeta=\lambda,\,\tilde{\chi}=0$.}
\item{unstable FP at $\tilde{y}=2.98..,\,K=\frac{3}{4},\,\zeta=0,\,\tilde{\chi}=\pm0.274...$.}
\end{itemize}
As discussed below, the first one is a stable fixed point that characterizes the quasi-localized phase, and the second one is 
a saddle-point fixed point that characterizes a phase transition between quasi-localized and metal phases. Our extensive numerical investigations show no other fixed points with finite $\tilde{y}$. In the following, we will explain these fixed points in detail. 
\begin{enumerate}
\item The stable FP at $(\tilde{y},K,\zeta,\tilde{\chi})=(3.64...,\,0,\,0,\,\pm1.10..)$ characterizes quasi-localized phase at $\chi_y\ne 0$. This point is obtained by a condition of $K=\zeta=0$, and $2-\frac{1}{2}\tilde{y}^2(2\tilde{B}_x+\tilde{B}_y)=0$ and $1-\frac{1}{2}\tilde{y}^2(\tilde{B}_x-\tilde{B}_y)=0$, which leads to $\tilde{y}^2\tilde{B}_x=2$ and $\tilde{y}^2\tilde{B}_y=0$. $\tilde{y}^2\tilde{B}_y=0$ at $\zeta=0$ and $\tilde{y}\ne 0$ 
gives $-1+13\tilde{\chi}^2-10\tilde{\chi}^4=0$, while finite 
$\tilde{y}$ is determined by $\tilde{y}^2 \tilde{B}_x=\frac{3 \tilde{y}^2 \tilde{\chi}^2}{(1+\tilde{\chi}^2)^4}=2$. The RG equations can be linearized 
around the fixed point, revealing four RG eigenvalues around the fixed point as 
$-2.51.. - i3.77.., -2.51.. + i3.77.., -2, -1$.  
$``-2"$ is a scaling dimension for $\zeta$, and $``-1"$ is a scaling dimension for $K$. The linearized RG equations form an attractive flow with a swirl toward the fixed point in the $\tilde{y}$-$\tilde{\chi}$ plane. Around the fixed point, both $\zeta$ 
and $\sigma_x$ vanish with a same asymptotic form, $e^{-2l}$, because 
\begin{align}
\frac{\diff \ln \zeta}{\diff l} = \frac{\diff \ln \sigma_x}{\diff l} = - \tilde{y}\tilde{B}_x = -2. 
\end{align}
Thus, the conductivity $\sigma_y$ along the topological direction takes a finite value; $\sigma_y = \sigma_x/\zeta$. 
This stable fixed point characterizes a quasi-localized phase. 

\item The saddle-point fixed point at $(\tilde{y},K,\zeta,\tilde{\chi})=(3.58...,\,3/4,\,0,\,\pm1.22...)$ characterizes a phase 
transition between the quasi-localized phase and metal phase. The fixed point is 
obtained from a condition of $\zeta=0$, and $2-K-\frac{1}{2}\tilde{y}^2(2\tilde{B}_x+\tilde{B}_y)=0$, $\tilde{y}^2 (\tilde{B}_x + 3\tilde{B}_y)=0$, and $1-\frac{1}{2}\tilde{y}^2(\tilde{B}_x-\tilde{B}_y)=0$, which leads to $\tilde{y}^2\tilde{B}_x=\frac{3}{2}$ and $\tilde{y}^2\tilde{B}_y=-\frac{1}{2}$, $K=3/4$. $\tilde{B}_x+3 \tilde{B}_y=0$ at $\zeta=0$ gives $9\tilde{\chi}^4-14\tilde{\chi}^2+1=0$, while finite $\tilde{y}$ is determined by  
$\tilde{y}^2\tilde{B}_x=\frac{3}{2}$. The RG equations are linearized around the 
fixed point with RG eigenvalues; $-1.55.. + 3.41..i$, $-1.55.. - 3.41.. i$, $-2$, $0.819..$. $``-2"$ is a scaling dimension for $\zeta$. Notably, the fixed point has  
one positive RG eigenvalue (``0.819.."), which stands for a scaling dimension of a relevant scaling 
variable. The relevant scaling variable is given by a linear superposition of $\delta \tilde{y}/\tilde{y}_{\rm FP}$, $\delta K/K_{\rm FP}$, and $\delta \tilde{\chi}/\tilde{\chi}_{\rm FP}$, while linear coefficients suggest that it is mainly $\delta K/K_{\rm FP}$. 
The critical exponent of the phase transition 
is given by $\nu=1/0.819.. = 1.22..$.

\item The saddle-point fixed point at $(\tilde{y},K,\zeta,\tilde{\chi})=(3.34..,\,0,\,0,\,\pm0.287..)$ is from the same condition as the 
stable fixed point, $\tilde{B}_y=0$ and $\tilde{y}^2 \tilde{B}_x=2$,  
while it is from the other solutions of $-1+13\tilde{\chi}^2-10\tilde{\chi}^4=0$. RG eigenvalues around the fixed point are $-5.73..,\, -2,\, -1,\, 7.33..$, suggesting that it characterizes a phase transition with an extremely small critical exponent $1/7.33..=0.13..$. The exponent clearly breaks the Chayes
inequality between the critical exponent $\nu$ and spatial 
dimension $d=2$ ($\nu d>2$)~\cite{Chayes86}. A numerical examination 
of the differential equations shows that the fixed point characterizes a phase transition between quasi-localized and a strong-coupling regime with infinitely large $\tilde{y}$. Since the perturbative RG 
equation can be applicable only for weak $\tilde{y}$ region, we regard 
the strong-coupling phase as well as this phase transition as unphysical.

\item The unstable fixed point at $(\tilde{y},K,\zeta,\tilde{\chi})=(2.98..,\,3/4,\,0,\,\pm0.274..)$ is obtained from the same condition as the  
saddle-point fixed point at $K=3/4$, $\tilde{y}^2\tilde{B}_x=3/2$ and $\tilde{y}^2 \tilde{B}_y=-1/2$, while it is from the other solutions of 
$9\tilde{\chi}^4-14\tilde{\chi}^2+1=0$. RG eigenvalues around the fixed point 
contain two positive values, $5.75..,\, -4.77..,\, -2,\, 0.966..$. concluding that  
it characterizes neither the criticality of a generic   
phase transition nor the thermodynamic phase.   

\item The unstable fixed point at $(\tilde{y},K,\zeta,\tilde{\chi})=(\frac{2(1+\lambda)}{\sqrt{3}},\,0,\,\lambda,\,0)$ satisifies $\tilde{B}_x-\tilde{B}_y =0$, $\tilde{\chi}=K=0$, 
and 
$4-\tilde{y}^2(2\tilde{B}_x+\tilde{B}_y)
=0$. RG eigenvalues around this fixed point contain two positive values; $-4,\,-\frac{8}{3},\,\frac{4}{3},\,1$. $``1"$ is the scaling dimension of one of the two relevant scaling variables, $\tilde{\chi}$, and $``\frac{4}{3}"$ is the scaling dimension of the other relevant scaling variable, $\delta \zeta \equiv \zeta - \lambda$. Within a 2D space of $\tilde{\chi}=0$ (no weak topological term) and $\zeta=\lambda$, this fixed point is stable, and it characterizes the Anderson insulator phase~\cite{konigMetalinsulatorTransitionTwodimensional2012}. However, the fixed point is unstable against an introduction of small $\delta \zeta$ as well as small $\chi$. The numerical examination shows that the small $\delta \zeta$ drives the RG flow into the strong coupling region with large $\tilde{y}$. 
		
\end{enumerate}

\begin{figure}[htb]
    \centering
    \includegraphics[width=0.6\textwidth]{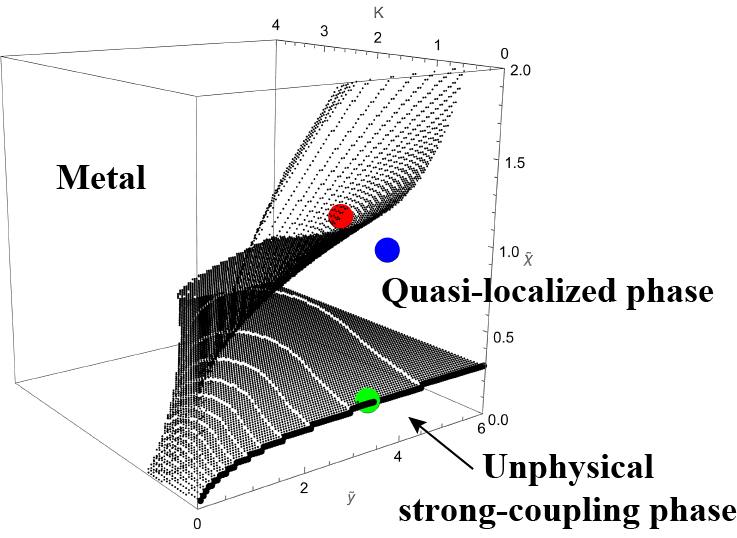}
    \caption{3D RG phase diagram of $N\to0$ case in the subspace $\zeta=0$ and $\lambda=1$. Blue and red points are stable fixed points for the quasi-localized phase, and saddle-point fixed points for a transition between quasi-localized and topological metal phases,  respectively. The green dot 
    is a saddle-point fixed point for a transition between quasi-localized and unphysical strong-coupling phase.
    The black dots form the boundary of different phases. The phase diagram is constructed by a numerical solution of the RG equations.  For given initial values of $(\tilde{y}(l=0),(l=0),\tilde{\chi}(l=0))$, we solve the differential equations numerically and determine the infrared(IR)-limit values of $\tilde{y}(l)$, $K(l)$, and $\tilde{\chi}(l)$ for sufficiently large $l$. Parameter points in the metal phase reach the stable fixed region with $\tilde{y}^2\tilde{B}_x=0$, Parameter points in the quasi-localized phase reach the stable point with $\tilde{y}^2\tilde{B}_x\ne 0$ (blue point). Parameter points in the unphysical phase reach a strong-coupling region with divergent $\tilde{y}$.}
    \label{fig:3D_RG}
\end{figure}

\begin{figure}[h!tb]
	\centering
    \subfloat[subspace $\tilde{\chi}=0$, $\zeta=\lambda=1$]{\includegraphics[width=0.39\textwidth]{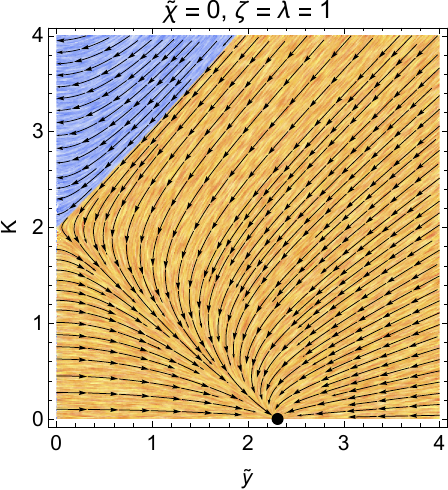}}
    \hspace{10mm}
    \subfloat[subspace $K=\zeta=0$, $\lambda=1$]{\includegraphics[width=0.4\textwidth]{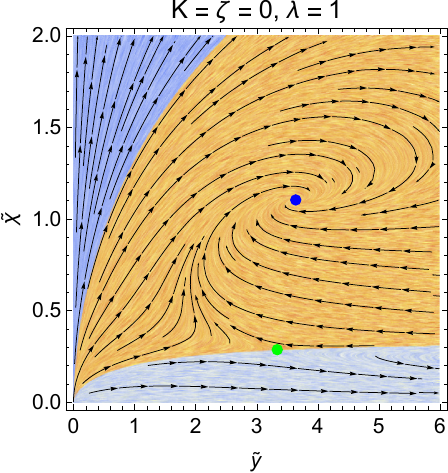}}
    \caption{2D cross-sections with RG flows of the 3D phase diagram in Fig.~\ref{fig:3D_RG}. (a) 2D RG phase diaram at $\tilde{\chi}=0$ and $\zeta=\lambda=1$. Blue and orange regions are nontopological metal and  Anderson insulator phases respectively. The black point stands for a fixed point of the Anderson insulator phase. (b) 2D RG phase diagram at $K=\zeta=0$, and $\lambda=1$. Blue, orange, and light blue regions are topological metal, quasi-localized, and unphysical strong-coupling phases, respectively. The blue point is the stable fixed point of the quasi-localized phase. The green point is a saddle-point fixed point that characterizes a transition between the quasi-localized phase and the unphysical strong-coupling region. The arrow stands for the RG flow in the infrared limit ($l\rightarrow \infty$).}
    \label{fig:RGflow_crosssection_N0}
\end{figure}

\subsection{Fixed points at \texorpdfstring{$\tilde{y}=0$}{}}
Let us consider the RG equations around infinitesimally small $\tilde{y}$ or at $\tilde{y}=0$. When linearized with respect to small $\tilde{y}$, the coupled RG equations reduce to 
\begin{align}
\frac{\diff \tilde{y}}{\diff l} = (2-K) \tilde{y}, \quad,
\frac{\diff \tilde{\chi}}{\diff l} = \tilde{\chi}, \quad, 
\frac{\diff K}{\diff l} = \frac{\diff \zeta}{\diff l}=0. 
\end{align}
The reduced RG equations have the following fixed regions, 
\begin{itemize}
\item{stable FP region at $(\tilde{y},K,\zeta,\tilde{\chi})=
(0,K,\zeta,\infty)$ with $K>2$ (metal phase).}
\item{unstable FP region at $(\tilde{y},K,\zeta,\tilde{\chi})=
(0,K,\zeta,\infty)$ with $K<2$.} 
\item{unstable FP region at $(\tilde{y},K,\zeta,\tilde{\chi})=
(0,K,\zeta,0)$.} 
\end{itemize}
The stable fixed region at $(\tilde{y},K,\zeta,\tilde{\chi})=
(0,K,\zeta,\infty)$ with $K>2$ form a two-dimensional (2D) region subtended 
by $K$ and $\zeta$. It is stable against small $1/\tilde{\chi}$ as well as small $\tilde{y}$. At and around this fixed point, $\tilde{B}_x$ and $\tilde{B}_y$ vanish as $1/\tilde{\chi}^6$, and $\tilde{y}$ go to zero. Thus, the  vanishing right hand side of Eq.~(\ref{eq:SM_lnsigma}) suggests that both $\sigma_x$ and $\sigma_y$ reach finite values and the fixed region describes 
(topological) metal phase with $\chi_y \ne 0$. 
A fixed region at $(\tilde{y},K,\zeta,\tilde{\chi})=
(0,K,\zeta,0)$ with $K<2$ characterizes a non-topological metal with 
$\chi_y=0$~\cite{konigMetalinsulatorTransitionTwodimensional2012}, 
while it is always unstable against small $\tilde{\chi}$. 

As discussed above, a general structure of the RG phase diagram in 
the $\tilde{y}$-$K$-$\zeta$-$\tilde{\chi}$ space has 
complications. Nonetheless, the phase diagram becomes simpler, when conductivity and Gade constant satisfy the following relatively generic conditions, 
\begin{align}
\frac{\sigma_x}{\sigma_y}=
\frac{\sigma_x+c_x}{\sigma_y+c_y},  
\end{align}
or equivalently $\zeta=\lambda$. This condition includes a case with 
isotropic conductivity 
and Gade constant; $\sigma_x=\sigma_y$ 
and $c_x=c_y$, When initial values of the constants satisfy this condition, the $K$-$\tilde{y}$-$\tilde{\chi}$ phase diagram at $\tilde{\chi}\ne 0$ comprises only of the quasi-localized and (topological) metal phases. As discussed above, these two phases are characterized 
by the stable fixed points at $(\tilde{y},K,\zeta,\tilde{\chi})=(3.64.., 0,\,0,\,\pm1.10...)$, 
and the stable 2D fixed region at $(\tilde{y},K,\zeta,\tilde{\chi})=(0, K,\,\zeta,\,\pm \infty)$,  respectively, while the phase transition between these two phases are controlled by the 
saddle-point fixed point at $(\tilde{y},K,\zeta,\tilde{\chi})=(3.56.., 0.75,\,0,\,\pm1.22...)$. In Fig.~\ref{fig:solutions}, 
we show how $y$, $\sigma_x$, $\sigma_y$, $c_{x}$ and $c_y$ 
are renormalized along the RG equations for typical cases with this condition. 
\begin{figure}[h!tb]
	\centering
\subfloat[A (topological) metal phase with $\chi_y \ne 0$]{\includegraphics[width=0.33\textwidth]{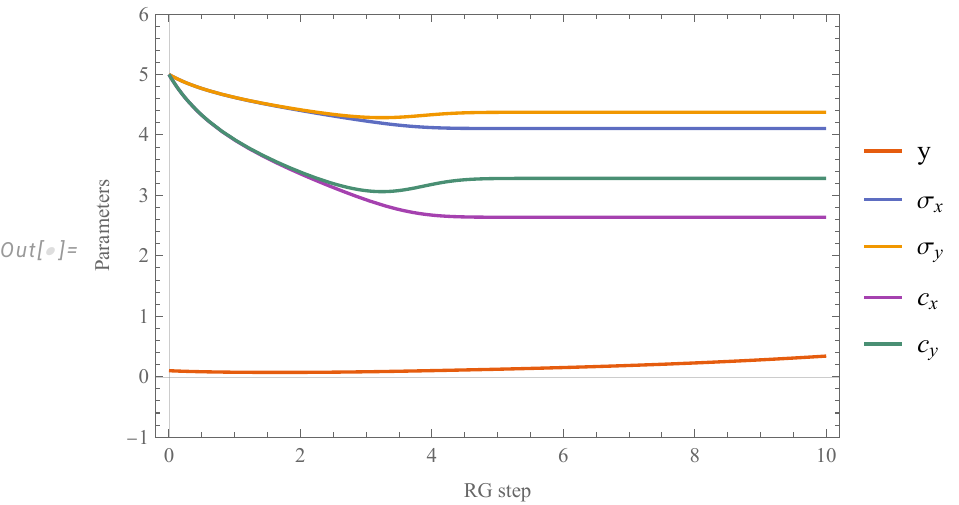}}
\subfloat[Quasilocalized phase]{\includegraphics[width=0.33\textwidth]{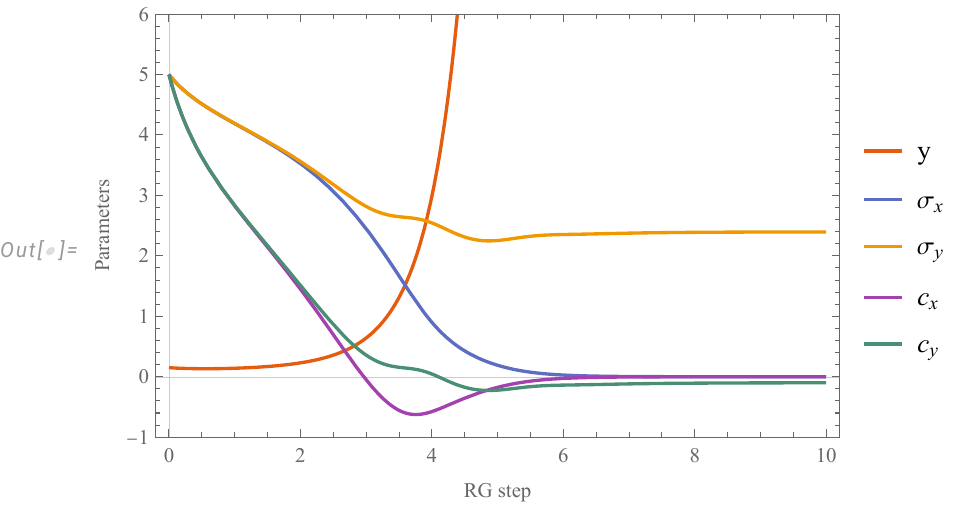}}
\subfloat[Another parameter for quasi-localized phase]{\includegraphics[width=0.33\textwidth]{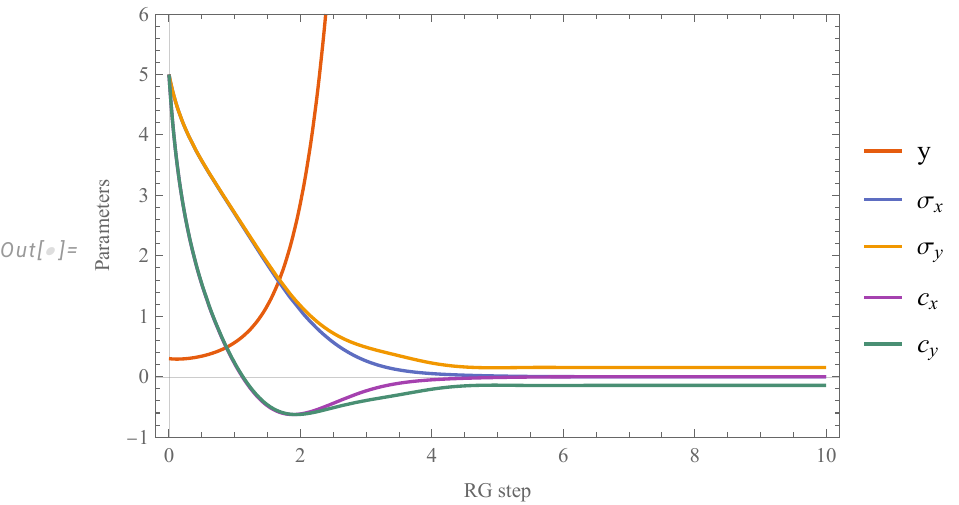}}
\caption{Numerical solutions of the RG equations. Initial values are chosen to be $\sigma_x=\sigma_y=c_x=c_y=5$, $\chi=0.1$ for all these three 3 figures, while (a) $y=0.1$, 
(b) $y=0.15$, and (c) $y=0.30$. The solution for (a) 
indicates that this parameter set is in the (topological) metal phase. The solution for (b) and (c) suggest that these parameter sets are in the quasi-localized phase.\label{fig:solutions}}
\end{figure}

\section{Numerical simulation}
	\subsection{Tight-binding models}

    \begin{figure}[htb]
        \centering
        \includegraphics[width=0.6\textwidth]{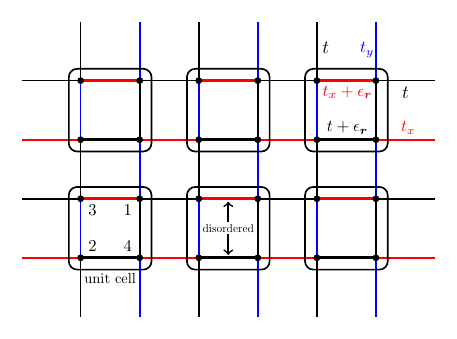} 
        \caption{The two-dimensional square-lattice model with four inequivalent sublattice sites. Each unit cell (rectangle with rounded corners) contains four sublattice sites, labeled as 1, 2, 3, 4. Different hopping strengths are shown with different colors. Thinner links represent uniform hoppings (hopping without the disorder) such as $t$, $t_y$, and $t_x$, while thicker links represent the hoppings with complex-valued random variables such as $t_x+\epsilon_{\bm r}$, $t+\epsilon_{\bm r}$. Note that ${\bm r}$ stands for coordinates of the unit cell, and the complex-valued random variable is added only within the intra-unit-cell hopping term.}
        \label{fig:lattice}
    \end{figure}

In the previous section, we show by the renormalization group analysis that the weak  
topological term ($\chi_y$) changes the nature of the localized phase 
proximate to the (critical) metal phase. Ref.~\cite{xiaoAnisotropicTopologicalAnderson2023} 
demonstrated numerically that a diffusive metal phase universally 
undergoes a phase transition to the quasi-localized phase in three-dimensional (3D) 
chiral symmetric models with weak topology. In this section, we will 
demonstrate numerically that this is also  the case with two-dimensional 
tight-binding chiral symmetric models with weak topology. We introduce a two-dimensional square-lattice 
model in chiral classes that has a weak topological index, i.e. one-dimensional 
topological winding number. The model in the clean limit was introduced previously~\cite {liTopologicalStatesTwoDimensional2022}. A unit cell of the model 
has four inequivalent sublattice sites, 1, 2, 3, and 4. They are divided into two sublattice groups; (1,2) and 
(3,4). The model has no on-site terms and it comprises only hopping terms. 
The hopping terms are only between the two sublattice groups; no hopping within the same sublattice group. 
	\begin{equation}
		\begin{aligned}
H&=\sum_{\bm{r}}\sum_{a,b=1,2,3,4} 
\ket{\bm{r},a}\bra{\bm{r},b}\begin{pmatrix}
				&   & t_x+\epsilon_{\bm{r}} & t \\
				&   & t_y & t+\epsilon_{\bm{r}} \\
				t_x+\epsilon_{\bm{r}}^* & t_y &   &   \\
				t & t+\epsilon_{\bm{r}}^* &   &  
			\end{pmatrix}_{a,b}\\
			&+\sum_{\bm{r},a,b}\ket{\bm{r},a}\bra{\bm{r}+\hat{\bm{x}},b}\begin{pmatrix}
				&   & t &  0 \\
				&    & 0  & 0  \\
				0 &  0 &   &   \\
				0 & t_x &   &  
			\end{pmatrix}_{a,b}
			+\sum_{\bm{r},a,b}\ket{\bm{r},a}\bra{\bm{r}-\hat{\bm{x}},b}\begin{pmatrix}
				&   &  0 & 0  \\
				&    &  0 & t_x \\
				t &  0 &   &   \\
				0  &  0 &   &  
			\end{pmatrix}_{a,b}\\
			&+\sum_{\bm{r},a,b}\ket{\bm{r},a}\bra{\bm{r}+\hat{\bm{y}},b}\begin{pmatrix}
				&   &  0 & t_y \\
				&    &  0 & 0  \\
				0  & t  &   &   \\
				0 &  0 &   &  
			\end{pmatrix}_{a,b}
			+\sum_{\bm{r},a,b}\ket{\bm{r},a}\bra{\bm{r}-\hat{\bm{y}},b}\begin{pmatrix}
				&   &  0 & 0 \\
				&    &  t & 0  \\
				0 &  0 &   &   \\
				t_y &  0 &   &  
			\end{pmatrix}_{a,b}.
		\end{aligned} \label{eq:SM_H_TB}
	\end{equation} 
Here ${\bm r}\equiv(n_x,n_y)$, ${\bm r}\pm \hat{\bm x}\equiv(n_x\pm 1,n_y)$, 
${\bm r}\pm \hat{\bm y}\equiv(n_x,n_y \pm 1)$ specify the unit cell of the model; unit cells 
also form a square lattice (see Fig.~\ref{fig:lattice}). ``$a$" and ``$b$" specify 
sublattice site within the unit cell, $a,b=1,2,3,4$. The first term in Eq.~(\ref{eq:SM_H_TB}) 
stands for intra-unit-cell hopping terms, while the others are inter-unit-cell hopping 
terms. A complex-valued random hopping $\epsilon_{\bm r}$ is introduced in the intra-unit-cell hopping term. Real and imaginary parts of $\epsilon_{\bm{r}}$ are uniformly 
distributed in a range of $[-W/2,W/2]$; $W$ stands for the disorder strength of the model. 
Other hopping terms take finite definite real values. 

The model has the chiral symmetry, $\sigma_3 H \sigma_3=-H$, where $\sigma_3$ takes +1 
for one of the two sublattice groups, e.g. (1,2), and takes $-1$ for the other sublattice 
group, (3,4). Since $H=H^*$, the model in the clean limit ($W=0$) belongs to the chiral 
orthogonal class (class BDI). In the presence of $\epsilon_{\bm r}$, the model 
breaks the time-reversal symmetry,  belonging to the chiral 
unitary class (class AIII).

Due to the chiral symmetry, $H$ takes a block-off-diagonal form, 
$H=\sigma_+\otimes h+\sigma_-\otimes h^\dagger$ where the off-diagonal block $h$ is 
a non-Hermitian matrix, 
\begin{align}
    			h&=\sum_{\bm{r},a,b} 
        \ket{\bm{r},a}\bra{\bm{r},b}
	   		\begin{pmatrix}
				t_x+\epsilon_{\bm{r}} & t \\
				t_y & t+\epsilon_{\bm{r}}
			\end{pmatrix}_{a,b} \nonumber \\
			&+\sum_{\bm{r},a,b}
   \ket{\bm{r},a}\bra{\bm{r}+\hat{\bm{x}},b}
			\begin{pmatrix}
				t & 0\\
				0 & 0
			\end{pmatrix}_{a,b}
			+\sum_{\bm{r},a,b}\ket{\bm{r},a}\bra{\bm{r}-\hat{\bm{x}},b}
			\begin{pmatrix}
				0 & 0 \\
				0 & t_x
			\end{pmatrix}_{a,b} \nonumber \\
			&+\sum_{\bm{r},a,b}\ket{\bm{r},a}\bra{\bm{r}+\hat{\bm{y}},b}
			\begin{pmatrix}
				0 & t_y \\
				0 & 0
			\end{pmatrix}_{a,b}
			+\sum_{\bm{r},a,b}\ket{\bm{r},a}\bra{\bm{r}-\hat{\bm{y}},b}
			\begin{pmatrix}
				0 & 0 \\
				t & 0
			\end{pmatrix}_{a,b}. \label{eq:SM_def_h}
	\end{align} 
 The weak topological index of $H$ is defined by $h$ with an attachment of uniform magnetic fluxes. Let us attach the magnetic flux $\phi_{\mu}$ into the hopping terms along $\mu$ of $h$; $h\rightarrow h(\phi_{\mu})$. In $h(\phi_{\mu})$, a particle that travels around the system with the periodic boundary condition along $\mu$ direction acquires a phase $e^{i\phi_{\mu}}$ for every round. In terms of such $h(\phi_{\mu})$, the winding number along $\mu$ is defined by~\cite{mondragon-shemTopologicalCriticalityChiralSymmetric2014,altlandQuantumCriticalityQuasiOneDimensional2014,claes2020}
\begin{align}
\nu_{\mu} \equiv \frac{i}{L} 
\int^{\pi}_{-\pi} 
\frac{d\phi_{\mu}}{2\pi} \partial_{\phi_{\mu}}{\rm Tr}
\big[\log [h(\phi_{\mu})]\big] = \frac{i}{L} 
\int^{\pi}_{-\pi} 
\frac{d\phi_{\mu}}{2\pi} \partial_{\phi_{\mu}} 
\log \big[\det [h(\phi_{\mu})]\big]. \label{nu-def}
\end{align}
Here $L$ is a linear dimension of the system size along the transverse direction to $\mu$. In the clean limit ($W=0$), $\nu_{\mu}$ can be also 
written in terms of the momentum integral with a single-particle Green's function $G_{0}({\bm k})$ for the zero-energy wavefunction;
\begin{align}
\nu_{\mu} = i \int^{\pi}_{-\pi} \int^{\pi}_{-\pi} \frac{dk_x dk_y}{(2\pi)^2} \!\ {\rm tr} \Big[G_{0}({\bm k}) \big(\partial_{k_{\mu}} 
G^{-1}_{0}({\bm k})\big)\left[\begin{array}{cc}
0 & 0 \\
0 & \mathds{1} \\
\end{array}\right] \Big], \label{1d-G0}
\end{align}
where $G_{0}({\bm k})$ is given by the a Fourier transform $h({\bm k})$ of $h$;
\begin{align}
G^{-1}_{0}({\bm k}) = \left[\begin{array}{cc} 
0 & h({\bm k}) \\
h^{\dagger}({\bm k}) & 0 \\
\end{array}\right], \quad G_{0}({\bm k}) = \left[\begin{array}{cc} 
0 & (h^{-1}({\bm k}) )^{\dagger} \\
h^{-1}({\bm k}) & 0 \\
\end{array}\right].
\end{align}
In the clean limit ($W=0$) and $|t_x+t_y|>2t$, $(\nu_x,\nu_y)=(-1,0)$ for $|t_x|>|t_y|$, and $(\nu_x,\nu_y)=(0,1)$ for $|t_x|<|t_y|$. In some regions of $t_x=t_y$ or $|t_x+t_y|<2t$, the model also has finite winding numbers, but the numbers are not quantized to integers. Note also that $\nu_{\mu}=0$ whenever $t_{\mu}=\pm t$ (see \cref{fig:cleanlimit_winding} for details). 


$\nu_{\mu}$ is zero at $t_{\mu}=\pm t$ for disordered systems,  
when $\nu_{\mu}$ is averaged over different disorder realizations. This is because an ensemble of $h$ with different disorder realizations is symmetric under certain symmetry operations at these special points~\cite{xiaoAnisotropicTopologicalAnderson2023}. For example, consider $t_x=t$, where
an ensemble of $h$ is symmetric under a transposition followed by $\tau_1$ 
and a mirror. $\tau_1$ exchanges the sublattice index of $h$ and the mirror exchange a pair of two sites, $r= (n_x,n_y)$ and 
${\bm r}^{\prime}= (n_x,-n_y)$;
\begin{align}
{\cal U}_{y,{(n_x,n_y|n^{\prime}_x,n^{\prime}_y)}} 
= \delta_{n_x,n^{\prime}_x} \delta_{n_y,-n^{\prime}_y}. 
\end{align}
The transposition of 
$h$ followed by $\tau_1$ and ${\cal U}_y$ is given by  
\begin{align}
{\cal U}_y\!\ \tau_1 h^T \tau_1 \!\ {\cal U}_y & 
= \sum_{\bm{r},a,b} 
    \ket{\bm{r},a}\bra{\bm{r},b}
			\begin{pmatrix}
				t+\epsilon_{(n_x,-n_y)} & t \\
				t_y & t_x+\epsilon_{(n_x,-n_y)}
			\end{pmatrix}_{a,b} \nonumber \\
& +\sum_{\bm{r},a,b}\ket{\bm{r},a}\bra{\bm{r}+\hat{\bm x},b}
			\begin{pmatrix}
				t_x & 0 \\
				0 & 0
			\end{pmatrix}_{a,b} + \sum_{\bm{r},a,b}
   \ket{\bm{r},a}\bra{\bm{r}-\hat{\bm x},b}
			\begin{pmatrix}
				0 & 0\\
				0 & t
			\end{pmatrix}_{a,b} \nonumber \\
&+\sum_{\bm{r},a,b}\ket{\bm{r},a}\bra{\bm{r}+\hat{\bm y},b}
			\begin{pmatrix}
				0 & t_y \\
			0 & 0
			\end{pmatrix}_{a,b}
			+\sum_{\bm{r},a,b}\ket{\bm{r},a}\bra{\bm{r}-\hat{\bm y},b}
			\begin{pmatrix}
				0 & 0 \\
				t & 0
			\end{pmatrix}_{a,b}. \label{eq:SM_after}
\end{align}
A statistical distribution of $\epsilon_{(n_x,n_y)}$ 
is same as that of $\epsilon_{(n_x,-n_y)}$. Thus, when $t_x=t$, an ensemble of the right hand side of Eq.~(\ref{eq:SM_after}) over different disorder realizations is same as that of Eq.~(\ref{eq:SM_def_h});
\begin{align}
\Big\{ \!\ {\cal U}_y \!\ \tau_1 h^T \tau_1 \!\ {\cal U}_y \!\ \Big| \!\ \epsilon_{\bm r} \in \big[-\frac{W}{2},\frac{W}{2}\big] \!\ \Big\} = \Big\{ \!\ h \!\  \Big| \!\ \epsilon_{\bm r} \in \big[-\frac{W}{2},\frac{W}{2}\big] \!\ \Big\}, \quad \quad @ \!\ t=t_x. \label{eq:SM_transpose}   
\end{align}
The relation is generalized into cases with the magnetic flux along $x$ and/or $y$, 
\begin{align}
\Big\{ \!\ {\cal U}_y \!\ \tau_1 h^T(\phi_x,\phi_y) \tau_1 \!\ {\cal U}_y \!\ \Big| \!\ \epsilon_{\bm r} \in \big[-\frac{W}{2},\frac{W}{2}\big] \!\ \Big\} = \Big\{ \!\ h(-\phi_x,\phi_y) \!\  \Big| \!\ \epsilon_{\bm r} \in \big[-\frac{W}{2},\frac{W}{2}\big] \!\ \Big\}, \quad \quad @ \!\  t=t_x.   
\end{align}
Here both $\phi_x$ and $\phi_y$ change their signs under the transposition, 
while only $\phi_y$ changes its sign under the mirror ${\cal U}_y$. 
Note that $\nu_{\mu}$ stands for a phase winding of the U(1) phase 
of $\det h(\phi_{\mu})$ from $\phi_{\mu} = -\pi$ to $\phi_{\mu} = \pi$. 
Thus, the statistical symmetry of the ensemble requires that the average of $\nu_x$ 
must be zero at $t_x=t$. Similarly, an ensemble of $h$ at $t_y=t$ is statistically 
symmetric under the transposition followed by a spatial mirror ${\cal U}_x$, 
\begin{align}
\Big\{ \!\ {\cal U}_x \!\ h^T(\phi_x,\phi_y) \!\ {\cal U}_x \!\ \Big| \!\ \epsilon_{\bm r} \in \big[-\frac{W}{2},\frac{W}{2}\big] \!\ \Big\} = \Big\{ \!\ h(\phi_x,-\phi_y) \!\  \Big| \!\ \epsilon_{\bm r} \in \big[-\frac{W}{2},\frac{W}{2}\big] \!\ \Big\}, \quad \quad @ \!\  t=t_y,   
\end{align}
with 
\begin{align}
{\cal U}_{x,{(n_x,n_y|n^{\prime}_x,n^{\prime}_y)}} 
\equiv \delta_{n_x,-n^{\prime}_x} \delta_{n_y,n^{\prime}_y}. 
\end{align}
The symmetry leads to $\nu_y=0$ at $t_y=t$. 
\begin{figure}[htb]
        \centering
        \includegraphics[width=0.8\textwidth]{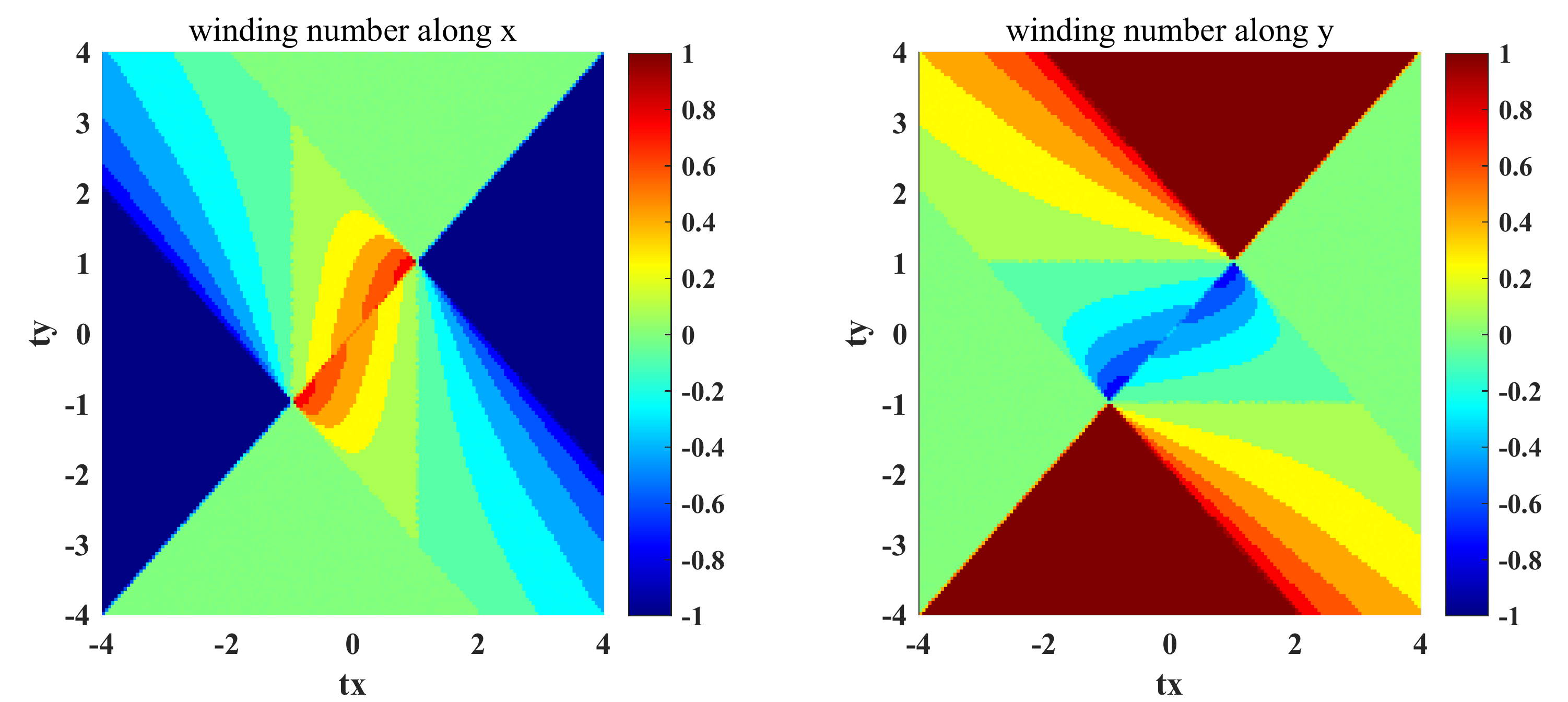}
        \caption{Distribution of one-dimensional winding numbers along $x$- and $y$-directions in a parameter space of $t_x$ and $t_y$ (the unit is $t$). Color represents the value of the winding numbers.
        \label{fig:cleanlimit_winding}}
    \end{figure}

In the following numerical simulation, we choose $t_x=t$ and $t_y \ne t$ with $\nu_y\ne 0$ and $\nu_x = 0$. This corresponds to the topological case with $\vec{\chi}=(0,\chi_y)$ in the previous section. We also study $t_x=t_y=t$ with $\nu_x=\nu_y=0$.  This point corresponds to the non-topological case ($\vec{\chi}=0$) in the chiral unitary class. 

\subsection{Transfer matrix}
The zero energy state of $H$ respects the chiral symmetry. An exponential localization length of the zero-energy state can be calculated via the transfer matrix method. To calculate the localization length $\xi_x$ along $x$, we consider a quasi-1D geometry with $L_x\times L_y$ unit cells and $L_x\gg L_y$; $L_{x}$ and $L_y$ are linear dimensions of the system size along $x$ and $y$ respectively. The quasi-1D system is regarded as $L_x$ slices of subsystems, and each subsystem has $L_y$ unit cells along $y$. The zero-energy state $\ket{\Psi}$ is 
partitioned into the $L_x$ slices, $\ket{\Psi}=[\cdots,\ket{\Psi}_{x-1},\ket{\Psi}_x,\ket{\Psi}_{x+1},\cdots]$, where $\ket{\Psi}_x$ is a part of $\ket{\Psi}$ at the $x$-th slice. The Schrodinger equation for $\ket{\Psi}$ can be rewritten into a recursive relation between $\ket{\Psi}_{x+1}$ and $\ket{\Psi}_x$ (or between $[\ket{\Psi}_{x+1},\ket{\Psi}_{x}]$ and $[\ket{\Psi}_{x},\ket{\Psi}_{x-1}]$ in other models); 
$\ket{\Psi}_{x+1}=T_{x\rightarrow x+1}\ket{\Psi}_x$. Here a matrix $T_{x\to x+1}$ is called a transfer matrix. Since the unit cell has four inequivalent sublattices, the transfer matrix thus introduced is $4L_y$ by $4L_y$. We calculate all singular values of a product of transfer matrices along $x$,
\begin{align}
\Omega \equiv \lim_{L_x\to\infty}\frac{1}{2 L_x}
  \ln(T_{1\to L_x}^\dagger T_{1\to L_x}),\quad T_{1\to L_x}
  \equiv \prod_{n=1}^{L_x-1}T_{n\to n+1}. 
\end{align}
The eigenvalues of $\Omega$ are called Lyapunov exponents (LEs); 
$\{\gamma_1,\gamma_2,\cdots,\gamma_{4L_y}\}$ with 
$\gamma_1<\gamma_2<\gamma_3<\cdots<\gamma_{4L_y-1}<\gamma_{4L_y}$. 
The Hermicity of $H$ ensures the eigenvalues are symmetric with respect to 
zero, $\gamma_{j}=-\gamma_{4L_y-(j-1)}$. Since the product relates 
the $\ket{\Psi}_{L_x}$ and $\ket{\Psi}_1$, the smallest nonnegative LE $\gamma\equiv\gamma_{NL+1}$ is related to the exponential localization length 
$\xi_x$ of the zero-energy eigenstate 
along $x$ direction; $\xi_x=\gamma^{-1}$. 
	
The zero-energy state of $H$ $\ket{\Psi}$ is given by zero-energy eigenstate $\ket{\Phi}$ of its non-Hermitian off-diagonal-block matrices $h$, 
$\ket{\Psi}=[0,\ket{\Phi}]^T$~\cite{feinberg97}. Also it is given by zero-energy eigenstate $\ket{\Phi}^{\prime}$ of the other off-diagonal-block $h^{\dagger}$, $\ket{\Psi}=[\ket{\Phi}^{\prime},0]^T$. Consequently, the transfer matrix of $H$ can be block diagonalized 
into transfer matrices of $h$ and $h^{\dagger}$, where the LEs of 
the zero-energy state of $H$ is the union of the LEs of the zero-energy 
states of $h$ and $h^\dagger$. 
The transfer matrix of $h$ is obtained from the 
Schrodinger equation for $\ket{\Phi}$;
\begin{align}\label{eq:SM_Schrodinger_Phi}
\left(\begin{array}{cc} 
t_x + \epsilon_{x,y} & t \\
t_y & t + \epsilon_{x,y} \\
\end{array}\right) \Phi_{x,y} + \left(\begin{array}{cc} 
0 & 0 \\
t & 0 \\
\end{array}\right) \Phi_{x,y-1} + \left(\begin{array}{cc} 
0 & t_y \\
0 & 0 \\
\end{array}\right) \Phi_{x,y+1}  +   \left(\begin{array}{cc} 
0 & 0 \\
0 & t_x \\
\end{array}\right) \Phi_{x-1,y} + \left(\begin{array}{cc} 
t & 0 \\
0 & 0 \\
\end{array}\right) \Phi_{x+1,y} = 0. 
\end{align}
Here $\ket{\Phi}$ is sliced into $L_x$ pieces; 
$\ket{\Phi}= [\cdots,\ket{\Phi}_{x-1},\ket{\Phi}_x,\ket{\Phi}_{x+1},\cdots]$, and $\Phi_{x,y}$ 
is the $y$-th component of $\ket{\Phi}_x$; 
$\ket{\Phi}_x= [\cdots,\Phi_{x,y-1},\Phi_{x,y},\Phi_{x,y+1},\cdots]$.
$\Phi_{x,y}$ has two components, 
$\Phi_{x,y}  \equiv [\psi_{x,y}, \phi_{x,y}]^T$. 
By solving the Schrodinger equation for $\psi_{x+1,y}$ and $\phi_{x+1,y}$ in terms of $\psi_{x,\cdots}$ and $\phi_{x,\cdots}$, one obtain 
\begin{align}
\psi_{x+1,y} &= -\frac{t_x+\epsilon_{x,y}}{t} \psi_{x,y} - \phi_{x,y} - \frac{t_y}{t} \phi_{x,y+1} \nonumber \\
\phi_{x+1,y}& = \frac{t_y(t_x+\epsilon_{x,y})}{t(t+\epsilon_{x+1,y})} 
\psi_{x,y} + \frac{2t_y - t_x}{t+\epsilon_{x+1,y}} \phi_{x,y} 
+ \frac{t_x+\epsilon_{x,y-1}}{t+\epsilon_{x+1,y}} \psi_{x,y-1} 
+ \frac{t}{t+\epsilon_{x+1,y}} \phi_{x,y-1} + \frac{t^2_y}{t(t+\epsilon_{x+1,y})} \phi_{x,y+1}.  \nonumber 
 \end{align}
The equation relates $\ket{\Phi}_{x+1}$ with $\ket{\Phi}_x$ 
by a transfer matrix $\hat{M}_{x\rightarrow x+1}$;
\begin{align}
\ket{\Phi}_{x+1} = M_{x\to x+1} \ket{\Phi}_{x} 
\end{align}
with 
\begin{align}
&[M_{x\to x+1}]_{(y,a|y^{\prime},b)} \nonumber \\
& \ \ = \left(\begin{array}{cc} 
-\frac{t_x+\epsilon_{x,y}}{t} & - 1 \\
\frac{t_y(t_x+\epsilon_{x,y})}{t(t+\epsilon_{x+1,y})} & \frac{2t_y - t_x}{t+\epsilon_{x+1,y}} \\ 
\end{array}\right)_{a,b} \delta_{y,y^{\prime}} 
+ \left(\begin{array}{cc} 
0 & -\frac{t_y}{t}  \\
0 & \frac{t^2_y}{t(t+\epsilon_{x+1,y})} 
\end{array}\right)_{a,b} \delta_{y+1,y^{\prime}}  
+ \left(\begin{array}{cc} 
0 & 0  \\
\frac{t_x+\epsilon_{x,y-1}}{t+\epsilon_{x+1,y}} 
& \frac{t}{t+\epsilon_{x+1,y}} \\ 
\end{array}\right)_{a,b} 
\delta_{y-1,y^{\prime}}.  
\end{align}
Similarly, for $h^{\dagger}$, the zero-energy state $\ket{\Phi}^{\prime}$ 
is sliced into the $L_x$ pieces, $\ket{\Phi}^{\prime} = [\cdots, \ket{\Phi}^{\prime}_{x+1},\ket{\Phi}^{\prime}_x, \ket{\Phi}^{\prime}_{x-1},\cdots]$, and $\ket{\Phi}^{\prime}_{x+1}$ and $\ket{\Phi}^{\prime}_x$ are recursively related by the 
transfer matrix of $h^{\dagger}$, 
\begin{align}
\ket{\Phi}^{\prime}_{x+1} = M^{\prime}_{x \rightarrow x+1} \ket{\Phi}^{\prime}_{x}.
\end{align}
The Schrodinger equation for $\ket{\Phi}'$ is
\begin{align}
\left(\begin{array}{cc} 
t_x + \epsilon_{x,y}^* & t_y \\
t & t + \epsilon_{x,y}^* \\
\end{array}\right) \Phi_{x,y}' + \left(\begin{array}{cc} 
0 & t \\
0 & 0 \\
\end{array}\right) \Phi_{x,y+1}' + \left(\begin{array}{cc} 
0 & 0 \\
t_y & 0 \\
\end{array}\right) \Phi_{x,y-1}'  +   \left(\begin{array}{cc} 
0 & 0 \\
0 & t_x \\
\end{array}\right) \Phi_{x+1,y}' + \left(\begin{array}{cc} 
t & 0 \\
0 & 0 \\
\end{array}\right) \Phi_{x-1,y}' = 0,
\end{align}
which lead to
\begin{align}
& [M^{\prime}_{x \rightarrow x+1}]_{(y,a|y^{\prime},b)} \\
&= \begin{pmatrix}
    \frac{t(2t_y-t_x)}{t_x(t_x+\epsilon_{x+1,y}^*)} & \frac{t_y(t+\epsilon_{x,y}^*)}{t_x(t_x+\epsilon_{x+1,y}^*)} \\
    -\frac{t}{t_x} & -\frac{t+\epsilon_{x,y}^*}{t_x}
\end{pmatrix}_{a,b}\delta_{y,y'}
+
\begin{pmatrix}
    \frac{t^2}{t_x(t_x+\epsilon_{x+1,y}^*)} & \frac{t(t+\epsilon_{x,y+1}^*)}{t_x(t_x+\epsilon_{x+1,y}^*)} \\
    0 & 0
\end{pmatrix}_{a,b}\delta_{y+1,y'}
+
\begin{pmatrix}
    \frac{t_y^2}{t_x(t_x+\epsilon_{x+1,y}^*)} & 0 \\
    -\frac{t_y}{t_x} & 0
\end{pmatrix}_{a,b}\delta_{y-1,y'},
\end{align}
In terms of $M_{x \rightarrow x+1}$ and $M^{\prime}_{x \rightarrow x+1}$, 
the transfer matrix of $H$ is given by the block-diagonal form, 
\begin{align}
T_{x\rightarrow x+1} = \left(\begin{array}{cc} 
M^{\prime}_{x\rightarrow x+1} &  0 \\
0 & M_{x \rightarrow x+1} \\
\end{array}\right).  
\end{align}
Due to the block-diagonal form, the LEs of $H$ are given by 
a union of the LEs of $h$ and $h^{\dagger}$. We confirmed numerically that the LEs of $M^{\prime}$ are exactly the opposite numbers of the LEs of $M$. We thus 
expect that $M^{\prime}_{x \rightarrow x+1}$ and $M^{-1}_{x \rightarrow x+1}$ are related to each other by a non-singular similarity transformation.

Likewise, To calculate the localization length $\xi_y$ along $y$, we consider a quasi-1D geometry with $L_x\times L_y$ unit cells and $L_y\gg L_x$. We slice the zero-energy state $\ket{\Phi}$ into $L_y$ pieces; 
$\ket{\Phi}= [\cdots,\ket{\Phi}_{y-1},\ket{\Phi}_y,\ket{\Phi}_{y+1},\cdots]$, and relate $\ket{\Phi}_{y-1}$ with $\ket{\Phi}_y$ 
by a transfer matrix $\hat{M}_{y\rightarrow y-1}$,
\begin{align}
\ket{\Phi}_{y-1} = M_{y\to y-1} \ket{\Phi}_{y} 
\end{align}
with 
\begin{align}
&[M_{y\to y-1}]_{(x,a|x^{\prime},b)} \nonumber \\ 
&  \ \ = \left(\begin{array}{cc} 
-\frac{t_y}{t} & - \frac{t+\epsilon_{x,y}}{t} \\
\frac{t_y(t_x+\epsilon_{x,y-1})}{t^2} & 
\frac{(t_x+\epsilon_{x,y-1})(t+\epsilon_{x,y})+(t_x-t_y)t}{t^2} \\ 
\end{array}\right)_{a,b} \delta_{x,x^{\prime}} 
+ \left(\begin{array}{cc} 
0 & 0  \\
\frac{t_y}{t} & \frac{t+\epsilon_{x+1,y}}{t} 
\end{array}\right)_{a,b} \delta_{x+1,x^{\prime}}  
+ \left(\begin{array}{cc} 
0 & -\frac{t_x}{t}  \\
0 & \frac{t_x(t_x+\epsilon_{x,y-1})}{t^2} \\ 
\end{array}\right)_{a,b} 
\delta_{x-1,x^{\prime}}.
\end{align}
In terms of these transfer matrices, we calculate 
the LEs of the zero-energy state of $h$ along $x$ and $y$  
from the eigenvalues of the following two matrices, 
\begin{align}
\Omega_{h,x} & \equiv \lim_{L_x \rightarrow \infty} 
\frac{1}{2L_x} \ln \Big(M^{\dagger}_{1\to L_x} 
M_{1\to L_x}\Big), \quad M_{1\to L_x} 
\equiv \prod^{L_x-1}_{n=1} M_{n \to n+1}, \nonumber \\ 
\Omega_{h,y} & \equiv \lim_{L_y \rightarrow \infty} 
\frac{1}{2L_y} \ln \Big(M^{\dagger}_{L_y\to 1} 
M_{L_y \to 1}\Big), \quad M_{L_y \to 1} 
\equiv \prod^{1}_{n=L_y-1} M_{n+1 \to n}, \nonumber 
\end{align}

For $t_x=t$ and $t_y \ne t$, an ensemble of $h$ has the transpositional 
symmetry accompanied by the mirror $U_y$, Eq.~(\ref{eq:SM_transpose}). Since 
the transposition changes signs of LEs along $x$ and the mirror does not,  
a set of the LEs of $h$ along $x$ become statistically symmetric with respect to zero; 
$\gamma_{x,1}<\gamma_{x,2}<\cdots<\gamma_{x,2L_y-1},\gamma_{x,2L_y}$ 
with $\gamma_{x,j}= -\gamma_{x,2L_y-(j-1)}$. Meanwhile, since both the 
transposition and the mirror change the sign along $y$, a set of the LEs of $h$ 
along $y$ are not necessarily symmetric with respect to zero. More generally, 
a number $N_{y,+}$ of positive LEs along $y$ can be different from a number 
$N_{y,-}$ of negative LEs along $y$, and their difference is related to the 
topological winding number along $y$ as $\nu_y = \frac{1}{2L_x}(N_{y,+}-N_{y,-})$~\cite{xiaoAnisotropicTopologicalAnderson2023} (see also Fig.~\ref{fig:WTI}(c)).

\begin{figure}[htb]
    \centering
    \includegraphics[width=1.0\textwidth]{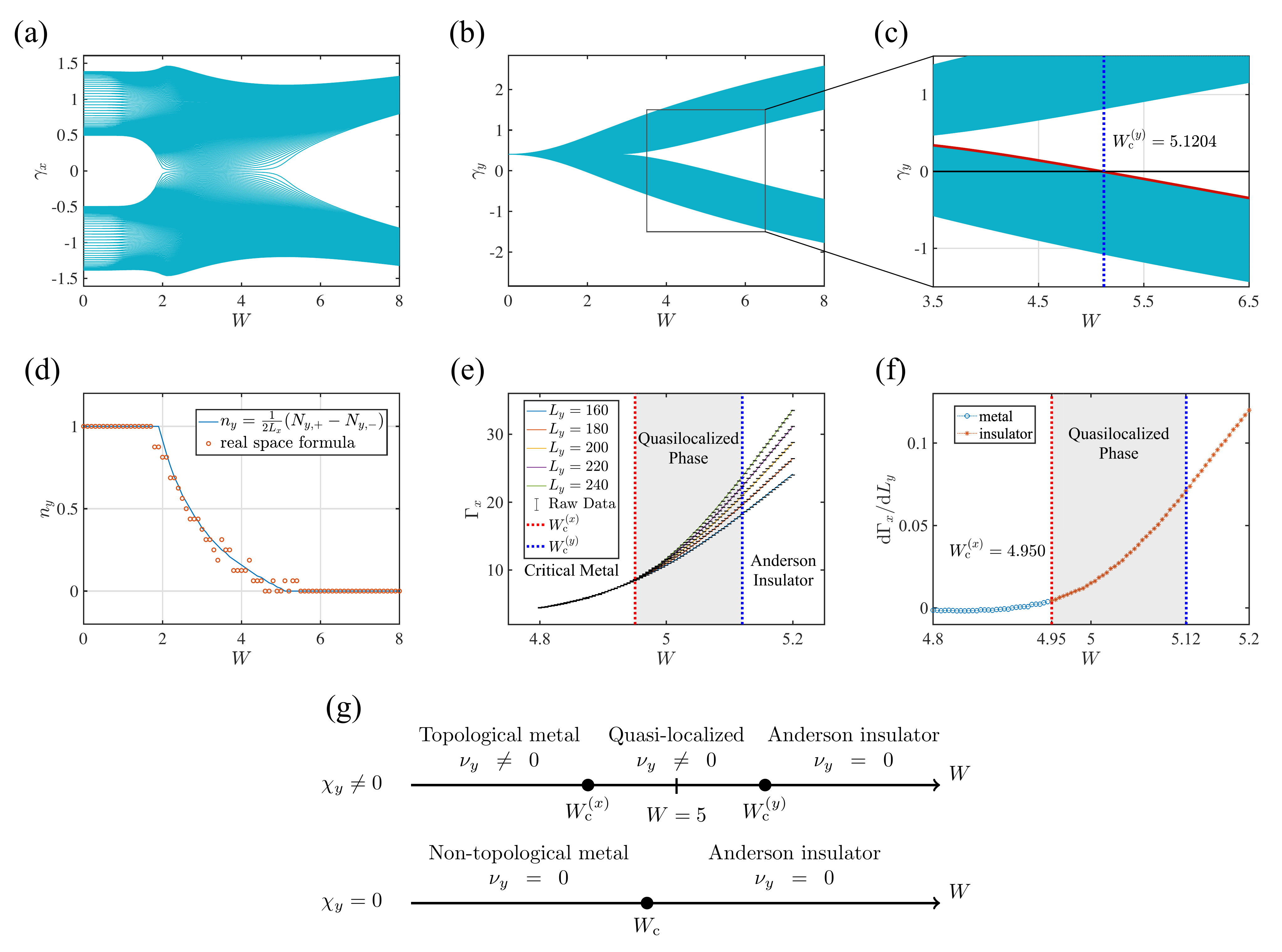}
    \caption{(a) and (b,c) Distributions of the LEs of the zero-energy state of $h$ as a function of the disorder strength $W$. (a) The LEs of $h$ along the $x$-direction. The distributions of the LEs along $x$ are symmetric with respect to zero. (b,c) The LEs of $h$ along the $y$-direction. The LEs along $y$ cross zero when $2<W<5.12$. In the thermodynamic limit ($L_x \rightarrow \infty$), these LEs form a continuous band, and the band covers the zero when $2<W<5.12$. (c) A zoom-in figure of the LEs of $h$ along the $y$-direction for $3.5<W<6.5$. The boundary of the region in which LEs cross zero is obtained by a linear fitting, $W_{\text{c}}^{(y)}=5.120$. (d) The one-dimensional topological winding number $\nu_y$ along $y$ as a function of the disorder strength $W$. The blue curve plots $\nu_y = \frac{1}{2L_x}(N_{y,+}-N_{y,-})$, where $N_{y,\pm}$ are calculated from the numbers of positive/negative LEs of $h$ along $y$, respectively. The orange dots are $\nu_y$ calculated by a real space formula of the winding  number~\cite{linRealspaceRepresentationWinding2021,mondragon-shemTopologicalCriticalityChiralSymmetric2014}. (e) Numerical data for normalized localization length $L_y/\xi_x \equiv \Gamma_x$ along $x$, as a function of disorder strength $W$. $W^{(x)}_{\text{c}}$ (red dotted line) and $W^{(y)}_{\text{c}}$ (blue dotted line) determined by finite size scaling analyses are shown for the eye guide. (f) A derivative of $\Gamma_x$ with respect to $L_y$ is shown around $W=5$. The slope $d\Gamma_x/dL_y$ is obtained by fitting $\Gamma_x$ at fixed $W$ as a linear function of $L_y$. (g) (upper) Schmatic phase diagram around $W=5$ for $(t_x,t_y)=t(1,1.5)$. (lower) Schematic phase diagram around $W=2.5$ for $(t_x,t_y)=t(1,1)$.}
    \label{fig:WTI}
\end{figure}

\cref{fig:WTI} shows numerical results of the transfer matrix calculation at 
$t_x=t$ and $t_y =1.5 t$, where the model is in the chiral unitary class 
with $\nu_x=0$ and $\nu_y\ne 0$. The LEs of the zero-energy state of $h$ along $x$ and $y$
are calculated with the quasi-1D geometry, $L_\perp\times L_\parallel$, where $L_\perp=60\sim120$ and $L_\parallel=10^6$. \cref{fig:WTI} shows distributions 
of all the LEs of $h$ along $x$ (a) and along $y$ (b,c). The LEs along 
$x$ are distributed symmetrically around zero, being consistent 
with the statistical symmetry of $h$ at $t_x=t$ and $t_y \ne t$. 
The LEs along $y$ are distributed asymmetrically around zero. Moreover, 
all the LEs along $y$ are positive near the clean limit, 
$N_{y,+}=2L_x$ and $N_{y,-}=0$, so that $\nu_y = (N_{y,+}-N_{y,-})/(2L_x)=1$ (b). 
Upon increasing the disorder strength $W$, some of the LEs along $y$ decrease 
and become negative for $W\gtrsim 2$ (b,c). For $W>W^{(c)}_{y}$, a lower half of the LEs 
along $y$ become negative, while the upper half of the LEs along $y$ 
are positive; $N_{y,+}=L_x$, $N_{y,-}=L_x$ and $\nu_y=(N_{y,+}-N_{y,-})/(2L_x)=0$.

The lower half of the LEs along $y$ form a continuous band in 
the thermodynamic limit $(L_x \rightarrow \infty)$~\cite{markosPhenomenologicalTheoryMetalinsulator1995}, 
and the band covers the zero for $2 \lesssim W < W^{(c)}_y$. Thereby, 
$0<N_{y.-}<L_x<N_{y,+}$ and $0<\nu_y<1$ [Fig.~\ref{fig:WTI}(d)]. 
A localization length of $H$ is an inverse of the smallest absolute 
value of the LEs of $h$. Thus, for $2\lesssim  W< W^{(c)}_y$, the 
exponential localization length of $H$ along $y$ is always divergent 
in the thermodynamic limit. The value of $W^{(c)}_{y}$ is determined by a 
fitting of the largest LE in the lower half of the LEs of $h$ along $y$;  
$W^{(c)}_{y}\approx 5.120 \pm 0.002$ [see a subsection of ``localization 
length $\xi_y$ along $y$ (topological direction)" in the next section]. 

Importantly, the localization length $\xi_x$
along $x$ undergoes a metal-insulator transition {\it within} a region of $2\lesssim W<W^{(y)}_{\rm c}$. In the insulator phase, $\xi_x/L_y\equiv 1/(\gamma_x L_y)$ decreases for increasing $L_y$, while in the (critical) metal phase, $\xi_x/L_y$ stays invariant for increasing $L_y$ [see Fig.~\ref{fig:WTI}(e,f)]. We determine a critical disorder 
strength $W^{(x)}_{\rm c}$ for this change of $\xi_x/L_y$ 
by a finite-size scaling analysis [see a subsection of 
``localization 
length $\xi_x$ along $x$ (non-topological direction)" in the next section for its details]. $W^{(x)}_{\rm c}$ thus determined turns out to be clearly smaller than $W^{(y)}_{\rm c}$, $W^{(x)}_{\rm c}=4.950\pm 0.001$. A discrepancy between $W^{(x)}_{\rm c}$ and $W^{(y)}_{\rm c}$ is about $3\%$ of the critical values, and 
it is far larger than the error bars of the fittings. 
This indicates a phase diagram with (i) critical metal phase $(W<W^{(x)}_{\rm c})$, (ii) quasilocalized phase ($W^{(x)}_{\rm c}<W<W^{(y)}_{\rm c}$), and (iii) Anderson localized phase ($W^{(y)}_{\rm c}<W$). The finite-size scaling analysis also estimates 
the critical exponent for the phase transition between critical metal and quasilocalized phases as $\nu=1.36\pm 0.01$. 

In summary, upon increasing $W$, the model undergoes a phase transition from topological metal phase ($\nu_y \ne 0$, $W<W^{(x)}_{\rm c}$) to quasi-localized phase ($\nu_y \ne 0$, $W^{(x)}_{\rm c}<W<W^{(y)}_{\rm c}$), and then undergoes another phase transition from quasi-localized phase to Anderson localized phase ($\nu_y=0$, $W^{(y)}_{\rm c}<W$) (upper panel of Fig.~\ref{fig:WTI}(g)). In the quasi-localized phase with $\nu_y \ne 0$, the exponential localization length along $y$ is divergent, while the localization length along $x$ is finite. 
 	
\cref{fig:NLM} shows numerical results of the transfer matrix calculation at $t_x=t_y=t$, where the model is in the chiral unitary class with $\nu_x=\nu_y=0$. \cref{fig:NLM}(b) shows that sets of the LEs along $x$ and along $y$ are distributed in the symmetry way around zero, being consistent with the statistical transpositional symmetry of $h$ at $t_x=t_y=t$. \cref{fig:NLM}(a) shows that the localization length along $x$ and that along $y$ 
diverges at the same critical disorder strength. Namely, the model in the chiral unitary class with $\nu_x=\nu_y=0$ undergoes a direct 
transition from the non-topological metal phase to the Anderson localized phase. 

	\begin{figure}[H]
		\centering
		\subfloat[]{\includegraphics[width=0.35\textwidth]{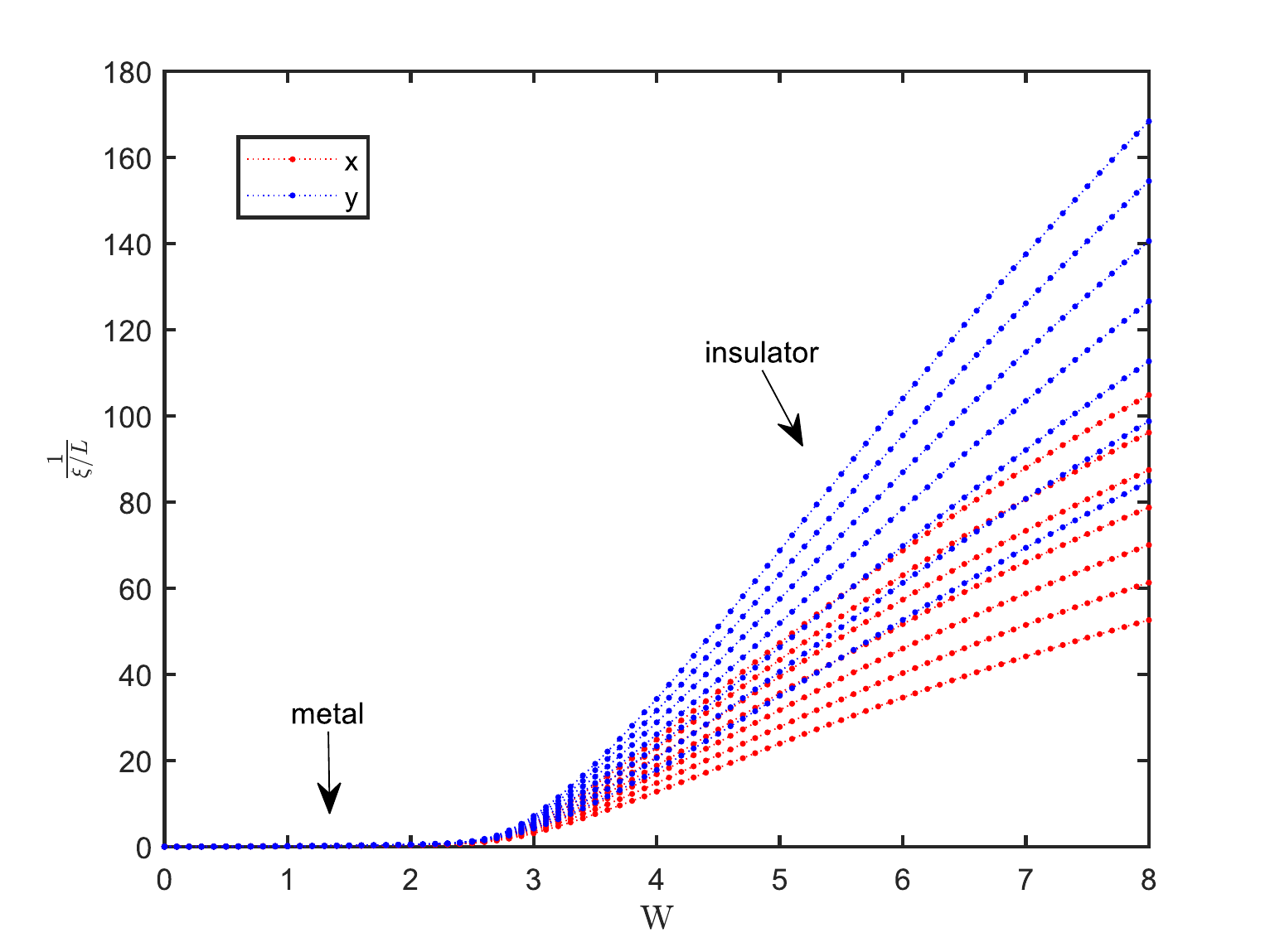}}
		\subfloat[]{\includegraphics[width=0.55\textwidth]{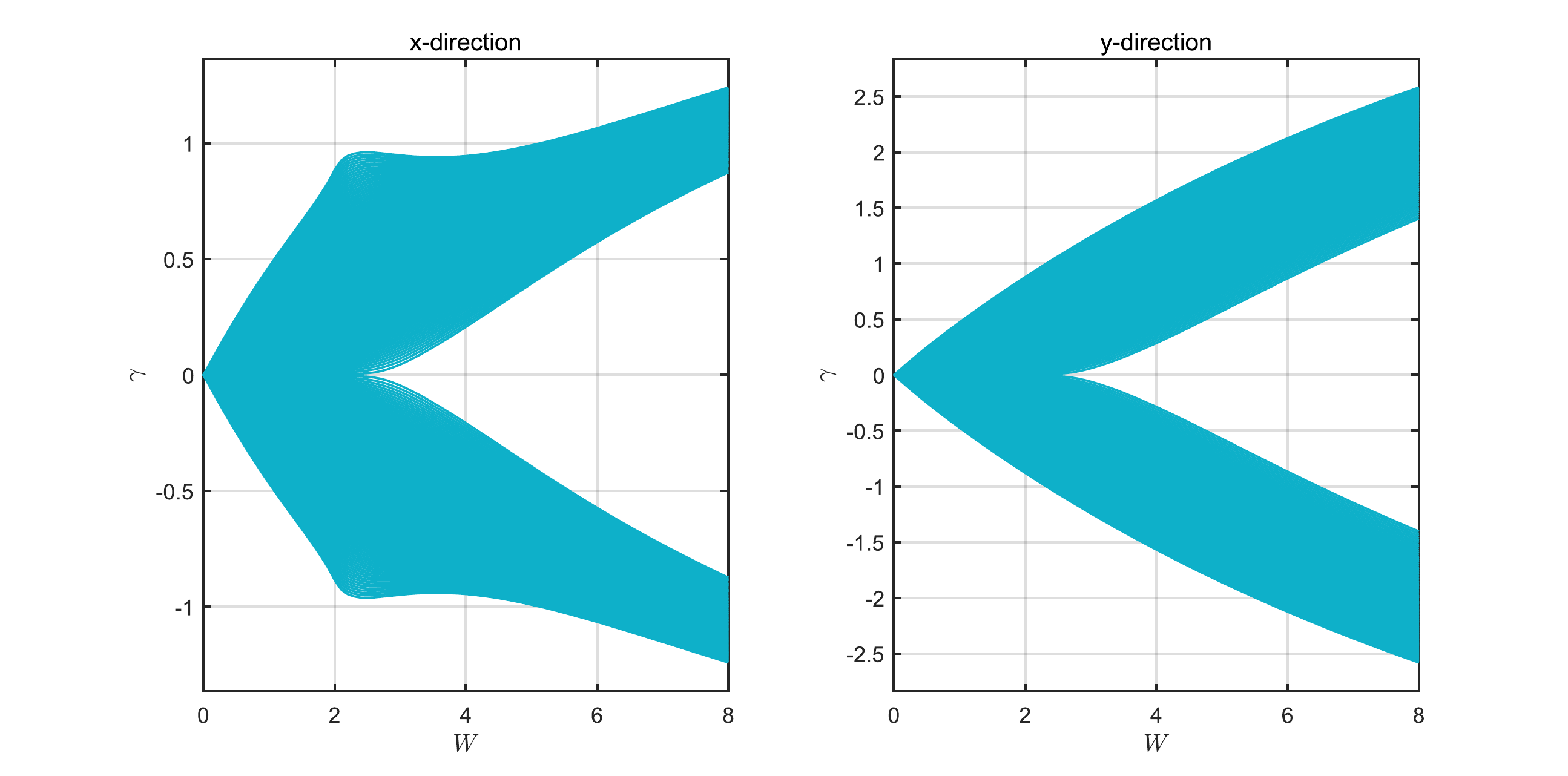}}
		\caption{(a) Inverse normalized localization length $L_{\perp}/\xi_{\mu}$ along $x$ ($\mu=x$ and $L_{\perp}=L_y$ : red) and $y$ ($\mu=y$ and $L_{\perp}=L_x$: blue) as a function of disorder strength $W$ at $t_x=t_y=t$. The system size $L_{\perp}$ varies from $L_{\perp}=60$ to $L_{\perp}=120$. The figure suggests that the exponential localization lengths $\xi_x$ and $\xi_y$ along $x$ and $y$ diverge at the same critical disorder strength. (b) Distributions of the LEs of zero-energy state of $h$ along $x$ and $y$ as a function of the disorder strength $W$ at $t_x=t_y=t$. \label{fig:NLM}}
	\end{figure}

\subsection{*Finite-size scaling analysis*}
We calculate numerically the Lyapunov exponents (LEs) of $H$ or $h$ along $x$ and $y$
with finite system sizes, and analyze the data based on finite-size scaling (FSS) theories. In the following, we will explain the FSS analyses for the LEs along $y$ and those along $x$.

\subsubsection{Localization length $\xi_y$ along $y$ (topological direction)}
 The LEs of $h$ along the topological direction ($y$) form two continuous bands for larger $L_x$. The lower continuous band covers the zero for $2\lesssim W<W^{(y)}_{\rm c}$. The smallest absolute value of the LEs of $h$ is an inverse of the localization length $\xi_y$ of $H$. Thus, $\xi_y$ is divergent in the region of $2\lesssim W<W^{(y)}_{\rm c}$. We determine the upper boundary $W^{(y)}_{\rm c}$ by a linear extrapolation of the largest LE of the lower continuous band; $\frac{1}{\xi_y(L_x)} = \frac{a}{L_x} + \frac{1}{\xi_y(\infty)}$~\cite{xiaoAnisotropicTopologicalAnderson2023,asada05}. The extrapolated values $\xi_y(\infty)$ are obtained for $W=4.90, 4.95, 5.00, \cdots, 5.25, 5.30$. A linear fitting line of the extrapolated values $\xi_y(\infty,W)$ crosses the zero at $W=W^{(y)}_{\rm c}=5.120\pm 0.002$.

\subsubsection{Localization length $\xi_x$ along $x$ (non-topological direction)}
The LEs of $h$ along the non-topological direction ($x$) are distributed symmetrically around zero [see Fig.~\ref{fig:WTI}(a)]. The smallest positive LE of $h$ is the inverse of the localization length $\xi_x$ of $H$. For $t_x=t$ and $t_y=1.5t$, the numerical data of $1/\xi_x$ are obtained from the LE of $h$ with the quasi-one-dimensional (Q1D) geometry, $L_x = 5\times 10^6$ and 
$L_y=160, 180, 200, 220, 240$ [Fig.~\ref{fig:WTI}(e),Fig.~\ref{fig:fitting_and_comparison}(b)], and  $L_x = 10^6$ and 
$L_y=60, 70, \cdots, 120$ [Fig.~\ref{fig:fitting_and_comparison}(a)].  

Fig.~\ref{fig:WTI}(e) and Fig.~\ref{fig:fitting_and_comparison} show that the localization length $\xi_x$ normalized by the finite system size $L_y$ becomes scale-invariant in the 2D critical metal phase, while it decreases for increasing $L_y$ in insulating regions. To determine a critical disorder strength $W^{(x)}_{\rm c}$ for the change of the $L_y$-dependence of $\xi_x/L_y$, we use the single-parameter scaling ansatz for numerical data of $\xi_x/L_y$ for $W>W^{(x)}_{\rm c}$,
\begin{align}
\Gamma_x \equiv L_y/\xi_x = f\Big(u(w) L^{1/\nu}_y\Big),
\end{align}
with the critical exponent $\nu$. Here a relevant scaling variable $u(w)$ is a function of 
the normalized disorder strength $w \equiv (W-W^{(x)}_{\text{c}})/W^{(x)}_{\text{c}}$. 
Generally, the single-parameter scaling function can be Taylor-expanded for smaller $u(w)$ with finite $L_y$, and the scaling variable can be expanded for smaller $w$,  
\begin{align}
    f(x) & = \sum_{m=0}^{N} a_{m} x^{m}, \quad u(w)=\sum_{n=1}^{M} b_n w^n. \label{u1} 
\end{align}
The expansions with small integers $N$ and $M$ are justified in a fitting model 
for the data near the critical disorder strength $W^{(x)}_{\rm c}$ with finite $L_y$. For a given $M$ 
and $N$, we minimize the following $\chi^2$ loss function with respect to 
$W^{(x)}_{\rm c}$, $\nu$, $a_0$, $a_1$, $\cdots$, $a_N$, $b_1$, $b_2$, $\cdots$, 
and $b_M$,
\begin{equation}
    \chi^2 \equiv \sum_{j=1}^{N_{\text{data}}}\frac{(f_j-\Gamma_j)^2}{\sigma_j^2}. \label{chi2}
\end{equation}

In Eq.~(\ref{chi2}), $\Gamma_j$ and $\sigma_j$ are the numerical value and error for 
$\Gamma_x$ at the $j$-th data point, respectively. $f_j$ is the fitting 
value for $\Gamma_j$ from the fitting model, Eq.~(\ref{u1}). The minimization of $\chi^2$ determines optimal parameters' values. An error bar for each fitting parameter is evaluated by 1000 sets of synthetic data. To be specific, we generate synthetic data $\tilde{\Gamma}_j$ around $\Gamma_j$ based on the Gaussian distribution with the standard error $\sigma_j$. Then we fit a set of the synthetic data points with the same fitting model, to obtain another set of the fitting parameters' values. We repeat this refitting 1000 times, to obtain statistics of the fitting parameters' values. The distribution of each fitting parameter should be Gaussian. From the distribution, we determine the $95\%$ confidence interval as the error bar of the fitting parameters.

\begin{figure}[b]
    \centering
    \includegraphics[width=1.0\textwidth]{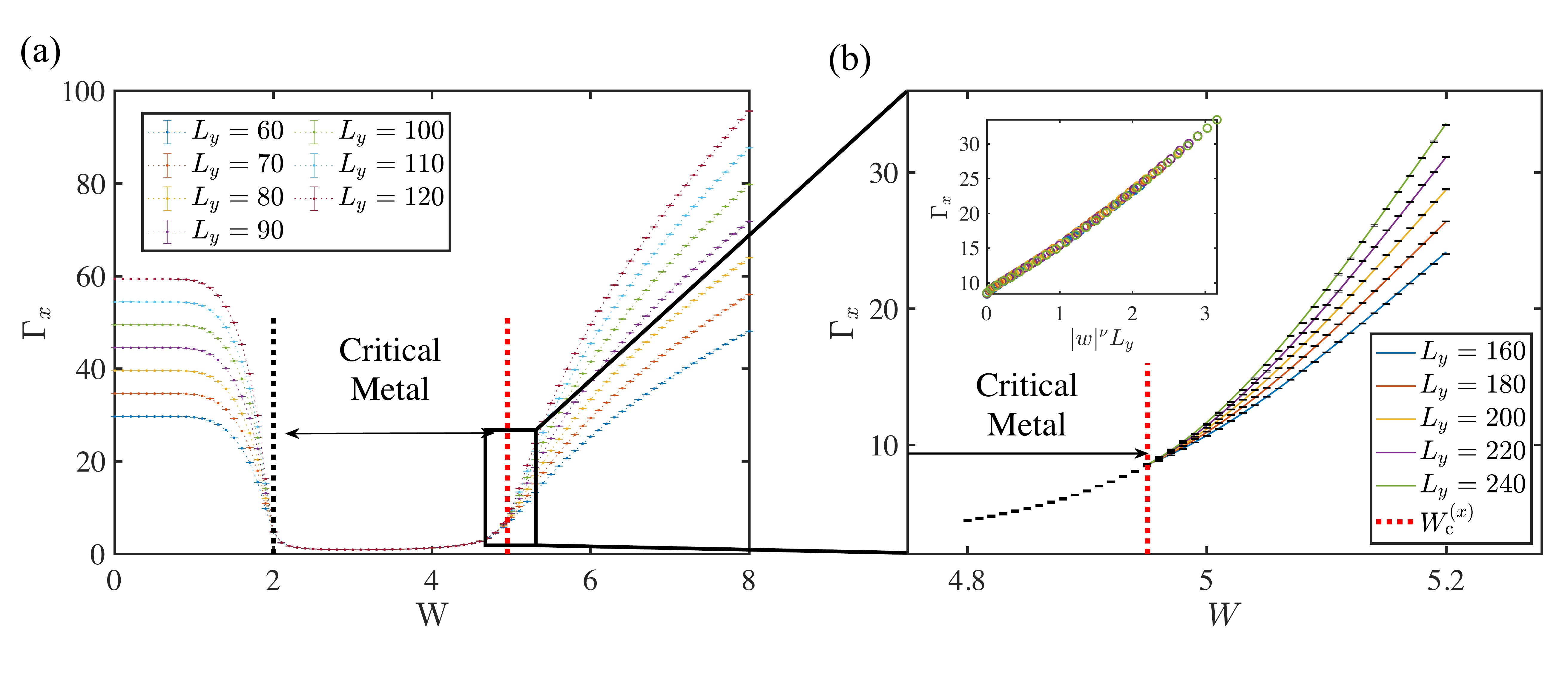}
    \caption{Inverse normalized localization length $\Gamma_x=\gamma_xL_y$ along $x$ for 
    $t_y=1.5t$ and $t_x=t$ with $L_x\times L_y$. (a) $L_x=1\times 10^6$ and $L_y=60\sim 120$ (b) $L_x=5\times 10^6$ and $L_y=160\sim 240$. In (a), the critical metal region is observed in $2\lesssim W<W_{\text{c}}^{(x)}$. In (b), raw numerical data are shown together with the fitting model of Eq.~(\ref{u1}), $(M,N)=(2,2)$, $W_x^{\text{(c)}}=4.950$, and $\nu=1.36$. Inset of (b): single-parameter data collapse of $\Gamma_x$ as a function of $|u_1(w)|^\nu L_y$ with $W_x^{\text{(c)}}=4.950$ and $\nu=1.36$.}
    \label{fig:fitting_and_comparison}
\end{figure}

The minimization of $\chi^2$ is carried out by the trust region reflective algorithm. $W^{(x)}_{\rm c}$ changes at every step of the optimization, and so does the range of the data points for the fitting. In practice, we first judge by eye that for $t_x=t$ and $t_y=1.5t$, $W^{(x)}_{\rm c}$ must be in a range of $4.8<W<5.2$ [see Fig.~\ref{fig:fitting_and_comparison}(a,b)]. Then, we prepare data points in the range and carry out the minimization of $\chi^2$. At every step in the optimization, $W^{(x)}_{\rm c}$ changes, and only the data points above $W^{(x)}_{\rm c}$ determine parameters' values at the next step. With $N=M=2$, we minimize the $\chi^2$ loss function by trying different initial parameters to obtain the best fitting, 
\begin{equation}
    W^{(x)}_{\text{c}}=4.950\pm0.001,\quad \nu=1.36\pm 0.01.
\end{equation}
With these two values, the data points in the range of $W=4.95\sim 5.2$ and 
$L_y=160 \sim 240$ collapse well into a single scaling function 
[see an inset of Fig.~\ref{fig:fitting_and_comparison}(b)]. 

Numerically, we have determined the three phases (critical metal, quasi-localized phase, and Anderson insulator). Their phase transitions are driven by the disorder strength $W$. Such a numerical phase diagram can be qualitatively explained by the RG phase diagram. In the RG phase diagram, the phases are controlled by fugacity $y$ and weak topological strength $\chi_y$. When $\chi_y\ne 0$ and $y$ are sufficiently small, the system is in a (topological) metallic phase; when $\chi_y\ne 0$ and $y$ are sufficiently large, the system transits to a quasi-localized phase; when $\chi_y =0$ and $y$ is sufficiently large, the system becomes an Anderson insulator. The fugacity $y$ and weak topological strength $\chi_y$ are both functions of the disorder strength $W$. When $W$ increases, $y$ increases and $\chi_y$ decreases, and finally $\chi_y$ ends up at zero. Then, the system goes through the metallic region, quasi-localized region, and Anderson localized region.


\section{*Derivation of nonlinear sigma model*}
In the paper, we analyzed the nonlinear sigma model (NLSM) in the chiral classes with the 1D weak topological term ($\vec{\chi}\ne 0$) and argued that the 1D weak topological term induces a spatial polarization of vortex-antivortex pairs along $\vec{\chi}$, and a proliferation of the polarized pairs leads to the emergence of the quasi-localized phase with finite conductivity along $\vec{\chi}$, and vanishing conductivity along the other direction. As a numerical test for this theory, the previous section studied a two-dimensional tight-binding model with 1D weak topology $\vec{\nu}$ [Eq.~(\ref{nu-def})], and demonstrated the emergence of the quasi-localized phase with the divergent localization length along $\vec{\nu}$ and finite localization length along the other direction. 

In this section, we derive the NLSM
with the 1D weak topological term $\vec{\chi}$ from the tight-binding model with $\vec{\nu}\ne 0$, and show that $\vec{\chi}$ and $\vec{\nu}$ are parallel to each other;
\begin{align}
\frac{\vec{\chi}}{8\pi} = \vec{\nu}. 
\end{align}
To this end, we choose $t_x=t$ and fix $\vec{\nu}$ along $y$ in the tight-binding model; $\vec{\nu}=(0,\nu_y)$. For $t_y>t$, the tight-binding model in the clean limit has an energy gap, and $\nu_y$ is quantized to an integer. For $t_y<t$, the gap closes and a pair of two gapless Dirac fermions appears at $(k_x,k_y)=(K_0,0)$ and $(k_x,k_y)=(-K_0,0)$ respectively. $\nu_y$ takes a non-integer value and changes its value with 
$K_0$ when $t_y$ changes.

Note that the numerical test in the previous section is applied for $t_y=1.5t$ and $t_x=t$, whose clean-limit Hamiltonian has the gap. A finite disorder $W$ renormalizes the tight-binding parameters, where the gap closes, and a pair of massless Dirac fermions emerge around the zero energy for $2\lesssim W \lesssim 5$. 
In fact, 
the disordered Hamiltonian is characterized by $0<\nu_y<1$ and $\nu_x=0$ [see Fig.~\ref{fig:WTI}(d)], and the same $(\nu_x,\nu_y)$ is realized by the clean-limit Hamiltonian at $t_y<-t$ and $t_x=t$ that has a pair of the massless Dirac fermions at $(k_x,k_y)=(K_0,0)$ and $(k_x,k_y)=(-K_0,0)$ 
[see Fig.~\ref{fig:cleanlimit_winding}].  
Thus, we regard an effective continuum model with 
the pair of two massless Dirac fermions as a good starting point 
for the derivation of the NLSM with the weak topological term $\vec{\chi}$. 

In the presence of a finite disorder, the gapless fermions at zero energy acquire a finite lifetime. The finite-life-time solution is a saddle-point (self-consistent Born) solution of an action for the disordered fermions, and a low-energy fluctuation around the saddle-point solution is described by the NLSM.
In the following, we first construct the effective continuum model for the tight-binding model at $t_y<t$ and $t_x=t$. The continuum model has the two massless Dirac fermions, and the random hoppings in the tight-binding model result in intravalley and intervalley scatterings between the two nodes. Using the replica method, we treat these single-particle scatterings and introduce a partition function for the disordered massless Dirac fermions. Based on the formulation, we clarify a saddle-point solution of an action for 
the partition function, and derive the NLSM as the low-energy effective theory around the saddle-point solution.

\subsection{Effective continuum model}
The tight-binding model at $t_x=t$ reduces to a 4 by 4 matrix in the momentum space, 
\begin{align}
H ({\bm k})= a_1({\bm k}) \sigma_1 \times \tau_0 + 
a_2({\bm k}) \sigma_2 \times \tau_3 + a_3({\bm k
}) \sigma_1 \times \tau_1 + a_4({\bm k}) \sigma_2 \times 
\tau_2 + a_5({\bm k}) \sigma_1 \times \tau_2 
+ a_6({\bm k}) \sigma_2 \times \tau_1, 
\end{align}
with $a_1 = t (1+\cos k_x)$, $a_2 = - t\sin k_x$, $a_3 = 
(t+t_y)/2 (1+\cos k_y)$, $a_4 = - (t-t_y)/2 (1-\cos k_y)$, 
$a_5 = - (t+t_y)/2 \sin k_y$, and $a_6 = (t-t_y)/2 \sin k_y$. 
The matrix has two pairs of chiral symmetric momentum-energy 
dispersions, 
\begin{align}
E_{\pm}({\bm k}) = \sqrt{|{\bm a}|^2 \pm 2\sqrt{(a^2_1+a^2_2)
(a^2_3+a^2_5) + (a_3 a_4 - a_5 a_6)^2}}. 
\end{align}
with $|{\bm a}|^2 \equiv a^2_1+a^2_2+a^2_3+a^2_4+a^2_5+a^2_6$. 
In the Brillouin zone $(k_x,k_y) \in [-\pi,\pi]\times [-\pi,\pi]$, 
$\pm E_{+}({\bm k})$ is gapped except for at $t_y=\pm t$. At $t_y=t$ and 
$t_y=-t$, the gap closes at $(k_x,k_y)=(\pi,\pi)$ 
and $(k_x,k_y)=(\pi,0)$, respectively. On the other hand, 
$\pm E_{-}({\bm k})$ form a pair of two massless Dirac fermions at 
$(k_x,k_y)=(-K_0,0) \equiv {\bm K}_0$ and $(K_0,0) \equiv -{\bm K}_0$ 
for $t_y<t$, and $K_0$ changes as a function of $t_y$. In this 
section, we consider the parameter region of 
$t_y<t$ ($t_y \ne -t$) and $t_x=t$, and derive 
an effective continuum model for the disordered massless Dirac fermions. 

Let us refer to the two zero modes at $(k_x,k_y)={\bm K}_0$ as 
$e^{i{\bm K}_0 {\bm r}_i} e_i$ $(i=1,2)$, and the two 
zero modes at $(k_x,k_y)=-{\bm K}_0$ as $e^{-i{\bm K}_0 
{\bm r}_i} e_i$ $(i=3,4)$. Since the tight-binding Hamiltonian 
is chiral symmetric, $\sigma_3 H \sigma_3 = - H$, these zero modes are eigenmodes of the chiral operator $\sigma_3$, $\sigma_3 e_{1} = e_1$, $\sigma_3 e_{2} = - e_2$, $\sigma_3 e_{3} = e_3$, $\sigma_3 e_{4} = - e_4$.

The effective continuum model for disordered massless Dirac fermions is 
derived from the original tight-binding Hamiltonian $H$ on the square lattice, 
\begin{align}
\sum_{{\bm r}^{\prime}}\sum_{b=1,2,3,4} 
(H)_{({\bm r},a|{\bm r}^{\prime},b)} 
\Phi({\bm r}^{\prime},b) = E \Phi({\bm r},a). 
\end{align}
We first expand an eigenstate $\Phi$ of the lattice model in 
terms of the Bloch wavefunctions near the two massless points;
\begin{align}
\Phi({\bm r},a) = \frac{1}{\sqrt{V}} \sum_{\bm k} e^{i{\bm k}{\bm r}} 
\Phi({\bm k},a) \simeq 
\frac{1}{\sqrt{V}} \sum_{|\bm q|\ll \Lambda^{-1}} 
e^{i({\bm K}_0+{\bm q}){\bm r}} \Phi({\bm K}_0+{\bm q},a) 
+ \frac{1}{\sqrt{V}} \sum_{|\bm q|\ll \Lambda^{-1}} 
e^{i(-{\bm K}_0+{\bm q}){\bm r}} \Phi(-{\bm K}_0+{\bm q},a). 
\end{align}
Here, $V$ is the total volume of the lattice model. 
The Bloch wavefunctions around the massless points can 
be expanded in terms of the zero modes,
\begin{align}
\Phi({\bm K}_0+{\bm q},a) = \psi_1({\bm q}) e_1(a) 
+ \psi_2({\bm q}) e_2(a), \quad  
\Phi(-{\bm K}_0+{\bm q},a) = \psi_1({\bm q}) e_3(a) 
+ \psi_2({\bm q}) e_4(a). 
\end{align}
Substitute the expansion into the eigenvalue problem on the lattice,  
and Taylor-expand the Fourier-transform of the tight-binding model 
in terms of small ${\bm q}$ [up to the first order]. Then, in the 
real-space representation, we get the effective continuum model ${\cal H}$ as,
\begin{align}
&{\cal H}\big({\bm r},\nabla_{\bm r}\big) \left(\begin{array}{c}
\psi_1({\bm r}) \\
\psi_3({\bm r}) \\
\psi_2({\bm r}) \\
\psi_4({\bm r}) \\
\end{array}\right) = E \left(\begin{array}{c}
\psi_1({\bm r}) \\
\psi_3({\bm r}) \\
\psi_2({\bm r}) \\
\psi_4({\bm r}) \\
\end{array}\right) \nonumber \\
&{\cal H}
\big({\bm r},\nabla_{\bm r}\big) \equiv  
\left(\begin{array}{cccc} 
 & &  \alpha_x i\partial_x + \alpha_y \partial_y & 0 \\
 & & 0 & - \alpha_x i\partial_x + \alpha_y \partial_y \\
 \alpha_x i\partial_x - \alpha_y \partial_y & 0 & & \\
  0 & - \alpha_x i\partial_x - \alpha_y \partial_y & & \\
\end{array}\right) + \nonumber \\
& \hspace{1cm} + \epsilon^{\prime}_{{\bm r}}
\left(\begin{array}{cccc}
 & &  \beta  & e^{-2i{\bm K}_0 {\bm r}} \\
 & &  e^{2i{\bm K}_0 {\bm r}} & \beta \\
 \beta & e^{-2i{\bm K}_0 {\bm r}} & & \\
  e^{2i{\bm K}_0 {\bm r}} & \beta & & \\
\end{array}\right) 
+ i\epsilon^{\prime\prime}_{{\bm r}}
\left(\begin{array}{cccc}
 & &  \beta  & e^{-2i{\bm K}_0 {\bm r}} \\
 & &  e^{2i{\bm K}_0 {\bm r}} & \beta \\
 -\beta & -e^{-2i{\bm K}_0 {\bm r}} & & \\
  -e^{2i{\bm K}_0 {\bm r}} & -\beta & & \\
\end{array}\right). \label{effective-2}  
\end{align}
Here, $\psi_j(\bm r)$ ($i=1,2,3,4$) stand for slowly-varying envelope functions for the massless Dirac fermions, 
\begin{align}
\psi_j({\bm r}) \equiv \frac{1}{\sqrt{V}} 
\sum_{{\bm q}\ll \Lambda^{-1}} 
e^{i{\bm q}{\bm r}} \psi_j(q).
\end{align}
$\psi_1({\bm r})$ and $\psi_2({\bm r})$ are for the  
Dirac fermion at ${\bm k}={\bm K}_0$, and $\psi_3({\bm r})$ and 
$\psi_4({\bm r})$ form the other Dirac fermion at ${\bm k}=-{\bm K}_0$. 
Real $\alpha_x$ and $\alpha_y$ are velocities of 
the massless Dirac fermions along $x$ 
and $y$ directions respectively, and their values depend on $t_y$. 
$\epsilon^{\prime}_{\bm r}$ and $\epsilon^{\prime\prime}_{\bm r}$ 
are real and imaginary parts of $\epsilon_{\bm r}$ in the tight-binding model [see Eq.~(\ref{eq:SM_H_TB})]. Real $\beta$ represents the intravalley scattering strength, while the intervalley scattering acquires a phase factor $e^{\pm i 2{\bm K}_0 {\bm r}}$, that depends on the distance $2{\bm K}_0$ between the two valleys in the momentum space. Note that 
the intravalley scattering is always smaller than 
the intervalley scattering, $\beta<1$. $\beta$ appoarches zero when 
$t_y\rightarrow -t$. At $t_y=-t$, the two massless Dirac fermions coalesce at 
the Brillouin zone boundary. Upon a local gauge 
transformation with ${\bm K}_0 \equiv (K_0,0)$, 
\begin{align}
\psi_{1,2}({\bm r}) \rightarrow e^{-i{\bm K}_0 {\bm r}} \psi_{1,2}({\bm r}), \quad \psi_{3,4}({\bm r}) \rightarrow e^{i{\bm K}_0 {\bm r}} \psi_{3,4}({\bm r}),
\end{align}
the phase factor in the intervalley scattering term can be gauged away, 
and the kinetic energy term along $x$ acquires a vector potential,
\begin{align}
&{\cal H}
\big({\bm r},\nabla_{\bm r}\big) 
\rightarrow 
\left(\begin{array}{cccc} 
&  &  \alpha_x i\partial_x + \alpha_x K_0 + \alpha_y \partial_y & 0 \\
&  & 0 & - \alpha_x i\partial_x + \alpha_x K_0 + \alpha_y \partial_y \\
\alpha_x i\partial_x + \alpha_x K_0 - \alpha_y \partial_y & 0 &  & \\
0  & - \alpha_x i\partial_x + \alpha_x K_0 - \alpha_y \partial_y & & \\
\end{array}\right) \nonumber \\
& \hspace{2cm} + \epsilon^{\prime}_{{\bm r}}
\left(\begin{array}{cccc}
 & &  \beta  & 1 \\
 & &  1 & \beta \\
 \beta & 1 & & \\
  1 & \beta & & \\
\end{array}\right) 
+ i\epsilon^{\prime\prime}_{{\bm r}}
\left(\begin{array}{cccc}
 & &  \beta  & 1 \\
 & &  1 & \beta \\
 -\beta & -1 & & \\
  - 1 & -\beta & & \\
\end{array}\right) \nonumber \\ 
& \hspace{2cm} \equiv \alpha_x i\partial_x \sigma_1 \times \tau_3 + \alpha_x K_0 \sigma_1 \times \tau_0 
+ i \alpha_y \partial_y \sigma_2 \times \tau_0 + 
\epsilon^{\prime}_{\bm r} \sigma_1 \times \hat{a} - \epsilon^{\prime\prime}_{\bm r} 
\sigma_2 \times \hat{a} \equiv \left(\begin{array}{cc}
0 & {\sf h} \\
{\sf h}^{\dagger} & 0 \\
\end{array}\right). 
\end{align}
Here, diagonal and off-diagonal matrix elements of $\hat{a}$ represent intravalley and intervalley scattering 
strengths, respectively;
\begin{align}
\hat{a} = \left(\begin{array}{cc} 
\beta & 1 \\
1 & \beta \\
\end{array}\right) = \beta \tau_0 + \tau_1. 
\end{align}

Like the original tight-binding model, the effective continuum model 
belongs to the chiral unitary class with the chiral symmetry, 
$\sigma_3 {\cal H}\sigma_3=-{\cal H}$. Though the time reversal 
symmetry of ${\cal H}$ is broken by $\epsilon^{\prime\prime}_{\bm r}$,  
an ensemble of ${\cal H}$ is statistically symmetric under the 
time-reversal operation, 
\begin{align}
\sigma_0 \times \tau_1 \!\ \{{\cal H}^*\} \!\ \sigma_0 \times 
\tau_1 &= 
\{{\cal H}\}. \label{T}
\end{align}
An ensemble of ${\cal H}$ has also 
following statistical mirror and inversion 
symmetries;
\begin{align}
\sigma_0 \times \tau_1 \!\ \{{\cal H}(x,\partial_x)\} \!\ 
\sigma_0 \times \tau_1 &= 
\{{\cal H}(-x,-\partial_x)\}, 
\quad \big(
{\rm i.e.}, \!\ \tau_1 \!\ \{{\sf h}(x,\partial_x)\} \!\ \tau_1 = 
\{{\sf h}(-x,-\partial_x)\} \!\ \big), \label{Px1} \\
 \sigma_1 \times \tau_1 \!\ \{{\cal H}^*(y,\partial_y)\} 
 \!\ \sigma_1 
 \times \tau_1 &= 
\{{\cal H}(-y,-\partial_y)\}, \quad 
\big({\rm i.e.}, \!\ \tau_1 \!\ \{{\sf h}^{T}(y,\partial_y)\} \!\ \tau_1 = 
\{{\sf h}(-y,-\partial_y)\} \!\ \big), \label{Px2} \\
\sigma_1 \times \tau_0 \!\ 
\{{\cal H}^*({\bm r},\nabla_{\bm r})\} \!\ 
\sigma_1 \times \tau_0 
&= 
\{{\cal H}(-{\bm r},-\nabla_{\bm r})\}, 
\quad \big({\rm i.e.}, \!\ \{{\sf h}^T({\bm r},\nabla_{\bm r})\}  = 
\{{\sf h}(-{\bm r},-\nabla_{\bm r}\} \!\ \big). \label{C2z} 
\end{align}
The statistical mirror symmetries, Eqs.~(\ref{Px1}) and 
(\ref{Px2}), require $\nu_x=0$. While non-zero $\nu_y$ is  
allowed by these statistical symmetries (see also below).

\subsection{partition function and action} 
The action for zero-energy states of 
the chiral symmetric Hamiltonian ${\cal H}$ is given by 
\begin{align}
\langle \!\ \ln Z  \!\ \rangle  = 
\bigg\langle \!\ \ln \bigg[
\int D\psi^{\dagger} D\psi \!\ e^{-i \int d^2{\bm r} 
\psi^{\dagger} 
(\pm i0 + {\cal H}) \psi} \bigg] \bigg\rangle, 
\end{align}
Here 8 Grassmann 
variables form two independent vectors, $\psi \equiv (\psi_1,\psi_3,\psi_2,\psi_4)^T 
\equiv (\psi_{+},\psi_{-})^T$, $\psi^{\dagger} \equiv 
(\psi^{\dagger}_1,\psi^{\dagger}_3,
\psi^{\dagger}_2,\psi^{\dagger}_4) 
\equiv (\psi^{\dagger}_{+},\psi^{\dagger}_{-})$. 
$``\pm"$ stands for the chiral index. 
$\langle \cdots \rangle$ denotes 
quenched disorder average over $\epsilon_{\bm r} \equiv 
\epsilon^{\prime}_{\bm r}+ i\epsilon^{\prime\prime}_{\bm r}$. 
In the numerical simulations, the real and imaginary parts of $\epsilon_{\bm r}$ 
were distributed uniformly within $[-W/2,W/2]$, with the disorder strength $W$. 
For convenience, we replace this average with an average with the Gaussian distribution, 
\begin{align}
\langle \ln Z \rangle = 
\frac{1}{\cal N}\int {\cal D}\epsilon_{\bm r} \!\ \ln Z \!\ 
e^{-\frac{1}{g} \int d^2{\bm r}  \!\ |\epsilon_{\bm r}|^2.}, \label{gaussian} 
\end{align}
with a normalization factor ${\cal N}$ and disorder 
strength $g$. The disorder average can be evaluated in terms of 
the replica method~\cite{efetov80}, 
\begin{align}
\langle \ln Z \rangle = \lim_{N\rightarrow 0}\frac{1}{N} 
\big(\langle Z^N \rangle - 1 \big) 
\end{align}
with  
\begin{align}
\langle Z^N \rangle = 
\int D \psi^{\dagger}_{\alpha} D \psi_{\alpha}  
\exp \Big[ -i\int d^2{\bm r}  \psi^{\dagger}_{\alpha} 
(\pm i 0 + {\cal H}_0) \psi_{\alpha} 
- g \int d^2 {\bm r}  
\big(\psi^{\dagger}_{+,\beta} \hat{a} \psi_{-,\beta}\big)_{{\bm r}} \big(\psi^{\dagger}_{-,\alpha} \hat{a} \psi_{+,\alpha}\big)_{{\bm r}} \Big], \label{znav}
\end{align}
and 
\begin{align}
{\cal H}_0 \equiv \alpha_x i\partial_x \sigma_1 \times \tau_3 
+ \alpha_x K_0 \sigma_1 \times \tau_0 + i\alpha_y \partial_y 
\sigma_2 \times \tau_0 \equiv \left(\begin{array}{cc} 
0 & {\sf h}_0 \\
{\sf h}^{\dagger}_0 & 0 \\
\end{array}\right).    
\end{align}
Here $\psi^{\dagger}_{\alpha}$ and $\psi_{\alpha}$ ($\alpha=1,2,\cdots,N$) 
are $N$-copies of $\psi^{\dagger}$ and $\psi$ vector fields, 
respectively. The Gaussian average leads to a 
quatric term in the right-hand side of Eq.~(\ref{znav}).
Following the convention, we rename the 
Grassmann variables as, $\psi^{\dagger}_{\pm,\alpha} = \overline{\phi}_{\mp,\alpha}$,  $\psi_{\pm,\alpha} = \phi_{\pm,\alpha}$, 
and rewrite the quartic term as,
\begin{align}
- g \int d^2 {\bm r} 
\big(\overline{\phi}_{-,\beta}({\bm r}) \hat{a} \phi_{-,\beta}({\bm r}) \big) \big(\overline{\phi}_{+,\alpha}({\bm r}) \hat{a} \phi_{+,\alpha}({\bm r})\big).  
\end{align}
The quartic term is further decoupled by an auxiliary complex 
matrix field $Q$ that acts on the replica space, 
\begin{align} 
&- g \int d^2 {\bm r} 
\big(\overline{\phi}_{-,\beta}({\bm r}) \!\ \hat{a} 
\!\ \phi_{-,\beta}({\bm r})
\big) \big(\overline{\phi}_{+,\alpha}({\bm r}) \!\ 
\hat{a} \!\ \phi_{+,\alpha}({\bm r})\big)  
\nonumber \\
& \rightarrow - g \int d^2 {\bm r} 
\big(\overline{\phi}_{-,\beta} ({\bm r}) \!\ \hat{a} \!\ \phi_{-,\beta} ({\bm r}) 
\big) \big(\overline{\phi}_{+,\alpha} ({\bm r})
\!\ \hat{a}  \!\ \phi_{+,\alpha} ({\bm r}) 
\big)  
- \frac{\gamma^2}{g}  
\sum_{\alpha,\beta=1}^N \sum_{i,j={\bm K}_0,-{\bm K}_0} 
\int d^2{\bm r}  \nonumber \\
& \bigg\{ 
[Q({\bm r})]_{(\alpha,i|\beta,j)} + \frac{g}{\gamma} \Big( \phi_{+,\alpha}({\bm r}) 
\overline{\phi}_{-.\beta}({\bm r}) \!\ \hat{a}\Big)_{ij} 
\bigg\} \bigg\{ 
[Q({\bm r})]^{*}_{(\alpha,i|\beta,j)} + \frac{g}{\gamma} \Big( \phi_{-,\beta}({\bm r}) 
\overline{\phi}_{+.\alpha}({\bm r}) \!\ \hat{a}\Big)_{ji} 
\bigg\}  \nonumber \\
& = - \frac{\gamma^2}{g} 
\int d^2{\bm r} \!\ {\rm Tr}
[Q({\bm r}) Q^{\dagger}({\bm r})] + 
\gamma 
\int d^2{\bm r} \!\  
\overline{\phi}({\bm r})
\left(\begin{array}{cc} 
\hat{a}\!\ Q^{\dagger}({\bm r}) & 0 \\
0 & \hat{a} \!\ Q({\bm r}) \\
\end{array}\right) \phi({\bm r}), 
\end{align}
with $\overline{\phi} \equiv 
(\overline{\phi}_{-},\overline{\phi}_{+})$, 
$\phi({\bm r}) \equiv 
(\phi_{+},\phi_{-})^T$. 
Here, $i,j=\pm {\bm K}_0$ denote the valley index, and the trace on the right-hand side stands for sums over the replica index ($\alpha,\beta$), and valley index ($i,j$). 
Including ${\cal H}_0$, we finally have the action  
for a pair of disordered massless Dirac fermions with the 
chiral symmetry;
\begin{align}
\langle Z^N \rangle &= 
\int DQ^{\dagger} DQ \!\ D\overline{\phi} D\phi \!\  
\exp\bigg[ - \frac{\gamma^2}{g} \int d^2 {\bm r
} \!\ {\rm Tr}\Big[Q(\bm r) Q^{\dagger}({\bm r})\Big] 
- i \int d{\bm r}^2 \!\ \overline{\phi}({\bm r}) 
\left(\begin{array}{cc} 
i \gamma \!\ \hat{a} \!\ Q^{\dagger} & {\sf h}_0 \mathds{1} \\
{\sf h}^{\dagger}_0 \mathds{1} & i\gamma \!\ \hat{a} \!\ Q \\
\end{array}\right) \phi({\bm r}) 
\bigg] \nonumber \\
&= \int DQ^{\dagger} DQ \!\  
\exp\Bigg[ - \frac{\gamma^2}{g} \int d^2 {\bm r
} \!\ {\rm tr}\Big[Q(\bm r) Q^{\dagger}({\bm r})\Big] 
+ {\rm Tr} \ln 
\left[\begin{array}{cc} 
i \gamma \!\ \hat{a} \!\ Q^{\dagger} & {\sf h}_0 \mathds{1} \\
{\sf h}^{\dagger}_0 \mathds{1} & i\gamma \!\ \hat{a} \!\ Q \\
\end{array}\right]  
\Bigg]. \label{FE}
\end{align}
The trace in the first term is a sum over the valley index ($i,j$), while  
the trace in the second term includes not only sum over 
the valley, replica ($\alpha,\beta$) indices, but also includes an integral over the 
space coordinates. 

\subsection{Self-consistent Born solution}

In Eq.~(\ref{FE}), $i\gamma\hat{a}Q$ and $i\gamma \hat{a} Q^{\dagger}$ play role of a self-energy of a single-particle Green's 
function for the disordered massless Dirac fermions. 
A saddle-point solution for the self-energy is  
obtained by a variation of the action   
with respect to $Q$ and $Q^{\dagger}$. We assume 
that the saddle-point solution is independent from ${\bm r}$, and it is 
diagonal in the replica indices;
\begin{align}
(\hat{a} \!\ Q)_{(\alpha,i|\beta,j)}
({\bm r}) =  
\mathds{1}_{\alpha,\beta} \!\ \hat{q}_{i,j}, \!\ \!\ 
(\hat{a} \!\ Q^{\dagger})_{(\alpha,i|\beta,j)}
({\bm r}) = 
\mathds{1}_{\alpha,\beta} \!\ \hat{q}^{\prime}_{i,j},
\end{align}
wth $i,j={\bm K}_0,-{\bm K}_0$. $\hat{q}$ and
$\hat{q}^{\prime}$ stand for 2 by 2 
matrix structures of the self-energy in the valley space. 
As $\hat{a}$ has both intervalley 
and intravalley components in general, 
$\hat{a} =\beta \tau_0 + \tau_1$, $\hat{q}$ and $\hat{q}^{\prime}$ 
are not necessarily the unit matrix;
\begin{align}
\hat{q} = \sum_{\mu=0,1,2,3} q_{\mu} \tau_\mu, \quad 
\hat{q}^{\prime} = \sum_{\mu=0,1,2,3} q^{\prime}_{\mu} \tau_{\mu}. 
\label{s200}
\end{align}
Complex ${q}_{\mu}$ and $q^{\prime}_{\mu}$
($\mu=0,1,2,3$) are related to each other by $q^*_{0}=q^{\prime}_0$, 
$q^{*}_1 = q^{\prime}_1$, and $(\beta \tau_0 + \tau_1)
(q^*_2 \tau_2+q^*_3 \tau_3)=(q^{\prime}_2 \tau_2 + q^{\prime}_3 
\tau_3) (\beta \tau_0 + \tau_1)$. The saddle-point solution of 
$\hat{q}$ and $\hat{q}^{\prime}$ are given by   
\begin{align}
i\gamma \hat{q} = - g \int \frac{d^2{\bm k}}{(2\pi)^2} \!\ 
\hat{a} \left[\begin{array}{cc} 
i\gamma \hat{q}^{\prime} & \alpha_x k_x \tau_3 + \alpha_x K_0 \tau_0 - i\alpha_y k_y \tau_0 \\
\alpha_x k_x \tau_3 + \alpha_x K_0 \tau_0 + i\alpha_y k_y \tau_0 & i\gamma \hat{q} \\
\end{array}\right]^{-1}_{(-|+)}  \hat{a},  \label{scb-1}
\end{align}
and 
\begin{align}
i\gamma \hat{q}^{\prime} = - g \int \frac{d^2{\bm k}}{(2\pi)^2} \!\ 
\hat{a} \left[\begin{array}{cc} 
i\gamma \hat{q}^{\prime} & \alpha_x k_x \tau_3 + \alpha_x K_0 \tau_0 - i\alpha_y k_y \tau_0 \\
\alpha_x k_x \tau_3 + \alpha_x K_0 \tau_0 + i\alpha_y k_y \tau_0 & i\gamma \hat{q} \\
\end{array}\right]^{-1}_{(+|-)}  \hat{a}. \label{scb-2}
\end{align}
Here $[\cdots]_{(\mp|\pm)}$ in the right hand sides 
mean 2 by 2 blocks of a 4 by 4 matrix (``$\cdots$") that are 
between $\overline{\phi}_{\mp}$ and $\phi_{\pm}$. 
When $q_2=q_3=0$, the integrands of the right-hand sides 
are symmetric under $k_x \rightarrow -k_x$ and a $\pi$ rotation 
around $\tau_1$, $(\tau_1,\tau_2,\tau_3)
\rightarrow (\tau_1,-\tau_2,-\tau_3)$. Thus, $\tau_2$ and $\tau_3$ components 
of the integrands are odd functions in $k_x$, justifying the following solution with $q_2=q_3=0$ self-consistently,   
\begin{align}
\hat{q}=q_0 \tau_0 + q_1 \tau_1, \quad \hat{q}^{\prime} = \hat{q}^{\dagger}
\label{scb}
\end{align}
Complex $q_0$ and $q_1$ are determined by coupled equations;
\begin{align}
\gamma q_0 = g \int \frac{d^2{\bm k}}{(2\pi)^2} \frac{(\beta^2+1)t_0({\bm k}) - 2\beta t_1({\bm k})}{t^2_0({\bm k}) - {\bm t}^2({\bm k})}, \quad 
\gamma q_1 = g \int \frac{d^2{\bm k}}{(2\pi)^2} \frac{ 
2\beta t_0({\bm k}) -  (\beta^2+1) t_1({\bm k})}{t^2_0({\bm k}) 
- {\bm t}^2({\bm k})}, \label{coupled}
\end{align}
where ${\bm t}^2({\bm k}) \equiv t^2_1({\bm k})+t^2_2({\bm k})+t^2_3({\bm k})$, and  
\begin{align}
t_0({\bm k}) &\equiv \gamma q^*_0 + \frac{1}{\gamma} \frac{q_0}{q^2_0-q^2_1} 
(\alpha^2_x k^2_x + \alpha^2_y k^2_y + \alpha^2_x K^2_0), \nonumber \\ 
t_{1}({\bm k}) &\equiv \gamma q^*_1 + \frac{1}{\gamma} 
\frac{q_1}{q^2_0-q^2_1} (-\alpha^2_x k^2_x + \alpha^2_y k^2_y + \alpha^2_x K^2_0), \nonumber \\
t_{2}({\bm k}) & \equiv \frac{1}{\gamma} \frac{q_1}{q^2_0-q^2_1} 2\alpha_x \alpha_y 
k_x k_y, \quad t_3({\bm k}) \equiv \frac{1}{\gamma} \frac{q_0}{q^2_0-q^2_1} 2\alpha^2_x K_0 k_x. 
\end{align}
Eq.~(\ref{coupled}) can be solved numerically with a proper ultraviolet cutoff for the momentum integral. $\hat{q}$ thus determined forms the self-energy part, and the self-consistent Born (SCB) single-particle Green function is given by,  
\begin{align}
\big[G({\bm k})\big]^{-1} = \left[\begin{array}{cc} 
i\gamma \mathds{1} \times \hat{q}^{\dagger} & \alpha_x k_x \tau_3 
+ \alpha_x K_0 \tau_0 - i\alpha_y k_y \tau_0 \\
\alpha_x k_x \tau_3 + \alpha_x K_0 \tau_0 + i\alpha_y k_y \tau_0 & i\gamma \mathds{1} 
\times \hat{q} \\ 
\end{array}\right] 
\equiv [G_0({\bm k})]^{-1}  
+ \left[\begin{array}{cc} 
i\gamma \mathds{1} \times \hat{q}^{\dagger} & 0 \\
0 & i\gamma \mathds{1} 
\times \hat{q} \\ 
\end{array}\right]. 
\label{scb-solution}  
\end{align}
Here $[G_0({\bm k})]$ stands for the Fourier transform of the 
single-particle Green's function in the clean limit in the effective continuum model. In the clean limit, the 1D weak 
topological index $\nu_{\mu}$ ($\mu=x,y$) is given by the single-particle Green's function 
[Eq.~(\ref{1d-G0})]. In the presence of disorder, it is natural to replace $G_0({\bm k})$ 
in Eq.~(\ref{1d-G0}) by the SCB Green's function, and define a generalized weak 1D topology as,  
\begin{align}
\nu_{\mu} \rightarrow \tilde{\nu}_{\mu} \equiv 
i \int_{[-\pi,\pi]\times [-\pi,\pi]} 
\frac{d^2{\bm k}}{(2\pi)^2} 
{\rm Tr}\bigg[ \big[G({\bm k})\big] 
\Big(\partial_{k_\mu} \big[G({\bm k})\big]^{-1}\Big)
\left[\begin{array}{cc} 
0 & 0 \\
0 & \mathds{1} \\
\end{array}\right]\bigg]. \label{nu-scb}
\end{align}

The SCB Green's function in Eq.~(\ref{scb-solution}) has the following symmetries,
\begin{align}
&\sigma_0 \times \tau_1 \!\ [G({k_x,k_y})] \!\ \sigma_0 \times \tau_1 = [G(-k_x,k_y)], \label{sym1} \\ 
&\sigma_3 \times \tau_0 \!\ [G(-k_x,-k_y;-K_0)] \!\ 
\sigma_3 \times \tau_0 = [G(k_x,k_y;K_0)], \label{sym2} \\
&\sigma_2 \times \tau_0 \!\ 
[G^T(k_x,-k_y)] \!\ \sigma_2 \times \tau_0 = - [G^*(k_x, k_y)].  \label{sym3}
\end{align}
Eq.~(\ref{sym1}) requires $\tilde{\nu}_x=0$, and   
Eq.~(\ref{sym2}) requires that $\tilde{\nu}_y$ 
is an odd function in $K_0$; $\tilde{\nu}_y \ne 0$ 
for $K_0 \ne 0$. In fact, a straightforward calculation shows that $\tilde{\nu}_y$ is 
linear in $K_0$.
Though the SCB Green's function $G({\bm k})$ is not a Hermitian matrix, 
Eq.~(\ref{sym3}) requires that $\tilde{\nu}_y$ is real. In the following, 
we derive a nonlinear sigma model based on the SCB Green's function. 

\subsection{Nonlinear sigma model}

Eq.~(\ref{FE}) implies that the SCB solution has degeneracy under those unitary rotations in the replica space that 
leave ${\sf h}_0 \mathds{1}$ and ${\sf h}^{\dagger}_0 \mathds{1}$ 
intact; 
\begin{align}
\left(\begin{array}{cc}
\overline{\phi}_{-} & \overline{\phi}_{+} \\
\end{array}\right) \rightarrow 
\left(\begin{array}{cc}
\overline{\phi}_{-} U^{\dagger} & \overline{\phi}_{+} V^{\dagger} \\
\end{array}\right), \quad 
\left(\begin{array}{c} 
\phi_{+} \\
\phi_{-} \\
\end{array}\right) \rightarrow 
\left(\begin{array}{c} 
V \phi_{+} \\
U \phi_{-} \\
\end{array}\right). 
\end{align}
The rotations generate a gauge group, rotating $\mathds{1} \times \hat{q}$ and $\mathds{1}\times \hat{q}^{\dagger}$  
by the two unitary matrices,
\begin{align}
\hat{a} \!\ Q^{\dagger} = \mathds{1} \times \hat{q}^{\dagger} \rightarrow 
\hat{U}^{\dagger} \hat{V} \times \hat{q}^{\dagger} 
\equiv \hat{T}^{\dagger} \times \hat{q}^{\dagger}, 
\quad \hat{a} \!\ Q = 
\mathds{1} \times \hat{q} \rightarrow 
\hat{V}^{\dagger} \hat{U} \times \hat{q} 
\equiv \hat{T} \times \hat{q},
\end{align}
An $N$ times $N$ unitary matrix $\hat{T}$ takes values 
from a symmetric subgroup 
${\rm U}(N) \times{\rm  U}(N)/{\rm U}(N)$ of the 
global gauge group. 

Consider the action of Eq.~(\ref{FE}) with the local unitary matrix $T({\bm r})$ that is slowly-varying in the spatial coordinate;
\begin{align}
- S [T^{\dagger}({\bm r}),T({\bm r})]\equiv 
-\frac{\gamma^2}{g} N V 
{\rm tr}\Big[ \hat{a}^{-1} \hat{q} \!\  \hat{a}^{-1} \hat{q}^{\dagger} \Big]  
+ {\rm Tr} \ln \left[\begin{array}{cc}
i \gamma T^{\dagger}({\bm r}) \times q^{\dagger}  & {\sf h}_0 \times \mathds{1} \\
{\sf h}^{\dagger}_0 \times \mathds{1} & i \gamma T({\bm r}) \times q \\
\end{array}\right], \label{free-energy-2}  
\end{align}
The trace in the first term is a sum over 
the valley index ($i,j$), and the trace in the second term includes the sum over the valley, replica  $(\alpha,\beta)$ indices, and the integral over the space coordinate. $V$ is the volume of the system, $N$ is the number of the replica fields; $\int d^2{\bm r} \equiv V$, and $\sum_{\alpha} \equiv N$.  
 In the second term, $T^{\dagger}({\bm r})$ and $T({\bm r})$ in 
the diagonal blocks can be gauged away by a local transformation, $\overline{\phi}_{-} \rightarrow \overline{\phi}^{\prime}_{-} = 
\overline{\phi}_{-} T({\bm r})$  
and $\phi_{-} \rightarrow \phi^{\prime}_{-} = T^{\dagger}({\bm r}) \phi_{-}$, 
while the transformation induces a spatial gradient term in the off-diagonal block,
\begin{align}
-S [T^{\dagger},T] = \cdots   
+ {\rm Tr} \ln \left[\begin{array}{cc}
i \gamma \mathds{1} \times q^{\dagger}  & 
{\sf h}_0 \times \mathds{1} 
+ T [{\sf h}_0 \times \mathds{1}, T^{\dagger}] \\
{\sf h}^{\dagger}_0 \times \mathds{1} & i \gamma \mathds{1} \times q \\
\end{array}\right]. \nonumber
\end{align}
The gradient term can be further expanded,  
\begin{align}
-S = -S_0 + {\rm Tr} [G \Delta] - \frac{1}{2} 
{\rm Tr} [G \Delta G \Delta] + \cdots. \label{expansion}
\end{align}
Here $S_0$ is the saddle-point value of the action $S$. 
$G$ is the SCB single-particle Green's function;
\begin{align}
- S_0 = -\frac{\gamma^2}{g} N V 
{\rm tr}\Big[ \hat{a}^{-1} \hat{q} \!\  \hat{a}^{-1} \hat{q}^{\dagger} \Big] 
+ {\rm Tr} \ln G^{-1}, \quad 
G^{-1} = \left[\begin{array}{cc}
i \gamma \mathds{1} \times \hat{q}^{\dagger}  & 
{\sf h}_0 \times \mathds{1}  \\
{\sf h}^{\dagger}_0 \times \mathds{1} & i \gamma \mathds{1} \times \hat{q} \\
\end{array}\right]. \nonumber 
\end{align}
$\Delta$ is the spatial gradient term in the off-diagonal block, $T \!\ [{\sf h}_0 \times \mathds{1}, T^{\dagger}]$, 
\begin{align}\label{eq:off-diagonal-block}
\Delta =  \left[\begin{array}{cc}
0  &  T \!\ [{\sf h}_0 \times \mathds{1}, T^{\dagger}] \\
0 & 0 \\
\end{array}\right] = \left[\begin{array}{cc}
0  &  i \alpha_x T\partial_x T^{\dagger} \times \tau_3 
+ \alpha_y  T \partial_y T^{\dagger} \times \tau_0 \\
0 & 0 \\
\end{array}\right]. 
\end{align}
The first-order gradient expansion 
in Eq.~(\ref{expansion})
yields the weak 1D topological term, 
and the second-order gradient term in Eq.~(\ref{expansion}) gives out the conductivity term in the non-linear sigma model; 
\begin{align}
&{\rm Tr}\big[G \Delta \big] = 
\int d^2{\bm r} \!\  \Big\{ \tilde{\nu}_x
{\rm Tr}\!\ \big[T\partial_x T^{\dagger}\big]  
+ \tilde{\nu}_y
\!\ {\rm Tr} \!\ 
\big[T \partial_y T^{\dagger}\big]\Big\}, \nonumber \\
%
&{\rm Tr}\big[G \Delta G \Delta] = \frac{1}{V} 
\sum_{\bm k} {\rm tr}\bigg[\big[G({\bm k})\big] 
\left[\begin{array}{cc}
0  &   \tau_3  \\
0 & 0 \\
\end{array}\right]\big[G({\bm k})\big] 
\left[\begin{array}{cc}
0  &   \tau_3  \\
0 & 0 \\
\end{array}\right] \bigg] \times (-\alpha^2_x) \int d^2{\bm r} 
\!\ {\rm Tr}\!\ \big[T(\partial_x T^{\dagger}) 
T(\partial_x T^{\dagger}) \big]  \nonumber \\
%
&\hspace{2cm} + \frac{1}{V} 
\sum_{\bm k} {\rm tr}\bigg[\big[G({\bm k})\big] 
\left[\begin{array}{cc} 
0  &   \tau_0  \\
0 & 0 \\
\end{array}\right]\big[G({\bm k})\big] 
\left[\begin{array}{cc}
0  &   \tau_0  \\
0 & 0 \\
\end{array}\right] \bigg] \times \alpha^2_y \int d^2{\bm r} 
\!\ {\rm Tr} \!\ \big[T(\partial_y T^{\dagger}) 
T(\partial_y T^{\dagger}) \big] \nonumber \\
%
&\hspace{2cm} + \frac{1}{V} 
\sum_{\bm k} {\rm tr}\bigg[\big[G({\bm k})\big] 
\left[\begin{array}{cc} 
0  &   \tau_0  \\
0 & 0 \\
\end{array}\right]\big[G({\bm k})\big] 
\left[\begin{array}{cc}
0  &   \tau_3  \\
0 & 0 \\
\end{array}\right] \bigg] \times 2i\alpha_x \alpha_y \int d^2{\bm r} \!\ 
{\rm Tr}\!\ \big[T(\partial_y T^{\dagger}) 
T(\partial_x T^{\dagger}) \big] + \cdots, \label{expansion2}
\end{align}
with the self-consistent Born Green's function $G({\bm k})$ 
in Eq.~(\ref{scb-solution}). 
$``{\rm Tr}"$ in the right hand side is 
a trace over $N\times N$ replica space, while $``{\rm tr}"$ stands for 
the trace over chiral and valley indices. Higher-order spatial gradient 
terms are omitted as $``\cdots"$ in Eq.~(\ref{expansion2}).

Due to the symmetries of the self-consistent Born Green's function, Eq.~(\ref{sym1}), (\ref{sym2}) and (\ref{sym3}), $\tilde{\nu}_x=0$, $\tilde{\nu}_y$ 
is real, and $\tilde{\nu_y}$ is an odd function in $K_0$. Eq.~(\ref{sym1}) also requires that the coefficient of 
${\rm Tr}[T(\partial_x T^{\dagger})T(\partial_y T^{\dagger})]$ is zero, 
and Eq.~(\ref{sym3}) guarantees the real numbers of coefficients 
of ${\rm Tr}[T(\partial_x T^{\dagger})T(\partial_x T^{\dagger})]$
and ${\rm Tr}[T(\partial_y T^{\dagger})T(\partial_y T^{\dagger})]$; 
\begin{align}
&\sum_{\bm k} {\rm tr} \bigg[\big[G({\bm k})\big]^* 
\left[\begin{array}{cc} 
0  &   \tau_i  \\
0 & 0 \\
\end{array}\right] \big[G({\bm k})\big]^* 
\left[\begin{array}{cc} 
0  &   \tau_i  \\
0 & 0 \\
\end{array}\right] 
\bigg] = \sum_{\bm k} {\rm tr} \bigg[\big[G(k_x,-k_y)\big]^T 
\left[\begin{array}{cc} 
0  &  0  \\
\tau_i & 0 \\
\end{array}\right] \big[G(k_x,-k_y)\big]^T 
\left[\begin{array}{cc} 
0  &  0  \\
\tau_i & 0 \\
\end{array}\right]\bigg]  \nonumber \\
&\hspace{7cm} = \sum_{\bm k} {\rm tr} \bigg[\big[G(k_x,k_y)\big]
\left[\begin{array}{cc} 
0  &   \tau_i  \\
0 & 0 \\
\end{array}\right] \big[G(k_x,k_y)\big] 
\left[\begin{array}{cc} 
0  &   \tau_i  \\
0 & 0 \\
\end{array}\right] 
\bigg]. 
\end{align}
We assume that the coefficients of $-{\rm Tr}[T(\partial_{\mu} T^{\dagger}) T(\partial_{\mu} T^{\dagger})]$ are all positive for $\mu=x,y$; 
\begin{align}
\tilde{\sigma}_{x} &\equiv \frac{\alpha^2_x}{2} \int \frac{{\rm d}^2{\bm k}}{(2\pi)^2}
\bigg[\big[G({\bm k})\big] 
\left[\begin{array}{cc} 
0  &   \tau_3  \\
0 & 0 \\
\end{array}\right]\big[G({\bm k})\big] 
\left[\begin{array}{cc}
0  &   \tau_3  \\
0 & 0 \\
\end{array}\right] \bigg] > 0, \label{bare-sigma-x} \\
%
%
\tilde{\sigma}_{y} &\equiv - \frac{\alpha^2_y}{2} \int \frac{{\rm d}^2{\bm k}}{(2\pi)^2}
\bigg[\big[G({\bm k})\big] 
\left[\begin{array}{cc} 
0  &   \tau_0  \\
0 & 0 \\
\end{array}\right]\big[G({\bm k})\big] 
\left[\begin{array}{cc}
0  &   \tau_0  \\
0 & 0 \\
\end{array}\right] \bigg] > 0. \label{bare-sigma-y} 
\end{align}
Otherwise, the spatially uniform SCB solution is not locally stable in the action. This leads to the nonlinear sigma model with the weak 1D topological term, 
\begin{align}
S[T^{\dagger},T] = S_0 - \int d^2{\bm r}  
\bigg\{ \tilde{\nu}_y {\rm Tr}\big[T\partial_y T^{\dagger}\big] 
+ \tilde{\sigma}_x {\rm Tr} 
\big[T\partial_x T^{\dagger}T\partial_x T^{\dagger}\big] 
+ \tilde{\sigma}_y {\rm Tr} \big[T\partial_y T^{\dagger}T\partial_y  T^{\dagger}\big] \bigg\}.
\label{nlsm-NoGade}
\end{align}
With $Q=T^{\dagger}$, $\tilde{\sigma}_{\mu} = 
\sigma_{\mu}/(8\pi)$ and $\tilde{\nu}_y= -\chi_y/(8\pi)$, 
we obtain Eq.~(1) with $c_x=c_y=0$ in the main text. 
One could also add the Gade term as in Eq.~(1) in the main text, by including a coupling between the first-order spatial gradient term $T\partial_{\mu}T^{\dagger}$, and a massive mode that is odd under $r_{\mu}\rightarrow - r_{\mu}$. In this section, we consider only the massless mode generated by the U$(N)$ rotations in the replica space, while the saddle-point solution generally has massive modes that take the form of the unit matrix in the replica space, e.g. fluctuations of $\hat{q}$ and $\hat{q}^{\dagger}$ in the self-energy part. Some of the massive modes are odd under the mirror operation or the spatial inversion. For example, in Eq.~(\ref{s200}), the 
self-energy part $\hat{q}$ and $\hat{q}^{\dagger}$ is expanded in the Pauli matrices, 
$\tau_{\mu}$ ($\mu=0,1,2,3$), where $q_2$ and $q_3$ components are odd under the 
mirror operation $(x,y) \rightarrow (-x,y)$. The self-consistent Born solution has chosen them to be zero, $q_2=q_3=0$, while one could consider $q_2$ and $q_3$ as massive 
fluctuations around the saddle-point solution. From a symmetry point of view, the 
action can have couplings between ${\rm Tr}[T\partial_{x}T^{\dagger}]$ and such massive modes [linear combinations of $q_2$ and $q_3$]. 
Thereby, the Gaussian integrations over the massive modes yield 
the Gade term, $({\rm Tr}[T\partial_{x}T^{\dagger}])^2$. Besides, the Gade term will be generated under the renormalization group even if starting from a model without the Gade term. This can be seen from the renormalization group section above.

\subsection{Anatomy of current operators and their correlations}
The above derivation suggests that the NLSM depends on $T\partial_{\mu} T^{\dagger}$ and a certain vector potential in a covariant way, and therefore correlations of a corresponding current in the lattice model could manifest the spatial polarization of vortex-antivortex pairs in the (critical) metal phase. To define this current operator, note first that a physical current operator $J_{\mu}$ ($\mu=x,y$) [see its definition below] as well as the Hamiltonian $H$ respects the chiral symmetry, $\sigma_3 J_{\mu} \sigma_3 = -J_{\mu}$. Thus, the current operator in the square-lattice model can be decomposed into the two non-Hermitian current operators, $J_{L,\mu}$ and $J_{R,\mu}$; $J_{L,\mu}$ is from the $\sigma_3=+1$ sublattice with its coordinate $(x,y)$ 
satisfying $x+y=$ even to the $\sigma_3=-1$ sublattice with its coordinate ($x,y$) 
satisfying $x+y=$ odd, and $J_{R,\mu}$ is from the $\sigma_3=-1$ sublattice to the $\sigma_3=+1$ sublattice, 
\begin{align}
J_{L,x} \equiv i \sum^{x+y={\rm even}}_{x,y} 
\Big(t^{(x)}_{x,y} c^{\dagger}_{x+1,y}c_{x,y} - 
t^{(x)\!\ *}_{x-1,y} c^{\dagger}_{x-1,y} c_{x,y}\Big), \quad 
J_{R,x} \equiv i \sum^{x+y={\rm odd}}_{x,y} 
\Big(t^{(x)}_{x,y} c^{\dagger}_{x+1,y}c_{x,y} - 
t^{(x)\!\ *}_{x-1,y} c^{\dagger}_{x-1,y} c_{x,y}\Big), \nonumber \\
J_{L,y} \equiv i \sum^{x+y={\rm even}}_{x,y} 
\Big(t^{(y)}_{x,y} c^{\dagger}_{x,y+1}c_{x,y} - 
t^{(y)\!\ *}_{x,y-1} c^{\dagger}_{x,y-1} c_{x,y}\Big), \quad 
J_{R,y} \equiv i \sum^{x+y={\rm odd}}_{x,y} 
\Big(t^{(y)}_{x,y} c^{\dagger}_{x,y+1}c_{x,y} - 
t^{(y)\!\ *}_{x,y-1} c^{\dagger}_{x,y-1} c_{x,y}\Big). \label{JLR} 
\end{align}
Here the physical current operator $J_{\mu}$ is given by a sum of these two, $J_{\mu}\equiv J_{L,\mu}+J_{R,\mu}$, and $\vec{J}_{L}$ and $\vec{J}_R$ are Hermitian conjugate to each other, $J^{\dagger}_{L,\mu}=J_{R,\mu}$. Vector potentials $A_{L,\mu}$ and  $A_{R,\mu}$ that are respectively coupled to $J_{L,\mu}$ and $J_{R,\mu}$ enter the hopping terms of the lattice model as the Aharonov-Bohm phase,
\begin{align}
H(\vec{A}_{L},\vec{A}_{R}) 
=& \sum^{x+y={\rm even}}_{x,y} \Big( 
t^{(x)}_{x,y} \!\ e^{iA_{L,x}} \!\ c^{\dagger}_{x+1,y} c_{x,y} + 
t^{(y)}_{x,y} \!\ e^{iA_{L,y}} \!\ c^{\dagger}_{x,y+1} c_{x,y} \Big)  \nonumber \\
 & + \sum^{x+y={\rm even}}_{x,y}
\Big(t^{(x)\!\ *}_{x-1,y} \!\ e^{-iA_{L,x}} \!\
c^{\dagger}_{x-1,y}c_{x,y} 
+ t^{(y)\!\ *}_{x,y-1} \!\ e^{-iA_{L,y}} \!\ c^{\dagger}_{x,y-1}c_{x,y} \Big) \nonumber \\
&  + \sum^{x+y={\rm odd}}_{x,y} \Big( 
t^{(x)}_{x,y} \!\ e^{iA_{R,x}} \!\ c^{\dagger}_{x+1,y} c_{x,y} + 
t^{(y)}_{x,y} \!\ e^{iA_{R,y}} \!\ c^{\dagger}_{x,y+1} c_{x,y} \Big)  \nonumber \\
 &  + \sum^{x+y={\rm odd}}_{x,y}
\Big(t^{(x)\!\ *}_{x-1,y} \!\ e^{-iA_{R,x}} \!\
c^{\dagger}_{x-1,y}c_{x,y} 
+ t^{(y)\!\ *}_{x,y-1} \!\ e^{-iA_{R,y}} \!\ c^{\dagger}_{x,y-1}c_{x,y} \Big). \label{HwithA}
\end{align}
The lattice constant of the square lattice is set to the unit. The derivative of $H[\vec{A}_{L},\vec{A}_R]$ with respect to the vector potentials gives the  current operators at $\vec{A}_L=\vec{A}_R=0$, 
\begin{align}
\frac{\partial H[\vec{A}_{L},\vec{A}_{R}]}{\partial A_{L,\mu}}=J_{L,\mu}, \quad  
\frac{\partial H[\vec{A}_L,\vec{A}_R]}{\partial A_{R,\mu}} = J_{R,\mu}, \quad \frac{\partial H[\vec{A},\vec{A}]}{\partial A_{\mu}} = J_{\mu} = J_{L,\mu} + J_{R,\mu}.  
\end{align}
In terms of the Peierls substitution [$k_{\mu} \rightarrow k_{\mu}+A_{L,\mu}$ in the upper right off-diagonal block, and $k_{\mu}\rightarrow k_{\mu}+A_{R,\mu}$ in the 
lower left off-diagonal block], these two vector potentials also enter 
the effective continuum model in Eq.~(\ref{effective-2}), 
\begin{align}
&{\cal H}({\bm r},\nabla_{\bm r}) 
\rightarrow {\cal H}({\bm r},\nabla_{\bm r};A_{L,\mu},A_{R,\mu}) 
= {\cal H}({\bm r},\nabla_{\bm r};\vec{A}_{L}=0,\vec{A}_R=0) \nonumber \\
&\hspace{2cm} 
+ \alpha_x A_{L,x} \sigma_{+} \times \tau_3 
+ \alpha_x A_{R,x} \sigma_{-} \times \tau_3 
+ \alpha_y A_{L,y} (-i) \sigma_{+} \times \tau_0 
+ \alpha_y A_{L,y} i \sigma_{-} \times \tau_0, 
\end{align}
as well as the action in Eq.~(\ref{free-energy-2}),
\begin{align}
-S[T^{\dagger},T] 
\rightarrow - S[T^{\dagger},T;\vec{A}_L,\vec{A}_R] 
= \cdots + {\rm Tr}\ln 
\left[\begin{array}{cc}
i\gamma T^{\dagger}({\bm r}) \times q^{\dagger} 
& {\sf h}_0 \times \mathds{1} + \alpha_x 
A_{L,x} \tau_3 - i\alpha_y A_{L,y} \tau_0 \\
{\sf h}_0 \times \mathds{1} + \alpha_x 
A_{R,x} \tau_3 + i\alpha_y A_{R,y} \tau_0 
& i\gamma T({\bm r}) \times q \\
\end{array}\right]. \label{stta}
\end{align}
Comparing Eq.~(\ref{eq:off-diagonal-block}) with Eq.~(\ref{stta}), notice  
that the gradient term $T\partial_{\mu} T^{\dagger}$ 
and the vector potentials appear in the action in a covariant way, e.g.  
\begin{align}
T\partial_{\mu}T^{\dagger} \rightarrow T\partial_{\mu} T^{\dagger} - 
iA_{L,\mu} \mathds{1}. \label{covariance}
\end{align}
In the following, using this covariance, we will argue that the correlation of the non-Hermitian current operator $J_{L,\mu}$ could be a direct measure of the polarizability  
of the vortex-antivortex pair in the lattice model.

To this end, let us set $\vec{A}_{R}=0$ for simplicity, and expand the action  $S[T,T^{\dagger};\vec{A}_L]$ in 
terms of the gradient terms $T\partial_{\mu} T^{\dagger}$ and the vector potential $\vec{A}_L$. 
Due to the covariant relation, one can readily see that the vector potential enters the NLSM in Eq.~(\ref{nlsm-NoGade}) as follows
\begin{align}
S[T,T^{\dagger};\vec{A}_L] = \int {\rm d}^2 {\bm r} 
\bigg\{ -\tilde{\nu}_y {\rm Tr}\big[T\partial_y T^{\dagger}\big] 
- \tilde{\sigma}_{\mu} {\rm Tr}\big[(T\partial_{\mu} T^{\dagger})^2\big] 
+ i \tilde{\nu}_y N A_{L,y} + 2i \tilde{\sigma}_{\mu} 
 A_{L,\mu} {\rm Tr}\big[T\partial_{\mu} T^{\dagger}\big] 
 + A^2_{L,\mu} \tilde{\sigma}_{\mu} N \bigg\}. 
 \end{align}
Since the covariant relation also applies to 
the derivation of the Gade terms discussed above, one could generally write down 
the NLSM in Eq.~(1) in the main text together with its coupling with $\vec{A}_{L}$ as follows, 
\begin{align}
S[Q,Q^{\dagger};\vec{A}_L] 
= \int \frac{{\rm d}^2{\bm r}}{8\pi} \Bigg\{
\sum_{\mu=x,y} \bigg[& - \sigma_{\mu} {\rm Tr}\big(Q^{-1}\partial_{\mu} Q\big)^2 
- c_{\mu} {\rm Tr}^2 \big(Q^{-1}\partial_{\mu} Q\big)
+ 2i (\sigma_{\mu} + N c_{\mu}) 
A_{L,\mu} {\rm Tr} \big(Q^{-1}\partial_{\mu} Q\big)
\nonumber \\ 
& + A^2_{L,\mu} (\sigma_{\mu} N + c_{\mu} N^2) 
\bigg] + \chi_{y} {\rm Tr}\big(Q^{-1}\partial_{y} Q\big) 
- i\chi_{y} N A_{L,y} \Bigg\}.  
\end{align}
Here we changed $T^{\dagger}$ by $Q$. Then, using 
$\langle Z^N[\vec{A}_L]\rangle = \int {\rm D}[Q] 
e^{-S[Q,Q^{\dagger};\vec{A}_L]}$, we can define   
the disorder-average of the the non-Hermitian current $\vec{J}_{L}$ and its correlation as follows, 
\begin{align}
\langle J_{L,\mu} \rangle_{\vec{A}_L=0} &\equiv 
\frac{\partial  \langle \ln Z[\vec{A}_L] \rangle}{\partial A_{L,\mu}} 
= \frac{\partial}{\partial A_{L,\mu}} 
\lim_{N\rightarrow 0} 
\frac{1}{N} \bigg( \langle Z^N[\vec{A}_L] \rangle -1\bigg) = -\lim_{N\rightarrow 0} 
\frac{1}{N} \int {\rm D}[Q]  \!\ 
\frac{\partial S[Q,Q^{\dagger};\vec{A}_L]}{\partial A_{L,\mu}} \!\ e^{-S[Q,Q^{\dagger};\vec{A}_L=0]} 
\nonumber \\
& = -\lim_{N\rightarrow 0}
\frac{1}{N} \int {\rm D}[Q] 
\int \frac{{\rm d}^2 {\bm r}}{8\pi} \bigg( -i\chi_y N \delta_{\mu,y} + 2i (\sigma_{\mu}+Nc_{\mu}) 
{\rm Tr}\big(Q^{-1}\partial_{\mu}Q\big) \bigg) \!\ e^{-S[Q,Q^{\dagger};\vec{A}_L=0]},\label{JLexp-1} \\
\langle (\delta J_{L,\mu})^2 \rangle_{\vec{A}_L=0} 
&\equiv \frac{\partial^2 \langle \ln Z[\vec{A}_L] \rangle }{\partial A^2_{L,\mu}} 
= \frac{\partial^2}{\partial A^2_{L,\mu}} 
\lim_{N\rightarrow 0} 
\frac{1}{N} \bigg( \langle Z^N[\vec{A}_L] \rangle -1\bigg)\nonumber \\ 
&= \lim_{N\rightarrow 0} 
\frac{1}{N} \int {\rm D}[Q]  \!\ \bigg( - 
\frac{\partial^2 S[Q,Q^{\dagger};\vec{A}_L]}{\partial A^2_{L,\mu}} + \Big(\frac{\partial S[Q,Q^{\dagger};\vec{A}_L]}{\partial A_{L,\mu}}\Big)^2 
\bigg) \!\ e^{-S[Q,Q^{\dagger};\vec{A}_L=0]} 
\nonumber \\
& = \lim_{N\rightarrow 0}
\frac{1}{N} \int {\rm D}[Q] 
\Bigg[ - \frac{V}{4\pi}(\sigma_{\mu} N + c_{\mu} N^2) 
\nonumber \\
&\hspace{1.5cm} + 
\Bigg\{ \int \frac{{\rm d}^2 {\bm r}}{8\pi} \bigg(- i\chi_y N \delta_{\mu,y} + 2i (\sigma_{\mu}+Nc_{\mu}) 
{\rm Tr}\big(Q^{-1}\partial_{\mu}Q\big) \bigg)\Bigg\}^2 
\Bigg]\!\ e^{-S[Q,Q^{\dagger};\vec{A}_L=0]}.\label{JLexp-2}
\end{align}
$V$ stands for the total volume $V$ of the system. $\langle J_{L,\mu}\rangle$ 
and $\langle (\delta J_{L,\mu})^2\rangle$ thus defined correspond to 
the 1D weak topological index and conductivity in the disorder systems. 

To see this, let us first consider a general $Q$-field configuration without vortex excitations, 
\begin{align}
Q ({\bm r})= \sum^N_{i=1} |p_i({\bm r})\rangle e^{i\phi_i({\bm r})} \langle p_i({\bm r})|. 
\label{no-vortex}
\end{align}
Here $\phi_j({\bm r})$ ($j=1,\cdots,N$) are smooth functions of ${\bm r}$ 
that have no phase windings. In the replica limit with $\lim_{N\rightarrow 0}\int {\rm D}[Q] e^{-S}= \lim_{N\rightarrow 0} 
\langle Z^N\rangle = 1$, the right hand side of Eqs.~(\ref{JLexp-1},\ref{JLexp-2}) can be evaluated exactly, 
\begin{align}
\langle J_{L,x} \rangle = 0, \quad 
\langle J_{L,y} \rangle = -i V\tilde{\nu}_y, \quad 
\langle (\delta J_{L,x})^2 \rangle = - 2V \tilde{\sigma}_x, \quad 
\langle (\delta J_{L,y})^2 \rangle = - 2V \tilde{\sigma}_y, 
\label{mf-result}
\end{align}
with $\frac{\chi_{y}}{8\pi}=-\tilde{\nu}_y$, $\frac{\sigma_{\mu}}{8\pi}
= \tilde{\sigma}_{\mu}$. Eq.~(\ref{mf-result}) means that 
without vortex excitations, $\langle J_{L,\mu}\rangle$ and $\langle (\delta J_{L,\mu})^2\rangle$ reproduce the bare values of the 1D weak topological index [Eq.~(\ref{nu-scb})], and the conductivity [Eqs.~(\ref{bare-sigma-x},\ref{bare-sigma-y})]. In terms of the self-consistent Born approximation, Eq.~(\ref{mf-result}) can be also directly obtained from $\partial  \langle \ln Z[\vec{A}_L] \rangle/ \partial A_{L,\mu}$, and $\partial^2 \langle \ln Z[\vec{A}_L] \rangle/ \partial A^2_{L,\mu}$. 

In the presence of vortex excitations, $\langle J_{L,\mu}\rangle$ 
and $\langle (\delta J_{L,\mu})^2\rangle$ 
acquire non-perturbative renormalizations~\cite{konigMetalinsulatorTransitionTwodimensional2012}. 
To see this, let us include the simplest $Q$-field 
configuration with a pair of vortex and antivortex into the $Q$-field integral  
in Eqs.~(\ref{JLexp-1},\ref{JLexp-2});
\begin{align}
Q({\bm r}) = 1 + |p\rangle (e^{i\phi({\bm r})}-1) \langle p|.
\label{ansatz2}
\end{align}
Here $\phi({\bm r})$ has vortex and antivortex at 
${\bm r} = {\bm r}_{\rm v}$ and ${\bm r}={\bm r}_{\rm av}$ respectively. In the metal phase, 
they are spatially proximate to each other, so that we could assume that they share  
a same eigenvector $|p({\bm r}_{\rm v})\rangle=|p({\bm r}_{\rm av})\rangle \equiv|p\rangle$, and that $|p\rangle$ in Eq.~(\ref{ansatz2}) is independent from ${\bm r}$. With this simplification, the additions to Eqs.~(\ref{JLexp-1},\ref{JLexp-2}) can be calculated 
for $\mu=x$ and $\mu=y$ as follows, 
\begin{align}
\langle J_{L,x} \rangle &= 
- 4 V \tilde{\sigma}_x \int {\rm d }^2{\bm m} \!\ 
m_y \!\ e^{-2 S_v -S_0[|m_x|,|m_y|] + 2\pi i \tilde{\nu}_y m_x} = 0,
\label{jlx} \\
\langle J_{L,y} \rangle & = -i V \tilde{\nu}_y  
+ 4 V \tilde{\sigma}_y \int {\rm d}^2{\bm m} \!\ 
m_x \!\ e^{-2S_v -S_0[|m_x|,|m_y|] + 2\pi i \tilde{\nu}_y m_x} \equiv -i V \tilde{\nu} ^{\prime}_y 
\ne 0,\label{jly} \\
\langle (\delta J_{L,x})^2 \rangle &= - 2V\tilde{\sigma}_x + 
 16\pi V \tilde{\sigma}^2_x \int {\rm d }^2{\bm m} \!\ 
m^2_y \!\ e^{- 2S_v -S_0[|m_x|,|m_y|] + 2\pi i \tilde{\nu}_y m_x} 
\equiv - 2V\tilde{\sigma}^{\prime}_x,
,\label{jlxjlx} \\
\langle (\delta J_{L,y})^2 \rangle & = - 2V\tilde{\sigma}_y + 16\pi V \tilde{\sigma}^2_y \int {\rm d}^2{\bm m} \!\ 
m^2_x \!\ e^{-2S_v-S_0[|m_x|,|m_y|] + 2\pi i \tilde{\nu}_y m_x} \equiv - 2V\tilde{\sigma}^{\prime}_y. 
\label{jlyjly}
\end{align}
Here the integral over $Q$ is replaced by an integral over $|p\rangle$, 
${\bm R} \equiv ({\bm r}_{\rm v}+{\bm r}_{\rm av})/2$, and 
${\bm m} \equiv {\bm r}_{\rm v}-{\bm r}_{\rm av}$ for those integrands 
with $(\int {\rm d}^2{\bm r} \!\ {\rm  Tr}[Q^{-1} 
({\bm r})\partial_{r_{\mu}}Q({\bm r})]) \!\ e^{-S}$ 
or $(\int {\rm d}^2{\bm r} \!\ {\rm  Tr}[Q^{-1}
({\bm r})\partial_{r_{\mu}}Q({\bm r})])^2 \!\ e^{-S}$. 
We also use Eq.~(\ref{eq:SM_volume}) with $\lim_{N\rightarrow 0} N\Gamma(N)=1$. 
$V$ in Eq.~(\ref{jlx}) as well as in the second terms of 
Eqs.~(\ref{jly},\ref{jlxjlx},\ref{jlyjly}) comes from the integral 
over ${\bm R}$. $e^{-Sv}$ stands for a fugacity term of vortex (or 
antivortex). $S_{0}[|m_x|,|m_y|]$ denotes a logarithmic 
attractive interaction between vortex and antivortex. 

The right hand side of Eq.~(\ref{jlx}) is zero, because 
a vortex-antivortex dipole can be polarized either parallel or anti-parallel to 
$\vec{\chi}=(0,\chi_y)$, and the parallel and anti-parallel configurations cancel each other in the right hand side. Generally, $\langle J_{L,x}\rangle=0$ because of the 
statistical mirror symmetry with $(x,y) \rightarrow (-x,y)$ at $t=t_x$.
$\langle J_{L,y}\rangle$ takes a pure imaginary number, 
and the correlations of the non-Hermitian current operators  
$\langle (\delta J_{L,\mu})^2 \rangle$ take real negative values in the metal phase 
with a smaller fugacity term. The real value of $\langle (\delta J_{L,\mu})^2 \rangle$ means that Hermitian part and anti-Hermitian part of $J_{L,\mu}$ have no correlation. The negative value of $\langle (\delta J_{L,\mu})^2 \rangle$ means that the fluctuation of the anti-Hermitian part of $J_{L,\mu}$ dominates over the fluctuation of the Hermitian part. 

The additions in Eqs.~(\ref{jlx}, \ref{jly}, \ref{jlxjlx}, \ref{jlyjly}) are 
nothing but the non-perturbative renormalizations by the vortex-antivortex pair~\cite{konigMetalinsulatorTransitionTwodimensional2012}. To be specific, the spatial 
polarization of the vortex-antivortex pair along a direction perpendicular to $\vec{\nu}$ 
takes a pure imaginary value, and its value renormalizes $\vec{\nu}$ into a smaller 
value as in Eq.~(\ref{jly}). Eqs.~(\ref{jlxjlx},\ref{jlyjly}) show that 
the polarizability of the vortex-antivortex pair along $x$ and $y$ direction respectively 
renormalize $\tilde{\sigma}_y$ and $\tilde{\sigma}_x$ into smaller values. Namely,  $\tilde{\sigma}_{\mu}>\tilde{\sigma}^{\prime}_{\mu}>0$ in the metal phase with the smaller fugacity term.
Thus, the spatial polarization of the vortex excitation 
along $\vec{\chi}=(0,\chi_{y})$ results in a larger reduction of 
$|\langle (\delta J_{L,x})^2 \rangle|$ than the reduction of 
$|\langle (\delta J_{L,y})^2 \rangle|$. Near the phase 
transition point between metal and quasilocalized phases, 
the fugacity term becomes larger, and the polarized vortex excitations are expected to 
yield $|\langle (\delta J_{L,x})^2 \rangle| \ll |\langle (\delta J_{L,y})^2 \rangle|$ in the lattice model. 
A numerical calculation of $|\langle (\delta J_{L,\mu})^2 \rangle|$ in the lattice model is feasible. It is an interesting future direction 
to demonstrate the proliferation of the polarized vortex excitations 
by observing $|\langle (\delta J_{L,x})^2 \rangle| \ll |\langle (\delta J_{L,y})^2 \rangle|$ near the transition point.


\end{widetext}

\end{document}